%% file: main.tex
\documentclass{article}

\usepackage{microtype}
\usepackage{graphicx}
\usepackage{subcaption}

\usepackage{arydshln}
\usepackage{booktabs}
\usepackage{makecell}
\usepackage{tabularray}

\usepackage{hyperref}

\usepackage{algorithm}
\usepackage{algpseudocode}
\usepackage{tabularx}
\usepackage{dblfloatfix}
\usepackage{graphics}

\usepackage[accepted]{icml2024}

\usepackage{amsmath}
\usepackage{amssymb}
\usepackage{mathtools}
\usepackage{amsthm}

\usepackage[capitalize,noabbrev]{cleveref}

\theoremstyle{plain}

\theoremstyle{definition}

\theoremstyle{remark}

\usepackage[textsize=tiny]{todonotes}

\usepackage{enumitem}
\newcolumntype{L}{>{\normalfont}l}
\newcolumntype{C}{>{\normalfont}c}
\newcolumntype{M}{>{\normalfont}m}
\newcolumntype{P}{>{\normalfont}p}

\newcommand{\algoname}[1]{%
    \ifthenelse{\equal{#1}{AND}}{\textsf{#1}}{\oldalgoname{#1}}%
}
\newcommand{\rulesep}{\unskip\ \vrule\ }

\newcommand\blfootnote[1]{%
  \begingroup
  \renewcommand\thefootnote{}\footnote{#1}%
  \addtocounter{footnote}{-1}%
  \endgroup
}

\icmltitlerunning{AND: Audio Network Dissection for Interpreting Deep Acoustic Models}

\begin{document}

\twocolumn[
\icmltitle{AND: Audio Network Dissection for Interpreting Deep Acoustic Models}



\icmlsetsymbol{equal}{*}

\begin{icmlauthorlist}
\icmlauthor{Tung-Yu Wu}{equal,yyy}
\icmlauthor{Yu-Xiang Lin}{equal,yyy}
\icmlauthor{Tsui-Wei Weng}{sss}
\end{icmlauthorlist}

\icmlaffiliation{yyy}{National Taiwan University, Taipei, Taiwan}
\icmlaffiliation{sss}{HDSI, UC San Diego,
CA, USA}

\icmlcorrespondingauthor{Tsui-Wei Weng}{lweng@ucsd.edu}

\icmlkeywords{Explainable AI, Neuron-level Interpretation, Audio Network}

\vskip 0.3in
]



\printAffiliationsAndNotice{\icmlEqualContribution} 
\input{0_Abstract}
\input{1_Introduction}

\input{2_Related_Work}
\input{3_Method}

\input{4_Experiments}
\input{5_Conclusion}

\section*{Impact Statement}
This paper presents work whose goal is to advance the field of Machine Learning. Particularly, this work targets acoustic model interpretability, with a network dissection tool developed, aiming to help the community gain deeper knowledge of the properties of acoustic neural networks.

\section*{Acknowledgement}
The authors thank the anonymous reviewers for insightful feedback and suggestions. 
This work was supported in part by National Science Foundation (NSF) awards CNS-1730158, ACI-1540112, ACI-1541349, OAC-1826967, OAC-2112167, CNS-2100237, CNS-2120019, the University of California Office of the President, and the University of California San Diego's California Institute for Telecommunications and Information Technology/Qualcomm Institute. Thanks to CENIC for the 100Gbps networks. T.-W. Weng is supported by National Science Foundation under Grant No. 2107189 and 2313105. T.-W. Weng also thanks the Hellman Fellowship for providing research support.

\bibliography{references}
\bibliographystyle{icml2024}

\newpage
\appendix
\onecolumn
\newpage
\definecolor{hyperblue}{RGB}{0,0,110}
\begin{enumerate}[label=\textcolor{hyperblue} {\textbf{\Alph*}}]
  \item \hyperref[sup sec: implementation details]{\textbf{Implementation Details}} \hspace*{\fill} \textbf{\pageref{sup sec: implementation details}}
    \begin{enumerate}[label=\textcolor{hyperblue}{A.\arabic*.}]
      \item \hyperref[sup subsec: implementation details of audio captioning models]{Audio Captioning Model}\dotfill \pageref{sup subsec: implementation details of audio captioning models}
      
      \item \hyperref[sup subsec: implementation details of LLMs]{Large Language Model} \dotfill \pageref{sup subsec: implementation details of LLMs}
      
      \item \hyperref[sup subsec: implementation details of in-context learning]{In-context Learning} \dotfill \pageref{sup subsec: implementation details of in-context learning}
      \item \hyperref[sup subsec: implementation details of others]{Others} \dotfill \pageref{sup subsec: implementation details of others}
    \end{enumerate}
  \item \hyperref[sup sec: Qualitative Results]{\textbf{Qualitative Results}} \hspace*{\fill} \textbf{\pageref{sup sec: Qualitative Results}}
    \begin{enumerate}[label=\textcolor{hyperblue}{B.\arabic*.}]
      \item \hyperref[sup subsec: dissection pipeline]{Dissection Pipeline}\dotfill \pageref{sup subsec: dissection pipeline}

    \item \hyperref[sup subsec: outputs from different modules]{Outputs from Different Modules}\dotfill \pageref{sup subsec: outputs from different modules}
    \end{enumerate}

  \item \hyperref[sup sec: middle layer analysis from basics]{\textbf{Middle Layer Analysis from Basic Acoustic Properties}} \hspace*{\fill} \textbf{\pageref{sup sec: middle layer analysis from basics}}
  \item \hyperref[sup sec: adj distribution]{\textbf{Adjective Distribution of Different Target Networks}} \hspace*{\fill} \textbf{\pageref{sup sec: adj distribution}}
  \item \hyperref[sup sec: detailed results of similarity functions]{\textbf{Detailed Results of Last Layer Dissection}} \hspace*{\fill} \textbf{\pageref{sup sec: detailed results of similarity functions}}
  
  \item \hyperref[sup sec: neuron interpretability]{\textbf{Audio Machine Unlearning Example}} \hspace*{\fill} \textbf{\pageref{sup sec: audio_machine_unlearning_example}}
  
  \item \hyperref[sup sec: neuron interpretability]{\textbf{Neuron Interpretability}} \hspace*{\fill} \textbf{\pageref{sup sec: neuron interpretability}}
    \begin{enumerate}[label=\textcolor{hyperblue}{G.\arabic*.}]
      \item \hyperref[sup subsec: interpretability algorithm]{Algorithm} \dotfill \pageref{sup subsec: interpretability algorithm}
      \item \hyperref[sup subsec: clustering]{Clusters in ESC50} \dotfill \pageref{sup subsec: clustering}
      \item \hyperref[sup subsec: interpretability threshold]{Measure Neuron Interpretability when adopting different $\tau$ and top-$K$} \dotfill \pageref{sup subsec: interpretability threshold}
      \item \hyperref[sup subsec: gtzan neuron Interpretability]{Neuron Interpretability under Different Training Strategies - GTZAN Music Genre} \dotfill \pageref{sup subsec: gtzan neuron Interpretability}
    \end{enumerate}
\end{enumerate}

\input{supp/A_implementation_details}

\clearpage
\newpage
\input{supp/B_qualitative_results}
\clearpage
\newpage
\input{supp/C_middle_layer_analysis_from_basic_acoustic_properties}
\input{supp/D_adjective_distribution_of_different_target_networks}
\input{supp/E_detailed_results_of_last_layer_dissection}

\input{supp/F_audio_machine_unlearning_example}
\clearpage
\newpage
\input{supp/G_neuron_interpretability}
\end{document}

%% file: 0_Abstract.tex
\begin{abstract}
Neuron-level interpretations aim to explain network behaviors and properties by investigating neurons responsive to specific perceptual or structural input patterns. Although there is emerging work in the vision and language domains, none is explored for acoustic models. To bridge the gap, we introduce \algoname{AND}, the first \textbf{A}udio \textbf{N}etwork \textbf{D}issection framework that automatically establishes natural language explanations of acoustic neurons based on highly-responsive audio. \algoname{AND} features the use of LLMs to summarize mutual acoustic features and identities among audio. Extensive experiments are conducted to verify  \algoname{AND}'s precise and informative descriptions. In addition, we demonstrate a potential use of  \algoname{AND} for audio machine unlearning by conducting concept-specific pruning based on the generated descriptions. Finally, we highlight two acoustic model behaviors with analysis by  \algoname{AND}: (i) models discriminate audio with a combination of basic acoustic features rather than high-level abstract concepts; (ii) training strategies affect model behaviors and neuron interpretability -- supervised training guides neurons to gradually narrow their attention, while self-supervised learning encourages neurons to be polysemantic for exploring high-level features.
\end{abstract}

%% file: 1_Introduction.tex
\section{Introduction}
Deep Neural Networks (DNNs) have achieved remarkable success across various tasks. However, their inherent nature of high non-linearity poses great challenges in understanding model behaviors and neuron functionalities. To tackle this longstanding issue, several lines of work have attempted to acquire deeper knowledge about DNNs, ranging from decision explanation for the input~\cite{simonyan2013deep, selvaraju2017grad, chattopadhay2018grad}, property observation of layer-wise features~\cite{pasad2021layer, pasad2023comparative}, to neuron-level interpretations~\cite{bau2017network, hernandez2021natural, oikarinen2022clip, bills2023language, lee2023importance, anonymous2024tellme, bai2024describe}.

Among them, neuron-level interpretation tools automatically analyze the functionality of the neurons inside a model. One classical strategy is to examine each neuron's activation to inputs with different perceptual, structural, and semantic patterns ~\cite{hernandez2021natural, oikarinen2022clip, kalibhat2023identifying}. Knowing neuron's responsive features brings benefits to understanding fine-grained model behaviors. For instance, FALCON~\cite{kalibhat2023identifying} observes that a group of neurons with varied responsive features may together build a more concise and interpretable feature; MILAN~\cite{hernandez2021natural}, which explains neurons with natural language descriptions, finds that visual neurons in shallow layers requires more adjectives to describe, and these neurons substantially affect model performance. In addition, neuron-level interpretability can also be applied to tasks that require neuron-level knowledge, such as LLM factual editing~\cite{meng2022locating}, anonymized models auditing and spurious features editing~\cite{hernandez2021natural}.

Nevertheless, current neuron-level interpretability frameworks are all designed for visual~\cite{bau2017network, hernandez2021natural, oikarinen2022clip, bai2024describe} and language modalities~\cite{bills2023language,lee2023importance}, rendering them incompatible with acoustic models. For example, some rely on external vision models such as image segmentation~\cite{bau2020understanding} and CLIP~\cite{oikarinen2022clip}. While sound event detection models~\cite{nam22_interspeech, shao2023fine} could offer timestamp annotations analogous to pixel-wise image segmentation, current models suffer from narrow detectable classes due to smaller dataset scale~\cite{turpault2019sound, serizel2020sound}. This limitation also hinders the development of text-audio constrastive models~\cite{guzhov2022audioclip, laionclap2023}, causing difficulties for the direct transfer of previous interpretability tools from vision modality to audio modality.

Hence, to fill in the gap, in this paper we present  \algoname{AND}\blfootnote{Our source code is available at \href{https://github.com/Trustworthy-ML-Lab/Audio_Network_Dissection}{https://github.com/Trustworthy-ML-Lab/Audio\_Network\_Dissection}}, an elegant and effective framework for \textbf{A}udio \textbf{N}etwork \textbf{D}issection using natural language descriptions. \algoname{AND} features an LLM-based pipeline, along with three specialized proposed modules, to capture responsive acoustic features of neurons in an audio network. The outputs of \algoname{AND}, including the closed-set concept, open-set concept, and LLM-generated summary, serve as mediums to inquire activities and properties of an audio network. Extensive experiments are presented in \cref{sec: experiments} to showcase the quality and usefulness of \algoname{AND}'s outputs. In particular, in \cref{subsec: Last Layer Dissection Accuracy}, we validate the efficacy of closed-concept identification by considering last-layer network dissection. In \cref{subsec:Human Evaluation}, we provide human evaluation to verify closed-concept identification and summary calibration. In \cref{subsec: Audio Machine Unlearning}, we discuss the middle-layer concept-specific pruning using \algoname{AND} by demonstrating a potential use of \algoname{AND} for audio machine unlearning, achieved by leveraging the open-concept identification module. Finally, we investigate critical properties of audio networks with \algoname{AND} in \cref{subsec: Analyzing Acoustic Feature Importance} and \cref{subsec: Training Strategy Affects Neuron Interpretability}. In particular we analyze feature importance of different acoustic features by looking into parts of speech (POS)~\cite{kumar2017discovering} in \cref{subsec: Analyzing Acoustic Feature Importance}, and argue the influence of different training strategies on neuron and model behaviors in \cref{subsec: Training Strategy Affects Neuron Interpretability}.

In conclusion, this work's contributions are threefold:
\begin{itemize}
    \item We present \algoname{AND}, the first automatic \textbf{A}udio \textbf{N}etwork \textbf{D}issection framework to provide natural language descriptions of acoustic neurons activities. Extensive experiments are conducted to verify \algoname{AND}'s efficacy.
    \item We showcase the effect of middle-layer concept-specific pruning conducted by \algoname{AND} and discuss its relations to audio machine unlearning as a potential use case of \algoname{AND}.
    \item We leverage \algoname{AND} to analyze the feature importance of different acoustic features and the influence of training strategy on model behaviors as well as neuron interpretability.
\end{itemize}

%% file: 2_Related_Work.tex
\section{Related Work}
\subsection{Neuron-level Interpretations}
\begin{figure*}[!tp]
  \centering
\includegraphics[width=\textwidth]{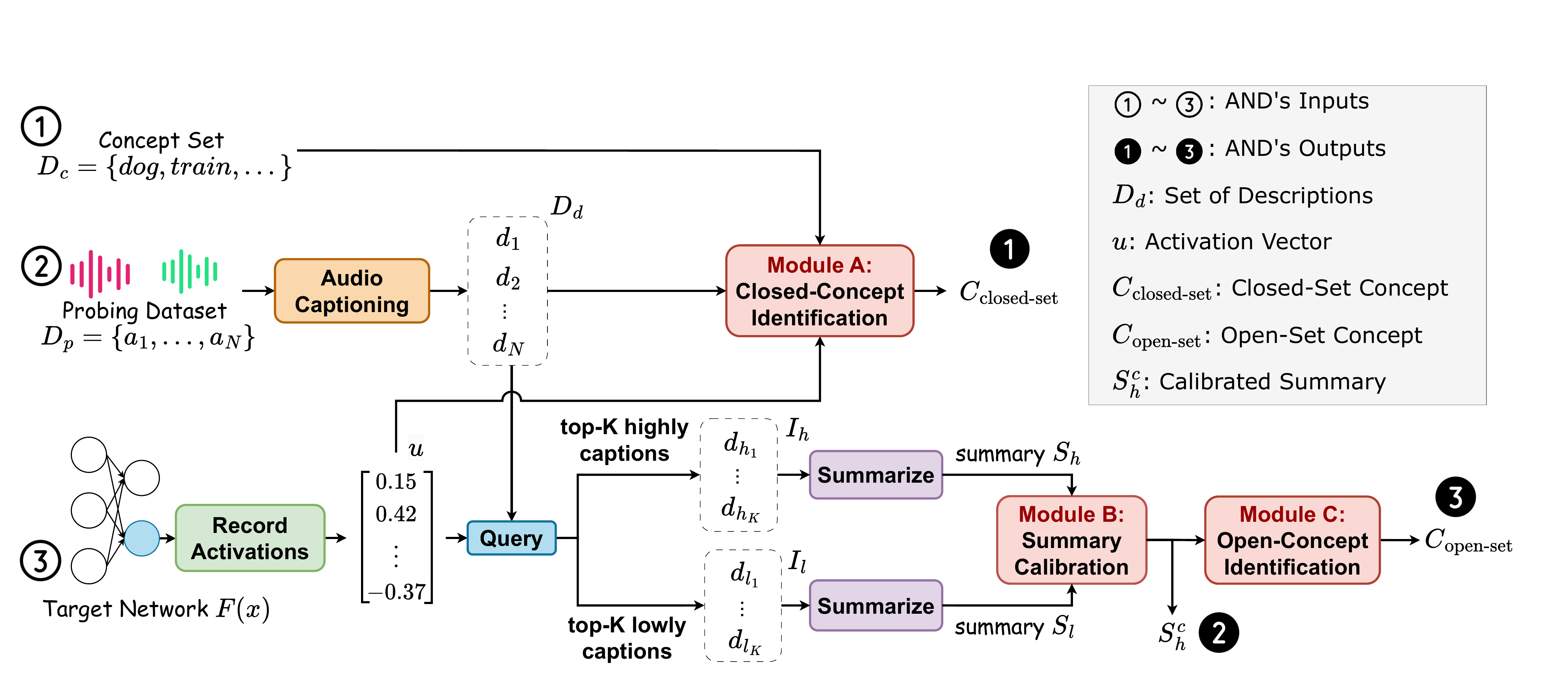}
  \caption{The proposed framework of \algoname{AND}. Taking concept set $D_{c}$, probing dataset $D_{p}$, and target network $F(\cdot)$ as inputs, \algoname{AND} employs a coarse-to-fine LLM-based pipeline to analyze each neuron's highly-responsive acoustic concepts by three specialized modules: (A) closed-concept identification, (B) summary calibration, and (C) open-concept identification. Closed-ended concept $C_{\text{closed-set}}$, calibrated summary $S^{c}_{h}$, and open-ended concept $C_{\text{open-set}}$ are generated as outputs of \algoname{AND}.}
  \label{fig:main framework}
\end{figure*}
Neuron-level Interpretations aim to investigate neurons' activities to acquire fine-grained knowledge of deep networks as well as high-level network properties by inquiring groups or layers of neurons. Network Dissection ~\cite{bau2017network} makes the first attempt to analyze visual models' behaviors by automatically exploring individual neurons' functionalities, with a line of follow-up work ~\cite{hernandez2021natural, oikarinen2022clip, kalibhat2023identifying, anonymous2024tellme,bai2024describe}. In particular, MILAN ~\cite{hernandez2021natural} utilizes LSTMs to directly generate natural language descriptions conditioned on a set of patched highly-activated images. To remove the need of concept labels in ~\cite{bau2017network, hernandez2021natural} and accelerate the dissection process, CLIP-Dissect ~\cite{oikarinen2022clip} identify neuron concepts by computing similarity between neuron activations and representative features of open-vocabulary concept sets through probing dataset and CLIP. FALCON ~\cite{kalibhat2023identifying} identifies critical concepts by scoring nouns, verbs and adjectives in predefined image descriptions of highly-activated images. Additionally, lowly-activated images are also utilized to address spurious and vague concepts, with overlapping concepts between the two sets being removed. Recently, LLM-based open-domain automatic description generation ~\cite{anonymous2024tellme, bai2024describe} is also proposed for interpreting visual models. Besides 
dissecting vision networks, there has being growing interest to interpret LLMs~\cite{meng2022locating, bills2023language, lee2023importance}. ROME~\cite{meng2022locating} employs causal mediation analysis (CMA) to locate neurons influential to specific factual knowledge. \cite{bills2023language} proposes to use LLMs to generate the explanations for each neuron in language models and assess dissecting qualities. \cite{lee2023importance} further improves~\cite{bills2023language} by proposing efficient prompting techniques to obtain high quality neuron descriptions at lower computation cost.

\subsection{Audio and Speech Network Interpretability}
Previous efforts for interpreting audio and speech models primarily focus on input-specific explanations~\cite{mishra2017local, becker2018interpreting} and layer-wise analysis~\cite{pasad2021layer, pasad2023comparative, li2023dissecting}. The former utilizes local interpretability tools, such as local interpretable model-agnostic explanations (LIME)~\cite{ribeiro2016should} and layer-wise relevance propagation (LRP)~\cite{bach2015pixel} to visualize models' attention to the MFCC or mel-spectrogram of a specific audio. The latter, mostly conducted on self-supervised learning (SSL) speech models, examines the acoustic, phonetic, and word-level properties encoded in the representations of each transformer layer by correlation-based analysis. Specifically, it is found that models pretrained with hidden discrete units prediction, such as HuBERT~\cite{hsu2021hubert} and WavLM~\cite{chen2022wavlm}, learn to encode richer phonetic and word information in deeper transformer layers~\cite{pasad2023comparative}. While layer-wise findings provide reliable guidance for utilizing the embeddings of pretrained SSL models to downstream speech tasks, they provide limited utility for neuron-level tasks, such as model editing and unlearning. On the other hand, previous work~\cite{kumar2017discovering} have pointed out relations between acoustic concepts and natural languages, with a large set of audio concepts such as ``glass breaking" and ``sound of honking cars" exploited through a designed pipeline including techniques such as part-of-speech (POS) tagging. Inspired by this, in contrast to existing research, we propose the first neuron-level description-based interpretability tool, \algoname{AND}, to understand audio network behaviors from a more fine-grained aspect but with informative natural language descriptions. Our framework is elegant and instructive, as described in~\cref{sec: main framework} and verified in~\cref{sec: experiments} with extensive experiments.

%% file: 3_Method.tex
\section{Audio Network Dissection (AND)}
\label{sec: main framework}
\subsection{Framework Overview}

\textbf{Inputs and Outputs} \enspace As shown in~\cref{fig:main framework}, \algoname{AND} takes a target network $F(\cdot)$, a predefined concept set $D_{c}$, $|D_{c}|=M$, with concepts $c_{1}, \ldots, c_{M}$, and a probing dataset $D_{p}$, $|D_{p}|=N$, with audio clips $a_{1}, \ldots, a_{N}$, as inputs. \algoname{AND} then dissects neurons by observing and summarizing the shared acoustic properties and identities among top-$K$ highly-activated audio. We design an LLM-based pipeline with three specialized modules in \algoname{AND}: (A) closed-concept identification, (B) summary calibration, and (C) open-concept identification. \algoname{AND} provides 3 types of output corresponding to each module: a set of closed-set concepts $C_{\text{closed-set}}$ selected out of $D_c$, a natural language summary $S^{c}_{h}$ describing commonalities among highly-activated audio, and a set of open-set concepts $C_{\text{open-set}}$ acquired from $S^{c}_{h}$. A dissection example of \algoname{AND} is provided in~\cref{sup sec: Qualitative Results}.

\textbf{Pipeline} \enspace  \algoname{AND} leverages closed-concept identification, summary calibration, and open-concept identification (marked as \textbf{Module A}, \textbf{Module B}, and \textbf{Module C} in~\cref{fig:main framework}), to acquire closed-concept $C_{\text{closed-set}}$, summary $S^{c}_{h}$, and open-concept $C_{\text{open-set}}$. Inputs of the three modules are acquired from concept set $D_{c}$, probing dataset $D_{p}$, and target network $F(\cdot)$, which are three original inputs of \algoname{AND}.

\begin{figure*}[htbp]
    \begin{subfigure}[b]{0.3\textwidth}
        \includegraphics[width=0.98\textwidth]{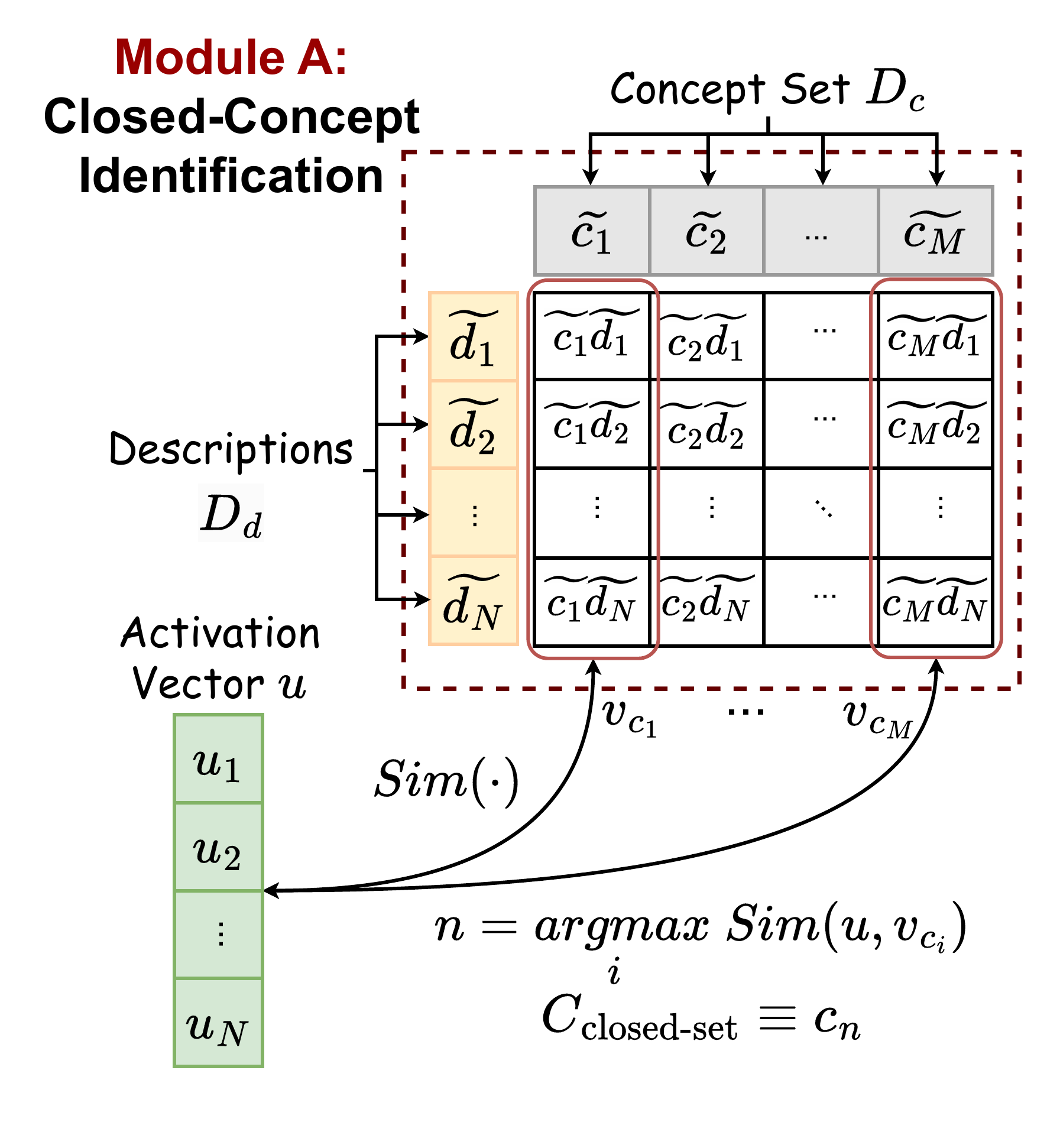}
        \subcaption{Illustration of closed-concept identification. Audio are represented as its descriptions rather than waveform feature embeddings in CLAP~\cite{laionclap2023}.}
        \label{subfig: closed concept identification}
    \end{subfigure}
    \hfill
    \rulesep
    \begin{subfigure}[b]{0.65\textwidth}
        \includegraphics[width=0.98\textwidth]{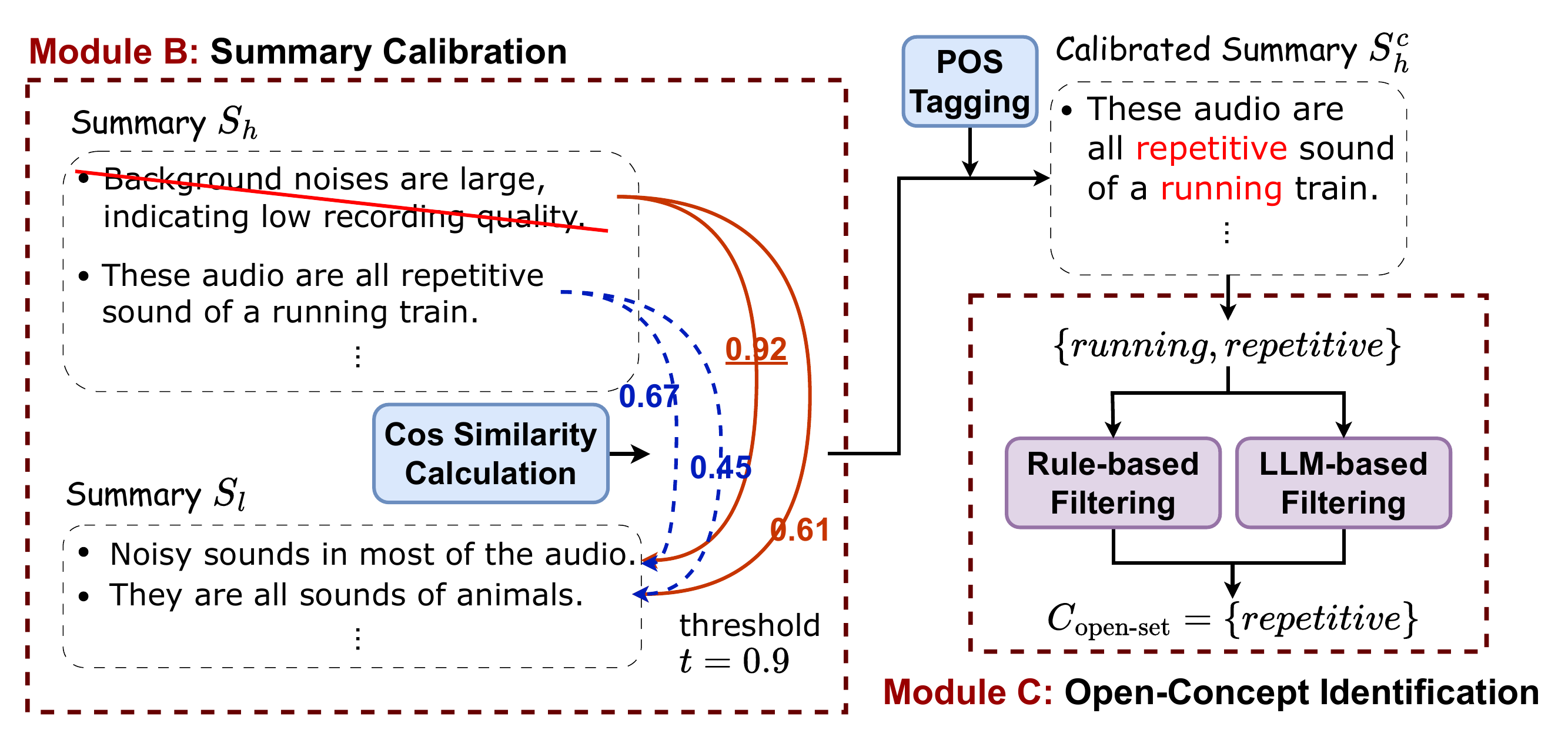}
        \subcaption{Illustration of summary calibration and open-concept identification. Pairwise sentence similarity between sentences of $S_{h}$ and $S_{l}$ is computed to remove homogeneous concepts and generate calibrated summary $S^{c}_{h}$. Critical concepts in $S^{c}_{h}$ are captured by filtering out concepts unrelated to acoustics (e.g. "running"). Adjectives are used as an example here.}
        \label{subfig: summary calibration and open-concept identification}
    \end{subfigure}

\caption{Detailed illustration of \algoname{AND}'s three specialized modules: (A) closed-concept identification, (B) summary calibration, and (C) open-concept identification.}
\label{fig: framework module}
\end{figure*}

Specifically, closed-concept identification (module A) takes $D_{c}$, activation vector $u=[u_{1}, \ldots, u_{N}]^T$, and descriptions $D_{d}=\{d_{1}, \ldots, d_{N}\}$ as inputs, where $d_{i}$ is the open-domain description of audio $a_{i}$ generated by an audio captioning model; $u_{i}$ is the activation value of interested neuron on audio $a_{i}$. Closed-concept identification generates $C_{\text{closed-set}}$ as the output. Summary calibration (module B) takes $S_{h}$ and $S_{l}$ as inputs, which are summaries of top-$K$ highly-activated and lowly-activated audio, with calibrated summary $S^{c}_{h}$ being the output. The retrieval of highly-activated and lowly-activated audio is achieved by querying $D_{d}$ based on the values in $u$. Finally, $S^{c}_{h}$ serves as inputs for open-concept identification (module C) to extract critical concepts as $C_{\text{open-set}}$. We elaborate each module in~\cref{sub sec: closed-concept identification}-\ref{sub sec: open-concept identification} respectively, with detailed illustration in \cref{fig: framework module}.


\subsection{Input Preprocessing for Each Module}
\label{sub sec: input preprocessing}
This section describes the preprocessing of \algoname{AND}'s inputs to generate inputs for three specialized modules. In particular we discuss the generation of $D_{d}$, $S_{h}$, and $S_{l}$.

To obtain caption $D_{d}$ of audio clips in $D_{p}$, we adopt SALMONN~\cite{tang2023salmonn} as the open-domain audio captioning model. We feed each audio $a_{i}$ into SALMONN to acquire $d_{i}$, forming $D_{d}=\{d_{1}, \ldots, d_{N}\}$. For $S_{h}$ and $S_{l}$, given an acoustic neuron $f(\cdot)$ in $F(\cdot)$, $u_i = f(a_i)$ is the activation value of the neuron for the audio clip $a_i$. It is the descriptions of audio with top-$K$ highest and lowest activation values that constitute the high-activation descriptions $I_{h}$:
 \[
 I_{h} = \{d_{i} \mid u_i \hspace{1.5mm} \textit{is of top-K highest value in u}\},
 \]
 and the low-activation descriptions $I_{l}$:
 \[
  I_{l} = \{d_{i} \mid u_i \hspace{1.5mm} \textit{is of top-K lowest value in u}\}.
 \]
$I_{h}$ and $I_{l}$ are subsequently used to generate the summary $S_{h}$ and $S_{l}$ by instructing Llama-2-chat-13B~\cite{touvron2023llama} to summarize and list down the commonalities among these audio descriptions in the set. The generated $D_{d}$, $S_{h}$, and $S_{l}$ serve as inputs for implementing closed-concept identification (module A), summary calibration (module B), and open-concept identification (module C), which are detailed in~\cref{sub sec: closed-concept identification},~\cref{sub sec: summary calibration}, and~\cref{sub sec: open-concept identification}.


\subsection{Module A: Closed-Concept Identification}
\label{sub sec: closed-concept identification}
Closed-concept identification takes concept set $D_{c}=\{c_{1},\ldots,c_{M}\}$, set of descriptions $D_{d}=\{d_{1},\ldots,d_{N}\}$, and activation vector $u$ as inputs. The output is a concept $C_{\text{closed-set}}$ from $D_{c}$ to label the interested neuron, similar to CLIP-Dissect's spirits. Note that the term "closed-set" is due to pre-defined concept set, but it can be an open vocabulary set as needed.

For each concept $c_{i}$, \algoname{AND} generates the representative feature vector $v_{c_{i}}$, where $[v_{c_{i}}]_j=\widetilde{c_{i}} \cdot \widetilde{d_{j}}, j = 1, \dots, N $, as shown in~\cref{subfig: closed concept identification}. CLIP's text encoder $E_{text}(\cdot)$ is used to encode concept $c_i$ into $\widetilde{c_{i}}$ and descriptions $d_j$ into $\widetilde{d_{j}}$. That is, $\widetilde{c_{i}} = E_{text}(c_i)$ and $\widetilde{d_{j}} = E_{text}(d_j)$. Then, a similarity function $Sim(\cdot)$, such as cosine similarity or weighted pointwise mutual information (WPMI), is applied to measure the similarity between $u$ and $v_{c_{i}}$. The concept with the highest similarity score is considered the most-matched closed-set concept, denoted as $C_{\text{closed-set}}$. We refer to this proposed approach as the description-based (DB) method.

In addition to DB, we present two straightforward alternative approaches than~\cref{subfig: closed concept identification} to achieve closed-concept identification. The first idea is using in-context learning (ICL) to query the LLM to directly select a concept from $D_{c}$ that best matches information in the calibrated summary $S^{c}_{h}$. We provide the instruction and examples for ICL in~\cref{sup subsec: implementation details of in-context learning} The second solution is to adopt a text-audio contrastive learning model, such as CLAP, and perform cross-modal retrieval as CLIP-Dissect~\cite{oikarinen2022clip} does but in a text-audio-based (TAB) manner. Notably, the TAB method can be considered as the variant of CLIP-Dissect for the acoustic model. Performance comparisons of ICL, DB, and TAB methods are presented in~\cref{subsec: Last Layer Dissection Accuracy} and~\cref{tab:last_layer_acc}.


\subsection{Module B: Summary Calibration}
\label{sub sec: summary calibration}
As shown in \cref{subfig: summary calibration and open-concept identification}, the summary calibration module takes high-activation summary $S_{h}$ and low-activation summary $S_{l}$ as inputs. $S_{l}$ is used to filter out spurious concepts that are ambiguous or hallucinating in $S_{h}$. For instance, in scenarios where all audio in $D_{p}$ contain no noise, $S_{h}$ and $S_{l}$ would hold sentences emphasizing the clarity of sound. We then remove this redundant information in $S_{h}$. In the calibration process, for each mentioned point $p$ in $S_{h}$, we compute the cosine similarity between its sentence feature~\cite{reimers-2019-sentence-bert} and those in $S_{l}$. If the similarity value between $p$ and any point in $S_{l}$ exceeds a predefined threshold $t$, we discard $p$. The calibrated $S_{h}$ is the output of this module, denoted as $S^{c}_{h}$.

\subsection{Module C: Open-Concept Identification}
\label{sub sec: open-concept identification}
As illustrated in~\cref{subfig: summary calibration and open-concept identification}, the open-concept identification takes calibrated summary $S^{c}_{h}$ as the input and extracts open-set concept $C_{\text{open-set}}$. Then naming of ``open-set'' is because $C_{\text{open-set}}$ is derived from the open-domain summary $S^{c}_{h}$.

\begin{figure}[h!]
\vskip 0.2in
\centerline{
\includegraphics[width=0.9\columnwidth]{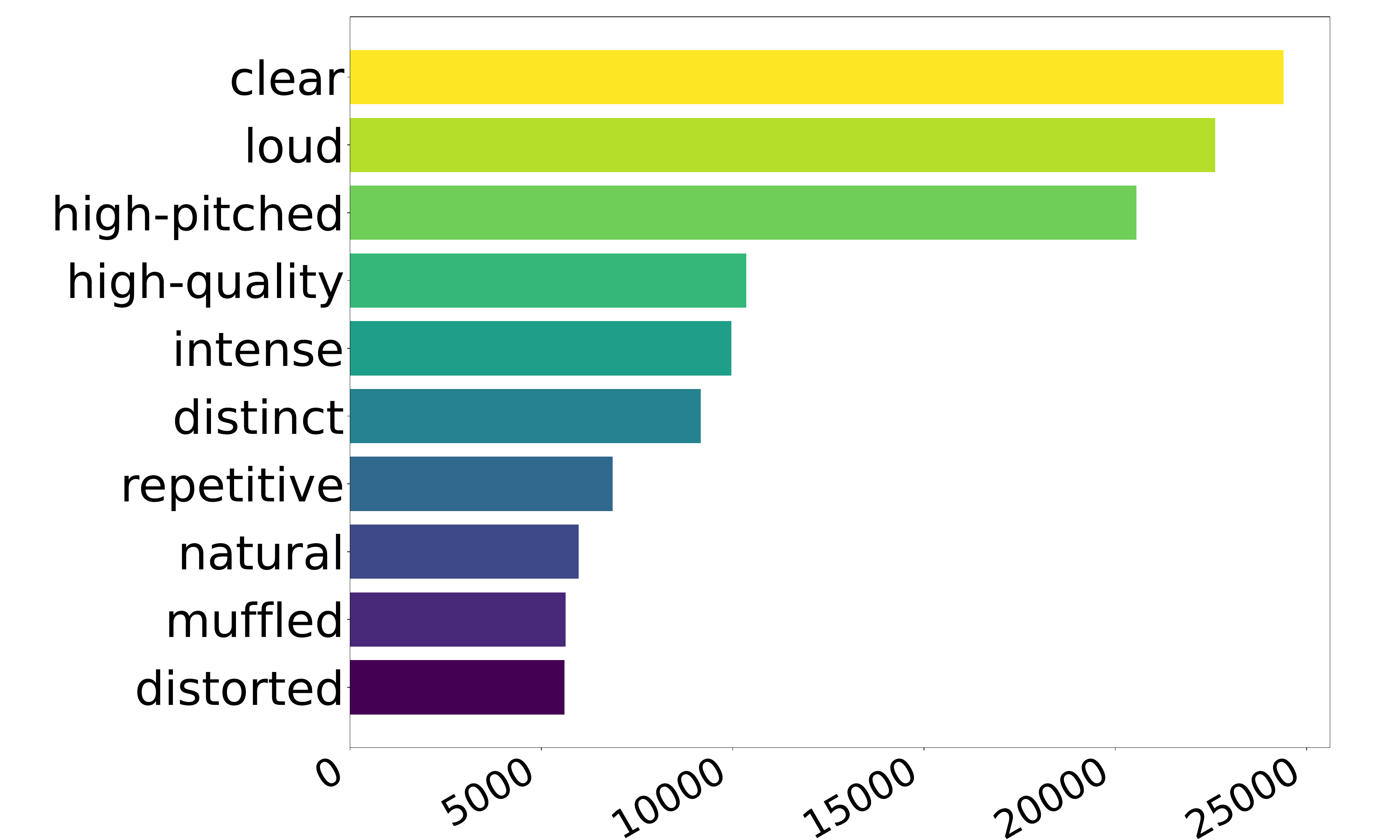}
}
  \caption{Counting of adjectives for AST's all linear layer neurons generated by open-concept identification module. We show the top-10 most-used adjectives here.}
  \label{fig: adj counting dist}
\end{figure}

Open-concept identification aims to identify critical concepts that are related to acoustics. This is implemented by applying designed filtering to $S^{c}_{h}$. Using adjectives as an example, we first apply POS tagging to draw out all adjectives from $S^{c}_{h}$, while some may be irrelevant to acoustics such as the word ``running". Second, we leverage rule-based and LLM-based filtering to remove trivial adjectives. The former eliminates common stop words, while the latter queries Llama-2-chat-13B to determine whether the adjective describes acoustic properties. For LLM-based filtering, a designed hard prompt is used\footnote{In practice, after applying POS tagging to all neurons' calibrated summaries, \algoname{AND} forms a universal set of adjectives. Llama-2-chat-13B is then used to examine adjectives in the set to prevent repeated queries to the same words.}. The resulting filtered set $C_{\text{open-set}}$ comprises sound-related concepts, reflecting shared acoustic properties among top-$K$ highly-activated audio.~\cref{fig: adj counting dist} displays the distribution of top-10 most common adjectives for all linear layer neurons within an Audio Spectrogram Transformer (AST)~\cite{gong21b_interspeech} trained on the ESC50~\cite{piczak2015esc} dataset. More results are in~\cref{sup sec: adj distribution}.

%% file: 4_Experiments.tex
\section{Experiments}
\label{sec: experiments}

\renewcommand{\arraystretch}{1.6}

\renewcommand{\arraystretch}{1}

\textbf{Overview} \enspace This section aims to verify the dissection quality of \algoname{AND} and investigate acoustic model behaviors using \algoname{AND}. First, we validate the dissection quality of \algoname{AND} in~\cref{subsec: Last Layer Dissection Accuracy} before diving into model analysis. Following CLIP-Dissect,~\cref{subsec: Last Layer Dissection Accuracy} measures last layer dissection accuracy as an indicator of the interpretation quality of \algoname{AND}'s closed-concept identification module (module A) because neurons in the output layer have 1-1 corresponding with classes in the training dataset and thus can serve as ground-truth. In~\cref{subsec:Human Evaluation}, we conduct a human evaluation to assess \algoname{AND}'s performance on middle-layer neurons with model-unseen concepts. Also ,we provide qualitative results  in~\cref{sup sec: Qualitative Results}.

Next, we verify $C_{\text{open-set}}$ from module C as well as introduce a use case of \algoname{AND} regarding machine unlearning~\cite{golatkar2020eternal, nguyen2022survey, zhang2022prompt} in~\cref{subsec: Audio Machine Unlearning}. Notably, classical strategies for machine unlearning involve ablating individual neurons that are influential to the target concept~\cite{pochinkov2024dissecting}. While there exist works focusing on unlearning vision and language models~\cite{meng2022locating, gandikota2023erasing, zhang2023forget}, audio machine unlearning remains relatively underexplored. To achieve this challenging task, we consider linguistically parsing each neuron's $S_{h}^{c}$ generated by \algoname{AND}'s module B and leverage acquired $C_{\text{open-set}}$ to conduct neuron pruning based on a given target concept. More concretely, we prune all of the neurons dissected with the target concept and examine audio models' perception abilities towards both related and unrelated concepts after ablation. Results are reported in~\cref{subsec: Audio Machine Unlearning}.

Finally,~\cref{subsec: Analyzing Acoustic Feature Importance} and~\cref{subsec: Training Strategy Affects Neuron Interpretability} are analyses and findings of acoustic model behaviors using \algoname{AND}. In~\cref{subsec: Analyzing Acoustic Feature Importance}, we follow MILAN's experimental settings but on acoustic networks with \algoname{AND} to investigate how it is different from or the same as vision network's observations discovered in MILAN. In~\cref{subsec: Training Strategy Affects Neuron Interpretability}, we analyze neuron polysemanticity (uninterpretability)~\cite{mu2020compositional, oikarinen2022clip, bills2023language}, i.e., neurons' diverse attention to seemingly unrelated features, affected by network training strategies.

\begin{table*}[tp]
\begin{center}
\begin{small}
\begin{sc}
\addtolength{\tabcolsep}{-1pt}
\caption{Training settings of audio models used in the experiments.}
\label{tab: Training settings of the audio models}
\vskip 0.1in
\begin{tabular}{C|C  C  C}
\toprule
\midrule
 & AST & BEATs-finetuned & BEATs-frozen\\
\midrule
Pretraining 
on Audioset  & Supervised  
(all layers) & Self-supervised  
(all layers) & Self-supervised  
(all layers) \\
\midrule 
\normalfont { Fine-tuning  
on ESC50 }& Supervised 
(all layers) & Supervised 
(all layers)
 & Supervised 
(final linear layer) \\ 
\midrule
\bottomrule
\end{tabular}
\end{sc}
\end{small}
\end{center}
\vskip -0.1in

\end{table*}

\begin{table*}[!tp]
\begin{center}
\begin{small}
\begin{sc}
\addtolength{\tabcolsep}{-1pt}
\caption{Last layer network dissection accuracy of AST, BEATs-frozen, and BEATs-finetuned on the ESC50 dataset, with the highest performance marked in bold. As discussed in~\cref{sub sec: closed-concept identification}, ICL refers to querying LLM to choose a best-matched concept for the summary, and TAB/DB refers to the text-audio-based/description-based method. DB achieves the best results among all metrics and models.}
\vskip 0.1in
\label{tab:last_layer_acc}
\begin{tabular}{L|CCC|CCC|CCC}
\toprule
\midrule
Model & \multicolumn{3}{C|}{AST} & \multicolumn{3}{C|}{BEATs-finetuned} & \multicolumn{3}{C}{BEATs-frozen}\\
\midrule
Method & Top-1 Acc & Top-5 Acc & Cos & Top-1 Acc & Top-5 Acc & Cos & Top-1 Acc & Top-5 Acc & Cos\\
\midrule
\algoname{AND} (module A: ICL)    & 72.0 & 60.0 & 0.83 & 52.0& 56.0 & 0.68& 16.0 & 16.0  & 0.37 
\\
\algoname{AND} (module A: TAB) & 96.0& \textbf{100.0} & 0.98& 74.0& 92.0& 0.82& 46.0& 72.0& 0.60
\\
\algoname{AND} (module A: DB)     & \textbf{100.0} & \textbf{100.0} & \textbf{1.00}& \textbf{76.0}& \textbf{100.0}& \textbf{0.83}& \textbf{58.0}& \textbf{82.0}& \textbf{0.69}\\
\midrule
\bottomrule
\end{tabular}
\end{sc}
\end{small}
\end{center}
\end{table*}

For experiments in~\cref{subsec: Last Layer Dissection Accuracy}-~\cref{subsec: Training Strategy Affects Neuron Interpretability}, we utilize the ESC50~\cite{piczak2015esc} as $D_{p}$ and consider all its 50 audio classes as $D_{c}$. The only exception is that we use open-ended acoustic concept set proposed by \citet{kumar2017discovering} as $D_{c}$ in ~\cref{subsec:Human Evaluation}. For the target networks $F(\cdot)$, as shown in~\cref{tab: Training settings of the audio models}, we select the Audio Spectrum Transformer (AST)~\cite{gong21b_interspeech} and BEATs~\cite{pmlr-v202-chen23ag}. BEATs is an SSL audio model pretrained with Masked Audio Modeling (MLM)~\cite{pmlr-v202-chen23ag}, which allows us to use either by finetuning the whole model (BEATs-finetuned) or training only the last linear layer (BEATs-frozen). We adopt both versions. These target networks are trained on the ESC50, achieving testing accuracies of 95.0\% for AST, 89.75\% for BEATs-finetuned, and 84.75\% for BEATs-frozen, respectively.



%

\input{4-1_Evaluating_Dissection_Quality}

\input{4-2_Human_Evaluation}
\input{4-3_Use_Case_Audio_Model_Unlearning}

\input{4-4_Analyzing_Acoustic_Feature_Inportance}
\input{4-5_Findings}

%% file: 4-1_Evaluating_Dissection_Quality.tex
\subsection{Evaluation by Last Layer Dissection}
\label{subsec: Last Layer Dissection Accuracy}

\textbf{Experimental Settings} \enspace We evaluate ICL, DB, and TAB methods for closed-concept identification (module A), as discussed in~\cref{sub sec: closed-concept identification}. For the ICL-based approach, we utilize ICL to query Llama-2-chat-13B to select one class from the given concept set that best matches the calibrated summary, with some hand-crafted examples provided. For TAB, we use CLAP to extract audio features in $D_{p}$ and concept features in $D_{c}$. For DB, we use CLIP to extract concept/description features.

We consider five similarity functions: cosine similarity, cubed cosine similarity, rank reorder, WPMI, and softWPMI, as discussed in CLIP-Dissect~\cite{oikarinen2022clip} and label-free concept bottleneck models~\cite{labelfreecbm}. Each method's best performance among the five functions is presented, with detailed results provided in~\cref{sup sec: detailed results of similarity functions}. We report the top-1 and top-5 classification accuracy as well as the CLIP-based cosine similarity between the predicted concept and corresponding ground-truth class for each final-layer neuron.

\textbf{Results} \enspace~\cref{tab:last_layer_acc} demonstrates DB's superiority compared with ICL and TAB across all three target models. Particularly, perfect dissection results are achieved on the AST. On BEATs-frozen, DB outperforms TAB by 12\% in top-1 accuracy and 10\% in top-5 accuracy. This underscores the advantages of similarity calculation within the text feature space, as proposed in DB to generate audio descriptions, mitigating potential imperfections in cross-modal feature projection between audio and text features that might arise from CLAP as in TAB.

Notably, the dissection accuracy typically decreases as the classification ability of the target models on the probing dataset declines. For instance, DB's top-1 accuracy drops from 100.0\% to 76.0\% and 58.0\%, corresponding to testing accuracies of 95.0\%, 89.75\%, and 84.75\% for AST, BEATs-finetuned, and BEATs-frozen, respectively. This decline is attributed to incorrect activation values (logits) of the output layer neurons for misclassified samples, which in turn leads to erroneous calculation of the similarity function. However, despite this decrease, DB still consistently outperforms ICL and TAB in all metrics.

%% file: 4-2_Human_Evaluation.tex
\subsection{Human Evaluation}
\label{subsec:Human Evaluation}

\textbf{Experimental Settings} This section evaluates \algoname{AND} on middle-layer neurons with model-unseen concepts. Specifically, we replace $D_{c}$ with a large-scale acoustic concept set proposed by~\citet{kumar2017discovering} to evaluate the quality of the calibrated summary $S_{h}^{c}$ from module B and the concept identified by DB and TAB in module A. Evaluating the middle-layer interpretation is challenging due to the lack of intrinsic labels for neurons in middle layers, driving us to conduct human evaluation. Following MILAN~\cite{hernandez2021natural} and CLIP-Dissect~\cite{oikarinen2022clip}, we ask human annotators to write summaries for top-$K$ high-activated audio as well as score the description generated by \algoname{AND}. Notably, previous data collected through small-scale experiments on Amazon Mechanical Turk are not satisfying despite providing instructions and examples. This is probably due to considerable challenges for people to precisely describe potential acoustic characteristics among a group of audio in natural language. As the middle-layer assessment requires professional writers, we turn to author-based experiments. For transparency and fairness, the human study results and the associated audio clips are provided in our code repository for the reader's reference.

We assess the dissection quality of 10 randomly selected neurons per layer in AST, with a total of 120 neurons selected. We as the human evaluators to (1) score whether the produced description matches the highly activated audio samples and (2) write the shared properties of highly activated audio samples. For each sampled neuron, five highly activated audio clips are given. For (1), We rate the quality of \algoname{AND}'s description on a scale of 1 to 5, with 1 being strongly disagree and 5 being strongly agree, in response to the question, ``Does the given description accurately describe most of these audio clips?" For (2), the collected summaries are evaluated through the semantic similarity (e.g. cos similarity or BERTScore~\cite{ZhangKWWA20bertscore}) with descriptions from \algoname{AND}.

\textbf{Results}
Firstly, scores across neurons are averaged to serve as a performance indicator. Second, we evaluate \algoname{AND} by computing cos similarity and BERTScore on the generated description, aiming to compare the acoustic similarity between the human-written summaries and descriptions. Results are shown in~\cref{tab:human_evaluation_similarity}. Calibrated summaries $S_{h}^{c}$ from module B achieve the highest quality among all methods, reflecting the effectiveness of open-domain interpretation.

\begin{table}[!hp]
    \vspace{-2mm}
    \begin{center}
    \begin{sc}
    \caption{Results of human evaluation to measure \algoname{AND}'s capability of dissecting middle-layer neurons.  Rating is the mean of scores (1-5) across neurons. Cos similarity and BERTScore are computed between \algoname{AND}'s descriptions and human's written descriptions.}
    \label{tab:human_evaluation_similarity}
    \vspace{1mm}
    \scalebox{0.9}{
    \begin{tabular}{L|C|CC}
    \toprule
    \midrule
     & Rating & Cos & BERTScore  \\ 
     \midrule
    \algoname{AND} (Moduel A: TAB)& 2.73 & 0.24 & 0.42  \\
    \algoname{AND} (Module A: DB) & 3.24 & 0.57 & 0.45 \\
    \algoname{AND} (Module B: SUM)& \textbf{3.49} & \textbf{0.72} & \textbf{0.46}  \\
    \midrule
    \bottomrule
    \end{tabular}
    }
    \end{sc}
    \end{center}
    \vspace{-2mm}
\end{table}

%% file: 4-3_Use_Case_Audio_Model_Unlearning.tex
\begin{table*}[!ht]
    \begin{center}
    \begin{small}
    \begin{sc}
    \caption{Averaged change of confidence after neuron ablation, with each class being the target concept. Avg, $\Delta$A, and $\Delta$R refer to the averaged pruned numbers of neurons, confidence change on ablating class samples, and confidence change on remaining class samples, respectively. OCP refers to open-concept pruning by leveraging $C_{\text{open-set}}$.}
    \vskip 0.15in
    \label{tab:confidencedroptable}
    \addtolength{\tabcolsep}{-2.5pt}
    \scalebox{0.98}{
    \begin{tabular}{L|CCCC|CCCC|CCCC}
    \toprule
    \midrule
    \multicolumn{1}{L|}{Model} &  \multicolumn{4}{C|}{AST} & \multicolumn{4}{C|}{BEATs-finetuned} & \multicolumn{4}{C}{BEATs-frozen} \\
    \midrule
    Method &  Avg& $\Delta$A &$\Delta$R &$\Delta$A - $\Delta$R $\uparrow$& Avg& $\Delta$A & $\Delta$R &$\Delta$A - $\Delta$R $\uparrow$& Avg& $\Delta$A & $\Delta$R &$\Delta$A - $\Delta$R $\uparrow$\\
    \midrule 
    Random baseline &   3000 & -13.52 & -13.48 & 0.04 & 3000 & -0.55 &  -0.55 & 0.00 & 3000 & -0.72 & -0.73 & -0.01\\

    \algoname{AND} (module A: TAB) &  3317& -4.19&-4.03 & 0.16  & 2765& -0.50&  -0.49& 0.01 & 2765& -0.42&  -0.51 & -0.09 \\

    \algoname{AND} (module A: DB) &  3317
    & -7.18& -7.41  & -0.23 & 2765& -0.53& -0.57& -0.04& 2765& -0.48 & -0.58  & -0.10\\
    
    \algoname{AND} (module C: OCP) &  2808& -10.73& -6.83 & \textbf{3.90} & 2677& -2.00& -0.51& \textbf{1.49} & 2906& -0.65& -0.55 & \textbf{0.10}
    \\
    \midrule
    \bottomrule
    \end{tabular}
    }
    \end{sc}
    \end{small}
    \end{center}
    \vskip -0.1in
\end{table*}

\subsection{Use Case: Audio Machine Unlearning}
\label{subsec: Audio Machine Unlearning}

\textbf{Experimental Settings} \enspace We observe models' change of confidence for samples in the ESC50 testing set after neuron ablation. For instance, upon choosing the class ``water drops" as the target concept, we prune out all neurons dissected to have ``water drops" as their responsive features. Then, samples of all 50 classes in the ESC50 testing set, such as ``pouring water" and ``cow", are fed to the pruned network to observe the change of logits. This process is repeated 50 times by using each of ESC50's classes as the target concept, and the values of confidence change on the target concept and non-target concepts are averaged.

There are several intuitive strategies to determine whether a neuron is relevant to the target concept. First method is to extract non-trivial open-domain entities in $S^{c}_{h}$ as $C_{\text{open-set}}$ (module C). If $C_{\text{open-set}}$ includes the target concept, we mask out this neuron. We refer to this method as open-concept pruning (OCP). The second method is using closed-set concept $C_{\text{closed-set}}$. Similarly, if the target concept is included in $C_{\text{closed-set}}$, we mask out this neuron. To align the pruning numbers between open-concept pruning and closed-concept pruning, we select the top-3 most matched concepts when constructing $C_{\text{closed-set}}$, i.e., $C_{\text{closed-set}}$ contains three most matched concepts drawn from $D_{c}$. Notably, ICL, DB, and TAB in module A are all potential options to construct $C_{\text{closed-set}}$, as discussed in \cref{subsec: Last Layer Dissection Accuracy}. Since DB and TAB are considerably better than ICL for last-layer dissection, we adopt DB and TAB in this experiment.

\textbf{Results} 
\enspace The effects of ablating neurons are displayed in~\cref{tab:confidencedroptable}. Results show that OCP based on module C is most efficient in middle-layer concept pruning. Although one-pass pruning may incur backup neurons in the network to show up and complement the pruned ability~\cite{mcgrath2023hydra}, OCP still imposes considerable effects on models' perceptions of the target concept. Specifically, for AST, the classification confidence drop of the target concept is higher than that of non-target concepts, with a gap of 3.9 achieved. This trend is also observed in BEATs-finetuned and BEATs-frozen. Notably, BEATs-frozen, whose model weights are from SSL pretraining, is much more robust to pruning. This may indicate the diverse attention of SSL pretrained neurons with stronger ``backup" abilities~\cite{mcgrath2023hydra} compared with models trained in a supervised manner.~\cref{fig:confidence-drop-figure} in~\cref{sup sec: audio_machine_unlearning_example} illustrates the change of confidence with ``water drops" as the target concept, BEATs-finetuned as the target network, and OCP as the ablating strategy. After pruning, BEATs-finetuned's classification abilities on water-related concepts, such as ``toilet flush" and ``pouring water", are heavily affected, with a smaller influence on other unrelated concepts.

%% file: 4-4_Analyzing_Acoustic_Feature_Inportance.tex
\subsection{Analyzing Acoustic Feature Importance}
\label{subsec: Analyzing Acoustic Feature Importance}

\begin{figure*}[!tp]
\vskip 0.2in
    \begin{subfigure}[b]{0.3\textwidth}
        \includegraphics[width=\textwidth]{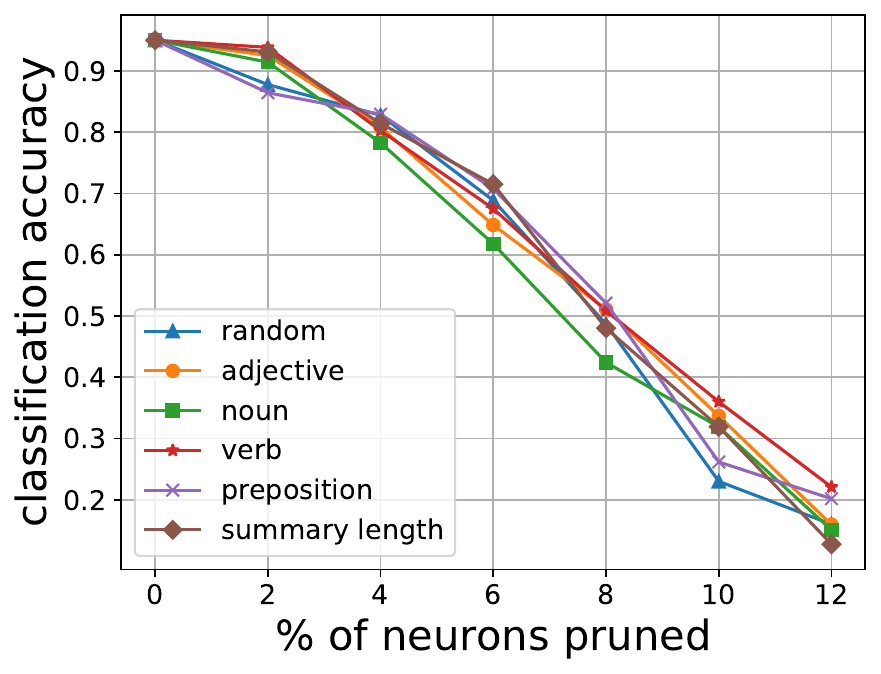}
        \caption{Ablating neurons with the highest number of POS and longest summary.}
        \label{all pos}
    \end{subfigure}
    \hfill
    \begin{subfigure}[b]{0.3\textwidth}
        \includegraphics[width=\textwidth]{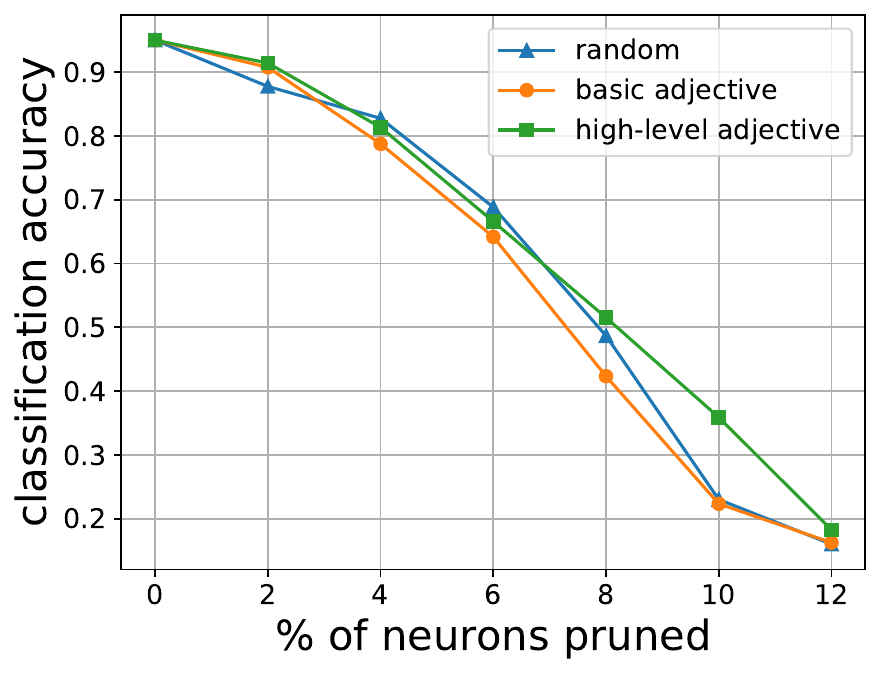}
        \caption{Ablating neurons with the highest number of basic and high-level adjectives.}
        \label{basicandhighleveladj}
    \end{subfigure}
    \hfill
    \begin{subfigure}[b]{0.3\textwidth}
        \includegraphics[width=\textwidth]{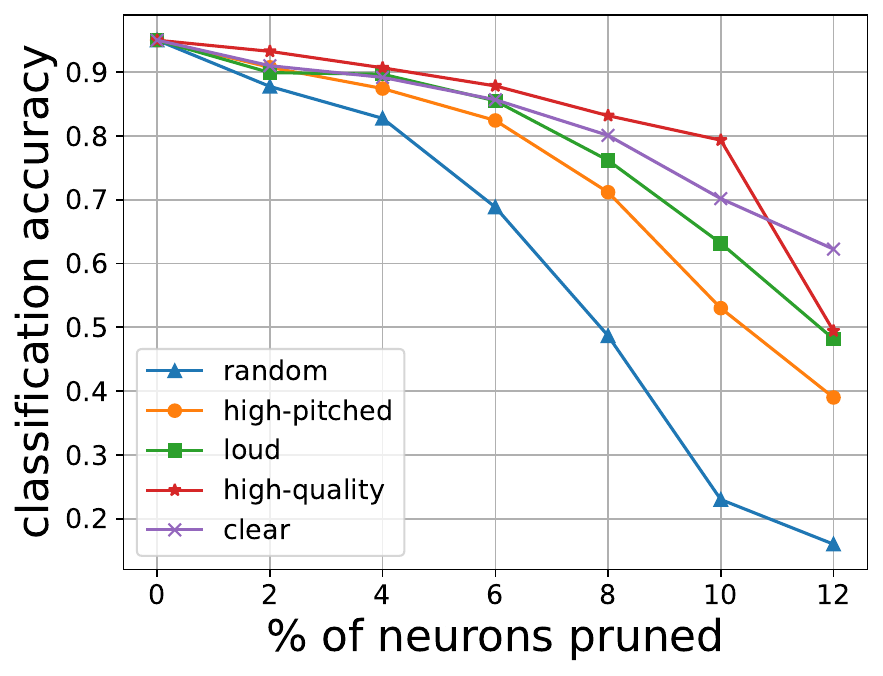}
        \caption{Ablating neurons if a certain basic adjective is contained in $C_{\text{open-set}}$.}
        \label{fourbasicadj}
    \end{subfigure}
    \caption{Feature importance analysis of AST on ESC50, measured by module C in \algoname{AND}. The x-axis is the percentage of ablated linear layer neurons, and the y-axis is the testing performance on ESC50 after pruning.}
    \label{Feature importance analysis}
\end{figure*}

\textbf{Experimental Settings} \enspace Similar to MILAN~\cite{hernandez2021natural}, we linguistically parse descriptions to fine-grained information and investigate factors influencing model performance by neuron ablation with certain criteria. Among all linear layer neurons of AST, we prune $r\%$ neurons with the highest number of nouns, adjectives, verbs, prepositions, and the longest calibrated summary, where $r$ is a parameter. Second, we categorize adjectives into two groups: basic adjectives and high-level adjectives. The former comprises ``high-pitched", ``high-quality", ``clear", and ``loud", which are common in generated descriptions and depict fundamental acoustic features such as frequency and amplitude. We group all the remaining adjectives as high-level ones with more abstract concepts such as ``dramatic" and ``repetitive". We again ablate neurons based on either the number of basic adjectives or high-level adjectives in their summaries. Finally, we study the individual role of four basic adjectives. For each adjective, we prune a neuron if it is in $C_{\text{open-set}}$. If the number of neurons meeting the criterion exceeds $r\%$ of the total neurons, we conduct random pruning among them. Experiments are conducted three times under three different random seeds and the numbers are averaged.

\textbf{Results} \enspace Our observations in acoustic neurons diverge from MILAN's findings in visual neurons, which observe that visual neurons with more adjectives in the descriptions are more influential to model performance. As shown in~\cref{all pos}, pruning neurons with most adjectives does not necessarily lead to a more significant performance drop. While pruning nouns exhibits a higher impact than random pruning for 4\%, 6\%, and 8\% neurons, it falls short for 2\% and 10\%. When further categorizing adjectives into basic and high-level ones, as illustrated in~\cref{basicandhighleveladj}, we find that basic adjectives are consistently more influential than high-level ones, with up to over 10\% performance difference for the case of 10\% pruning. Moreover, pruning basic adjectives incurs a slightly larger performance drop than random pruning, except for the 2\% case. This may indicate that acoustics are identified more with its fundamental features rather than abstract concepts such as repetitiveness and emotions, which is intuitive especially when the models are supervisedly trained to classify audio entities, such as AST on ESC50. Further examinations on the four individuals discover much smaller pruning effects for ablating neurons with a specific basic adjective, as shown in~\cref{fourbasicadj}. Comparing the results in~\cref{basicandhighleveladj}, we conclude that models distinguish audio more based on a combination of basic acoustic features rather than individual ones or high-level concepts, particularly for the task of entity classification.

\begin{figure}[tp]
\vskip 0.12 in
\centering
\includegraphics[height=140px]{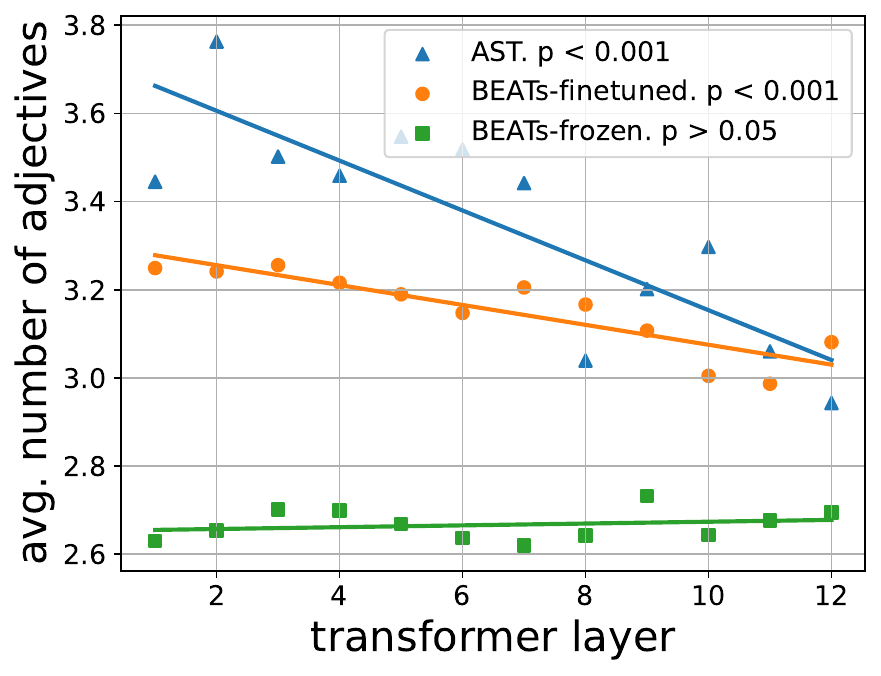}
  \caption{Number of averaged adjectives per neuron in different transformer blocks of AST, BEATs-finetuned, and BEATs-frozen.}
  \label{fig:adjtrend}
\end{figure}

As shown in~\cref{fig:adjtrend}, we also compute the averaged number of adjectives per linear layer neuron for each transformer block as in MILAN. In particular, we analyze AST, BEATs-finetuned, and BEATs-frozen, corresponding to three different training strategies: supervised training, SSL pretraining with whole-model supervised finetuning, and SSL pretraining with final layer finetuning. We observed that AST's attention to different acoustic features, represented as adjectives, narrows down in deeper transformer blocks. This trend is analogous to MILAN's observation for supervised ResNet18 on ImageNet. However, when considering SSL pretraining, the diverse-to-accordant trend is mitigated, as evidenced by the case of BEATs-finetuned. Moreover, for BEATs-frozen, all transformer layers exhibit the same level of attention to adjectives, and the overall numbers are lower. This suggests a potential effect of training strategies on model behaviors. Neurons in BEATs-frozen are more diverse and do not represent mutually similar concepts. While the supervised training leads to a more capable yet convergent neuron layout. Further discussions on the effects of training strategies are in~\cref{subsec: Training Strategy Affects Neuron Interpretability}.

%% file: 4-5_Findings.tex
\subsection{Training Strategy Affects Neuron Interpretability}
\label{subsec: Training Strategy Affects Neuron Interpretability}


\textbf{Experimental Settings} \enspace Pseudo code for the proposed pipeline is displayed in~\cref{sup subsec: interpretability algorithm}. We first pretrain a K-means model $K(\cdot)$ with sentences of audio descriptions $d_{i}$ for each audio $a_{i}$ in $D_p$. Number of clusters is set to 11, decided by the elbow method~\cite{thorndike1953belongs}. Then, for an interested neuron $f(\cdot)$, we acquire its $S_{h}$ and project sentences of each description to the 11 groups using $K(\cdot)$. We say a description is related to a cluster if at least one sentence within the description belongs to that cluster. If at least $\tau$ descriptions are related to the same cluster, i.e., descriptions of highly-activated audio share common information, we label the neuron as ``\textit{interpretable}". Otherwise, it is regarded as ``\textit{uninterpretable}". $\tau$ is a parameter. Higher $\tau$ denotes a stricter criterion. The semantics of found clusters are discussed in~\cref{sup subsec: clustering}.

\begin{figure}[!tp]
\vspace{-3mm}
\centering
\includegraphics[height=150px]{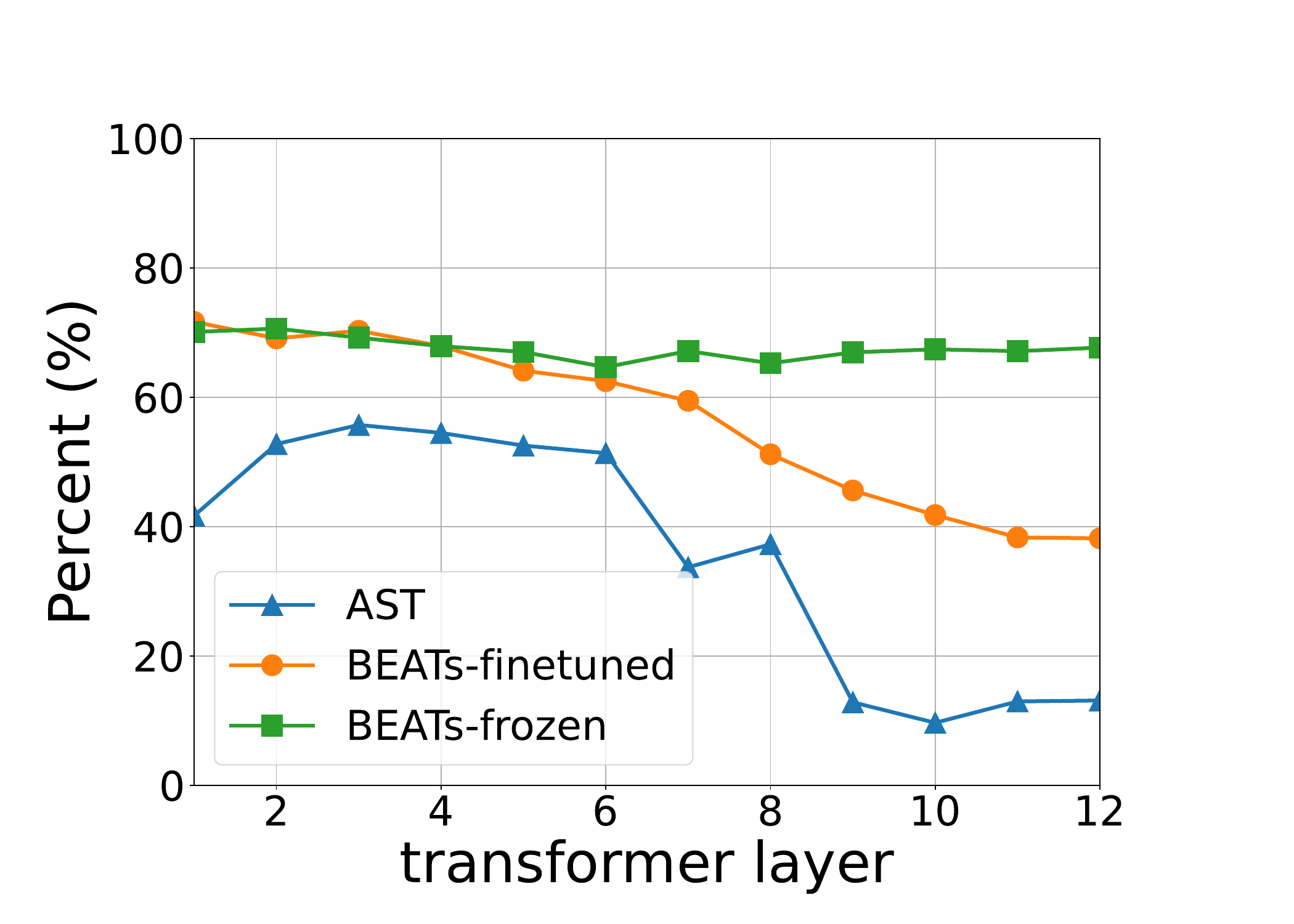}
  \caption{Percentage of uninterpretable neurons in different transformer blocks of AST, BEATs-finetuned, and BEATs-frozen. More results are provided in~\cref{sup sec: neuron interpretability}.}
  \label{fig:interpretability-mean}
\end{figure}

\textbf{Results} \enspace The percentages of uninterpretable neurons for each transformer block of three models are illustrated in~\cref{fig:interpretability-mean}, with $\tau=4$. AST demonstrates a decrease in the percentage of uninterpretable neurons from shallow to deep transformer blocks. Neurons in shallow layers are more diverse and gradually concentrate on certain concepts in deeper layers, with more overlapped information among highly-activated audio, to conduct the classification task. On the other hand, BEATs-frozen has consistent percentages of unexplainable neurons among all layers. From the scale of individual neurons, this manifests the effect of SSL pretraining on diversifying models' abilities. When finetuning the entire model, as in BEATs-finetuned, the supervision attempts to guide the model to converge its attention in deeper layers by narrowing neurons' responsive features. This finding links to the previously reported attentive overfitting of supervised models~\cite{Zagoruyko2017AT, ericsson2021well}, which SSL networks are less vulnerable to~\cite{ericsson2021well}. While previous discussions on attentive overfitting typically rely on observing performance metrics or using input-specific interpretation tools like Grad-cam~\cite{selvaraju2017grad}, we provide neuron-level explanations for this phenomenon using \algoname{AND}. Finally, this is also related to the averaged number of adjectives per neuron in~\cref{fig:adjtrend}, showing that a decrease in acoustic features extracted from natural language descriptions may represent a narrowed neuron polysemanticity. Figures under different top-$K$ and $\tau$ are presented in~\cref{sup subsec: interpretability threshold}, with similar trends observed. Similar experiments but adopting GTZAN Music Genre~\cite{1021072} as the probing and training dataset are presented in~\cref{sup subsec: gtzan neuron Interpretability}. 

%% file: 5_Conclusion.tex
\section{Conclusion}
In this work, we have introduced \algoname{AND}, the first \textbf{A}udio \textbf{N}etwork \textbf{D}issection framework, which applies an LLM-based pipeline incorporated with three specialized modules to identify the responsive features of acoustic neurons. Extensive experiments are conducted to verify \algoname{AND}'s dissection quality and discover acoustic model behaviors using \algoname{AND}. Specifically, last layer dissection and human evaluation demonstrate \algoname{AND}'s high-quality dissection. Second, we examine the effect of concept-specific pruning conducted using \algoname{AND} and discuss its potential use case of model unlearning. Third, we investigate the roles of acoustic features in the model's perception ability and make a comparison with previous observations in the vision domain. Finally, we discuss the impact of different training strategies on neuron-level model behaviors, where the results align with the analyses of acoustic features.

%% file: supp/A_implementation_details.tex
\section{Implementation Details}
\label{sup sec: implementation details}
\subsection{Audio Captioning Model}
\label{sup subsec: implementation details of audio captioning models}
To obtain natural language descriptions of audio, we utilize SALMONN~\cite{tang2023salmonn}, which is a pre-trained LLM-based multi-task audio model. One of SALMONN's pretraining tasks is audio captioning, endowing it with remarkable open-domain audio captioning abilities. Following its official guidance, we use the prompt \textit{``Please describe the audio in detail."} and employ SALMONN-13B. As SALMONN is newly proposed, its implementation requires further refinement, notably in the absence of batch inference capabilities. Captioning the entire ESC50 dataset takes approximately 40 hours on an NVIDIA-A6000 GPU, with batch size set to be 1. There might be a substantial efficiency gain once batch inference is available.

\subsection{Large Language Model}
\label{sup subsec: implementation details of LLMs}
We adopt Llama-2-chat-13B~\cite{touvron2023llama} for all LLM-related experiments. The vllm package~\citep{kwon2023efficient} is employed to boost the inference efficiency, which integrates efficient attention mechanism and other speed-up techniques to create a memory-efficient LLM inference engine. The summarization process for all linear layers in AST and BEATs takes around 12 and 10 hours respectively on an NVIDIA RTX A6000 GPU. This includes generating summaries for both highly and lowly activated samples.

For the description summarization, we instruct the LLM with prompt \textit{``Here are descriptions of some audio clips. Please summarize these descriptions by identifying their commonalities."} to ask LLM for summarization. Additionally, the LLM-based non-acoustic-word filtering is conducted with prompt \textit{``Can the adjective be used to describe the tone, emotion, or acoustic features of audio, music, or any other form of sound? Answer yes or no and give the reason."}. 

\clearpage
\newpage

\subsection{Instruction and Examples of In-context Learning }
\label{sup subsec: implementation details of in-context learning}
In~\cref{sub sec: closed-concept identification}, we introduce a method called ICL in \algoname{AND}'s module A, which uses LLM to interpret the calibrated summary $S_c^h$ and then explain neurons in target model $F(\cdot)$. This method is built upon in-context learning, enabling LLM to select a concept $C_{\text{close-set}}$ from $D_c$ as output. We have experimented with 1-shot and 2-shot learning. The instructions and examples used are provided in~\cref{tab:ICL example}.

\begin{table*}[!hp]
    \centering
    \caption{Instruction and examples used in ICL (module A in \algoname{AND}).}
    \vspace{1ex}
    \begin{tabular}{C|P{15cm}}
    \toprule
     Instruction    &  You have a set of object classnames:\newline 
[concept set $D_c$] \newline
    
The following is a description about some audio clips. Based on the description, select a classname out of the above classnames that matches the description most. \\
     \midrule
     Example 1  & \textbf{Description}: The audio features a car meowing: All of the clips contain the sound of a cat meowing. Loud sound: These clips are all of loud sound but with varying degrees of intensity. Repetitive barking: Clips 1 and 4 are repetitive, with the cat meowing multiple times in each clip. Poor audio quality: All clips have poor audio quality, with either distortion, muffling, or apparent background noises. 
    \newline
    \textbf{Response}: We know these clips are about the class ``cat'' in the concept set. We can get this answer since the description mentions All of the clips contain the sound of a cat meowing. 
    \newline
    \textbf{Answer}: cat 
    \newline
    \\ 
    Example 2 & \textbf{Description}: They all feature a person snoring loudly. The snoring starts off slow and gets louder over time. The audio is recorded in mono. There are no other sounds in the background. The snoring is described as loud and intense. The audio clips differ in the following ways. The first clip features a man snoring, while the second and fourth clips feature a person snoring (gender not specified). The third clip features a zombie growling and snarling, while the other clips only feature snoring. The third clip is described as scary and creepy, while the other clips are not. The third clip is intended for use in a horror movie or zombie video game, while the other clips do not have specific intended uses stated. The third clip is of poor quality, while the other clips are not specified as such. 
    \newline
    \textbf{Response}: Based on the description, the most suitable classname for the audio clips would be ``snoring'' or ``zombie growling and snarling''. Both of these classnames match the description of loud sounds with a strong emotional impact, specifically fear and terror. But ``zombie growling and snarling'' is not in the given classname set. So the answer is ``snoring'' 
    \newline
    \textbf{Answer}: snoring  \\
    \bottomrule
    \end{tabular}

    \label{tab:ICL example}
\end{table*}

\subsection{Others}
\label{sup subsec: implementation details of others}
For the CLIP model, we employ ViT-B/32. For the CLAP model, we utilize the 630k-audioset-best version. To capture the representation of textual artifacts in \algoname{AND}, we use all-MiniLM-L12-v2 pre-trained model provided by~\citet{reimers-2019-sentence-bert}. For all experiments, we use $K=5$ to select top-$K$ highly/lowly activated samples, $t=0.7$ to remove similar sentences in summary calibration module.

%% file: supp/B_qualitative_results.tex
\section{Qualitative Results}
\label{sup sec: Qualitative Results}

\subsection{Dissection Pipeline}
\label{sup subsec: dissection pipeline}
\begin{figure*}
    \begin{center}
    \centerline{\includegraphics[width=\columnwidth]{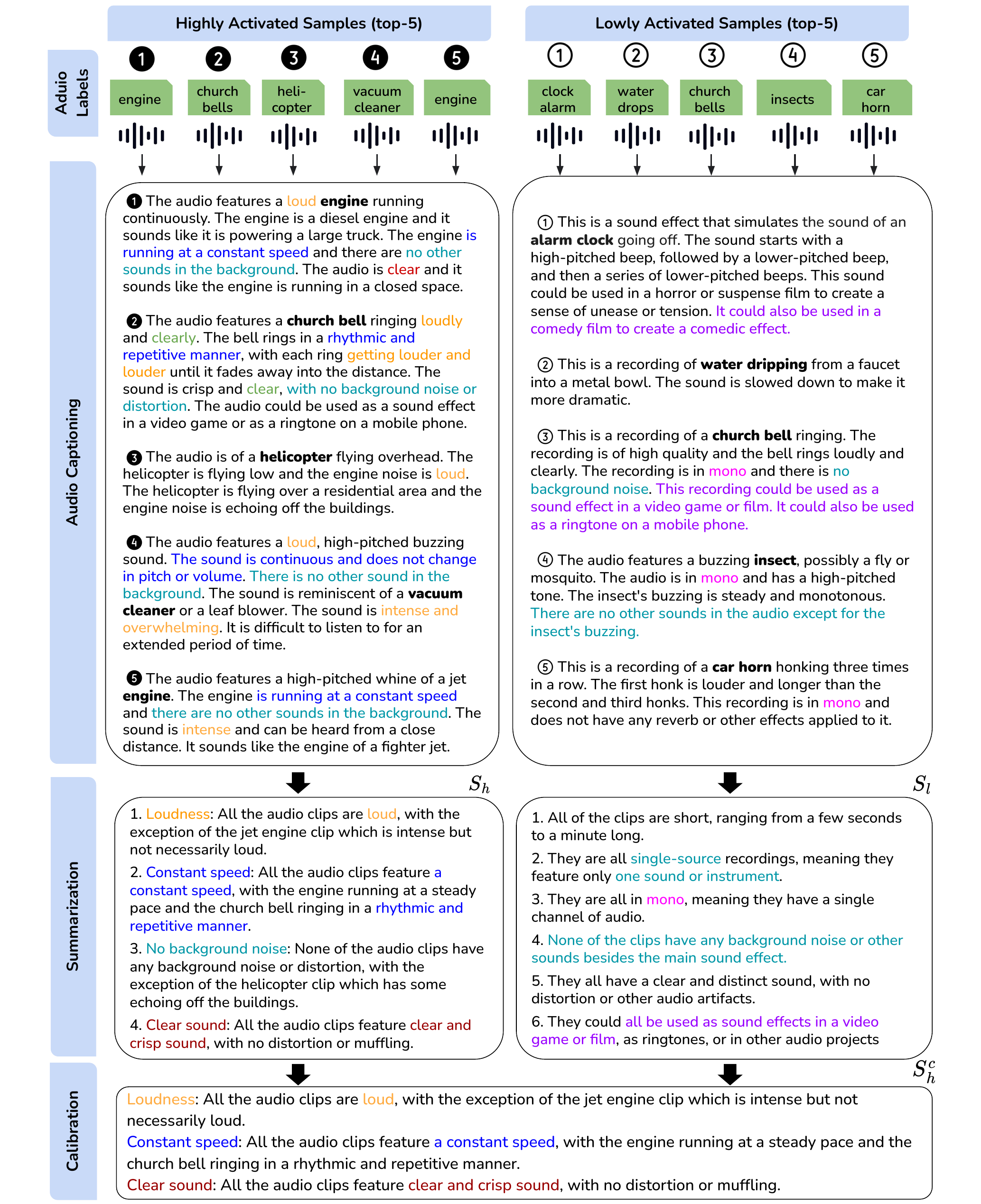}}
    \caption{The pipeline of \algoname{AND}'s module B. Texts with the same color means they refer to the same concept or property. In this example, a property \textit{``no background noise"} is removed during calibration because it appears in both summaries.}
    \label{fig:qualitative results}
    \end{center}
\vskip -0.2in
\end{figure*}

We provide an example to illustrate the pipeline in~\cref{fig:qualitative results}. We randomly select a neuron from the first transformer encoder's \textit{output} layer in AST with its corresponding top-5 highly and lowly activated audio samples. We highlight the predicted audio source with bold font and similar concepts with the same text color. SALMONN is shown to yield high-quality descriptions that capture correct audio sound source and its detailed acoustic properties for both top-5 highly and lowly activated audio samples. Llama-2-chat subsequently identifies mutual information among these textual representations. Notably, Llama-2-chat tends to produce some greeting messages due to its training objective, we remove these meaningless contents by rule-based filtering. Then, summary calibration utilizes summary of lowly-activated samples to adjust summary of highly-activated samples. For instance, the property \textit{``no background noise"} is removed due to its presence in both summaries. Finally, the calibrated summary of highly-activated samples serves as the output.

\subsection{Outputs from Different Modules in AND}
\label{sup subsec: outputs from different modules}

To make things clear, we randomly select some neurons in AST, using module A and B of \algoname{AND} to generate dissection descriptions. The results are shown in~\cref{tab: outputs from different modules}. We adopt  ESC50~\cite{piczak2015esc} as probing dataset $D_p$, and the open-domain acoustic concept set proposed by~\citet{kumar2017discovering} as $D_c$ here.

\begin{table*}[!hp]
    \centering
    \caption{Outputs from different modules in \algoname{AND}}
    \vspace{1ex}
    \begin{tabular}{P{3.3cm}||C|C|P{5.5cm}}
    \toprule
    \midrule
    Top-5 highly activated \newline audio label
    & AND (Module A: TAB) & AND (Module A: DB) & AND (Module B: SUM) \\
        \midrule
    \begin{itemize}
        \item church bells
        \item clock alarm
        \item clapping
        \item door wood creaks
        \item clapping
    \end{itemize}  
    & steam rushing & bell-ringers crucifying & 1. All of the clips are recordings of sounds: The first clip is a recording of a bell ringing, the second is an alarm clock ringing, the third is a frog croaking, and the fourth is a recording of a group of people clapping.\newline 2. High quality recordings: All of the clips are of high quality, meaning they are clear and well-recorded.\newline3. Loud and clear sounds: Each of the clips features loud and clear sounds that are easy to hear.\newline4. Gradual increase in volume: Three of the clips (the alarm clock, the frog, and the group of people clapping) feature a gradual increase in volume over time.\newline5. Potential use in video games or movies: All of the clips could be used as sound effects in video games or movies, based on their descriptions.
    \\
    \midrule
    \begin{itemize}
        \item glass breaking
        \item water drops
        \item glass breaking
        \item mouse click
        \item glass breaking
    \end{itemize}  

    &ribbon being& glass chattering  & 1. Loud and intense sounds: All of the audio clips feature loud and intense sounds, which suggests that they may be used to create a sense of drama or urgency in a video or film.\newline 2. High-pitched sounds: Four of the five audio clips feature high-pitched sounds, which could be used to create a sense of tension or excitement.\newline 3. Breaking or tapping sounds: Three of the audio clips feature sounds of breaking or tapping, which could be used to create a sense of impact or action in a video or film.\newline 4. Mono audio: Three of the audio clips are mono, which means they have a single audio channel and may be used to create a more intimate or focused sound.\newline 5. Short duration: Four of the audio clips are short, lasting only a few seconds, which could be used to create a quick impact or effect in a video or film.\newline Overall, these commonalities suggest that the audio clips could be used to create a sense of drama, tension, or action in a video or film, and could be particularly effective when used in quick succession or in combination with other audio or visual elements.
    \\
    \midrule
    \bottomrule
    \end{tabular}

    \label{tab: outputs from different modules}
\end{table*}

%% file: supp/C_middle_layer_analysis_from_basic_acoustic_properties.tex
\section{Middle Layer Analysis from Basic Acoustic Properties}
\label{sup sec: middle layer analysis from basics}
\begin{figure*}[htbp]
    \vskip -0.15in
    \begin{subfigure}[b]{0.5\textwidth}
    \centering
        \includegraphics[height=150px]{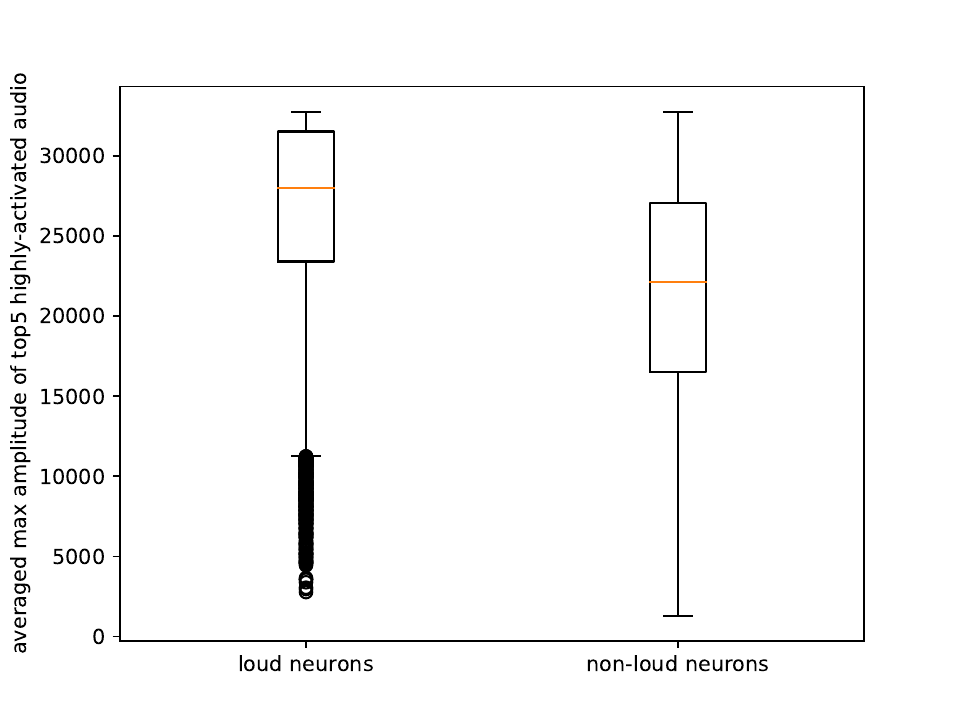}
        \caption{Amplitude.}
        \label{amplitude}
    \end{subfigure}
    \hfill
    \begin{subfigure}[b]{0.5\textwidth}
    \centering
        \includegraphics[height=150px]{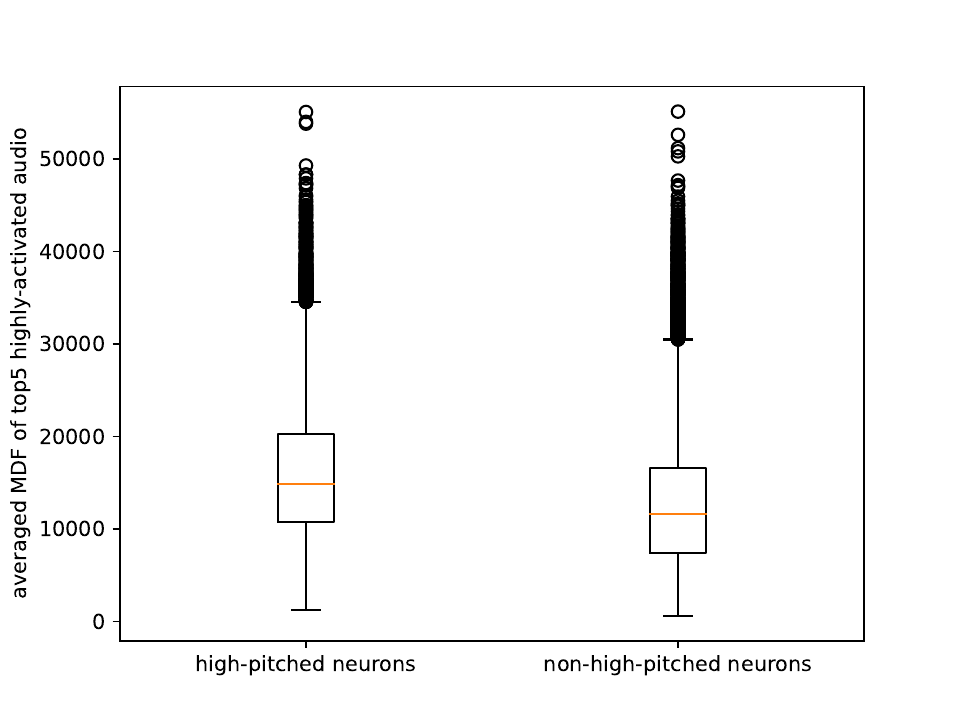}
        \caption{MDF.}
        \label{mdf}
    \end{subfigure}
    \caption{The averaged waveform amplitude and the median frequency (MDF) of top-$K$ highly-activated audio for all neurons in the AST.}
\end{figure*}

We measure the basic acoustic features of the top-$K$ highly-activated audio for each neuron in the AST, with ``loud" and ``high-pitched" adopted. Firstly, we group the neurons based on whether their $C_{\text{open-set}}$ include the word ``loud". There are 22612 and 32734 neurons dissected with and without this word respectively. Then, the averaged waveform amplitude of top-$K$ highly-activated audio for each neuron is calculated. As shown in~\cref{amplitude}, highly-activated audio of ``loud" neurons typically have larger sounds. On the other hand, we use median frequency (MDF) as a measure of audio's representative frequency. As shown in~\cref{mdf}, neurons dissected with the word ``high-pitched" tend to be more responsive to high-frequency audio. While the trend is not as significant as the case of ``loud" due to this indirect measure of overall frequency.

%% file: supp/D_adjective_distribution_of_different_target_networks.tex
\section{Adjective Distribution of Different Target Networks}
\label{sup sec: adj distribution}
\begin{figure*}[htbp]
    \vskip 0.15in
    \begin{subfigure}[b]{0.33\textwidth}
    \centering
        \includegraphics[height = 180px]{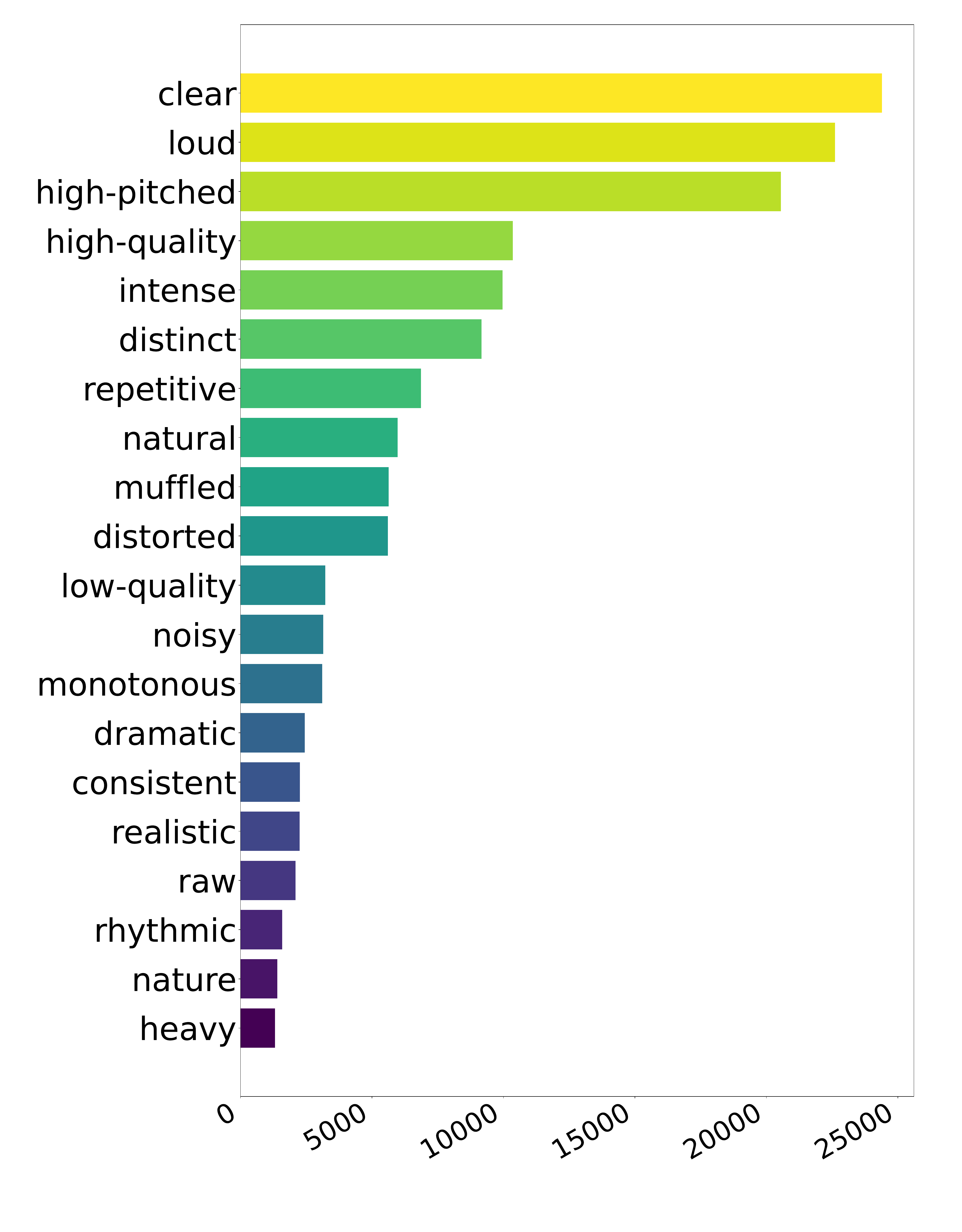}
        \caption{AST.}
    \end{subfigure}
    \hfill
    \begin{subfigure}[b]{0.33\textwidth}
    \centering
        \includegraphics[height = 180px]{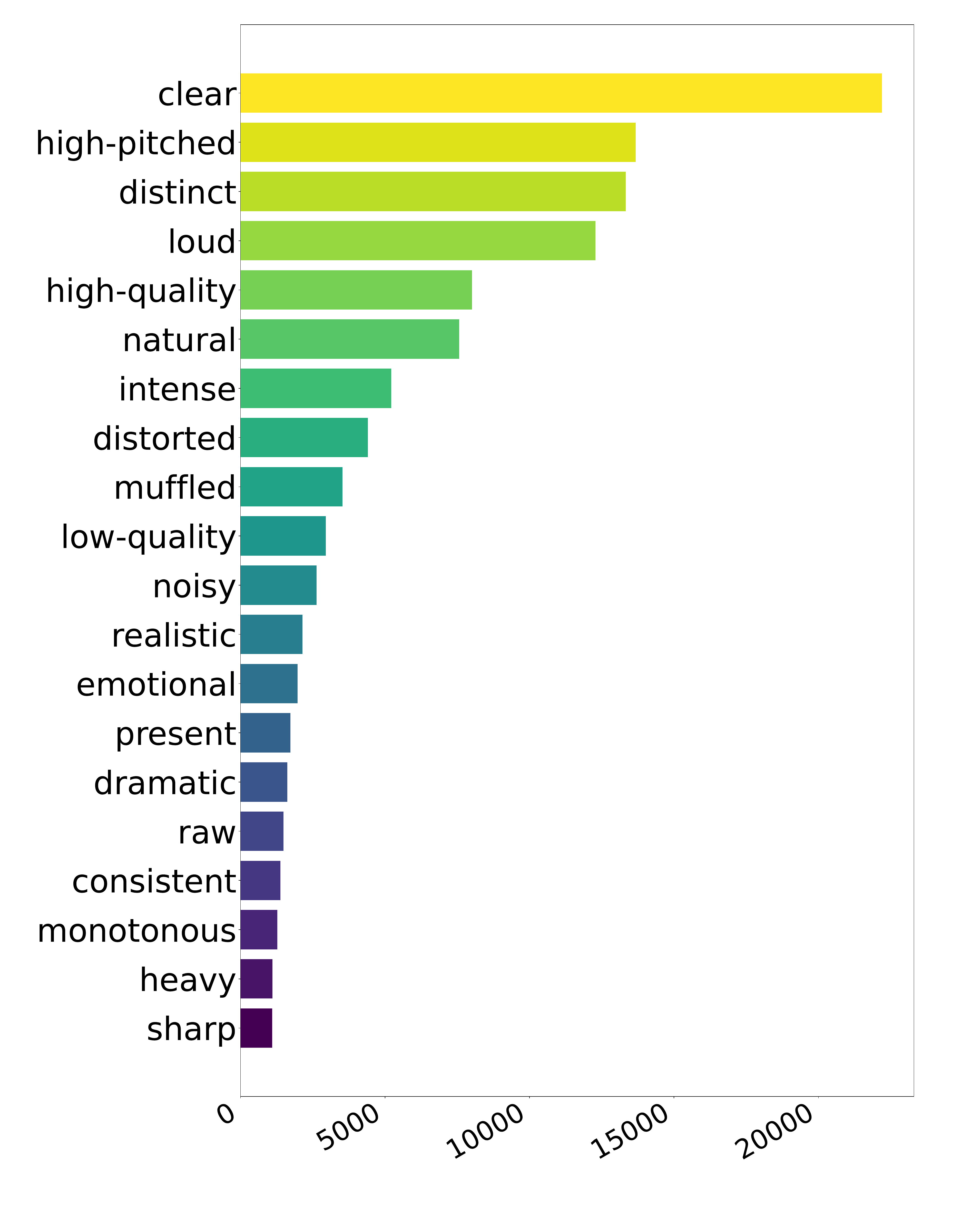}
        \caption{BEATs-finetuned.}
    \end{subfigure}
    \hfill
    \begin{subfigure}[b]{0.33\textwidth}
    \centering
        \includegraphics[height = 180px]{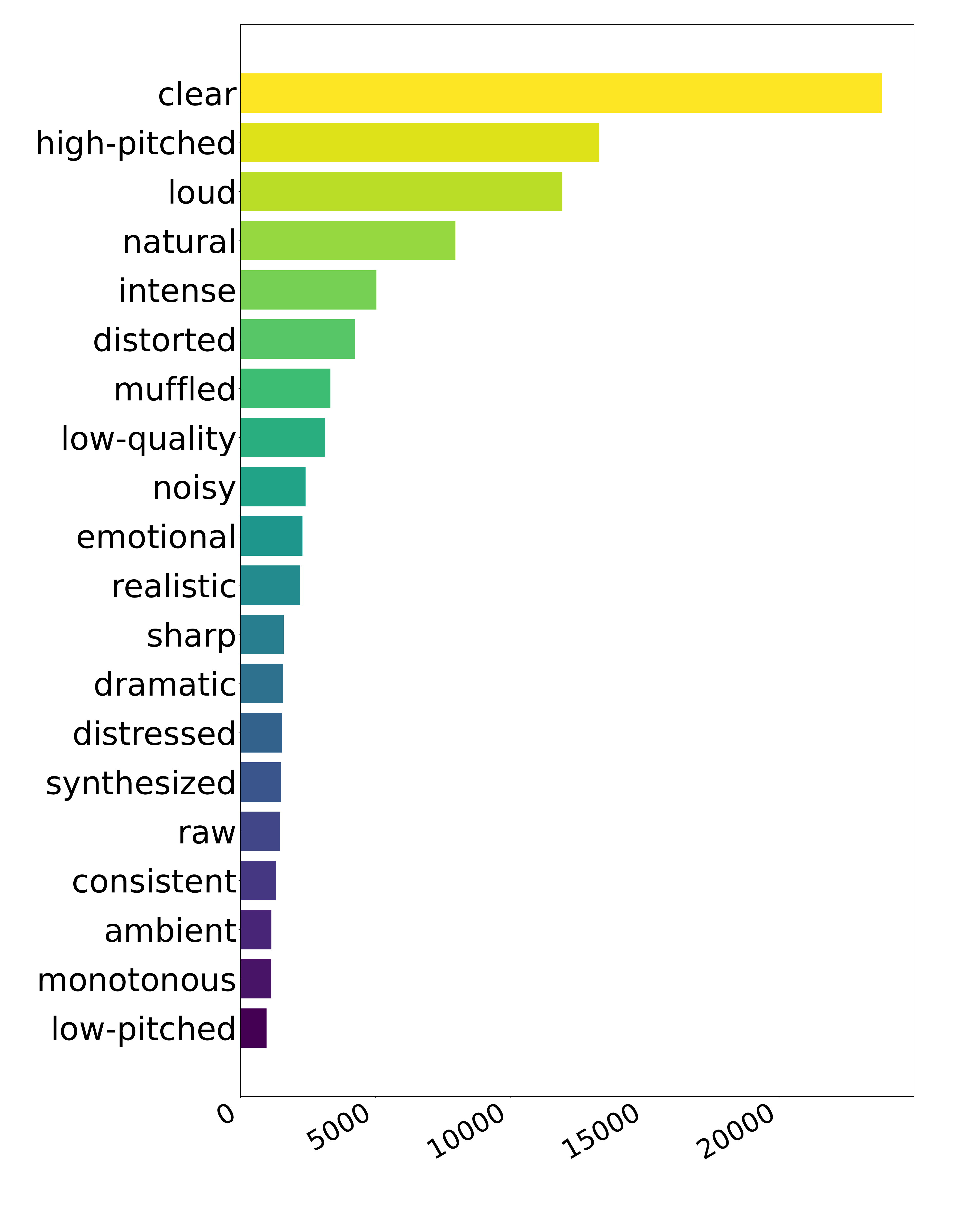}
        \caption{BEATs-frozen.}
    \end{subfigure}
    \caption{The distribution of top-20 most common adjectives across all linear layer neurons in AST, BEATs-finetuned, and BEATs-frozen on the ESC50. These adjectives are extracted from $C_{\text{open-set}}$}
    \label{adj dist. of three models}
\end{figure*}

\cref{adj dist. of three models} displays the distribution of top-20 most common adjectives of all neurons in AST, BEATs-finetuned, and BEATs-frozen on the ESC50, extracted from $C_{\text{open-set}}$. Several common words, such as ``loud" and ``high-pitched", can be observed across models. We provide an analysis on neuron's properties with respect to these two words as examples in~\cref{sup sec: middle layer analysis from basics}.

%% file: supp/E_detailed_results_of_last_layer_dissection.tex
\section{Detailed Results of Last Layer Dissection}
\label{sup sec: detailed results of similarity functions}

\cref{tab:last_layer_acc_full} presents detailed results of experiments in~\cref{subsec: Last Layer Dissection Accuracy}. In particular, we evaluate ICL, TAB, and DB of module A on five similarity functions on AST, BEATs-finetuned, and BEATs-frozen. Each method's performance is considered the best result achieved among all similarity functions and is reported in~\cref{tab:last_layer_acc}.

\begin{table}[hp]
\begin{center}
\begin{small}
\begin{sc}
\addtolength{\tabcolsep}{-1pt}
\caption{Last layer dissection accuracy of ICL, TAB, and DB of module A on AST, BEATs-frozen, and BEATs-finetuned when adopting five different similarity functions.}
\label{tab:last_layer_acc_full}
\vskip 0.15in
\begin{tabular}{P{3cm}|P{2.3cm}|P{2.5cm}|P{2.3cm}|P{2.3cm}|P{2.3cm}}
\midrule
     & cos similarity &  cos similarity cubed & rank reorder & wpmi & soft wpmi \\
\toprule
\midrule
 Model & \multicolumn{5}{C}{\textbf{AST}}
\\
\midrule
 \algoname{AND} (Module A: ICL) & \multicolumn{5}{C}{72 / 60 / 0.83} \\
 \midrule
  \algoname{AND} (Module A: TAB) & 2 / 16 / 0.23 & 92 / 98 / 0.97 & 84 / 92 / 0.91 & \textbf{96} / 100 / 0.98 & \textbf{96} / 98 / 0.98  \\
  \midrule
\algoname{AND} (Module A: DB) & 80 / 82 / 0.85 & \textbf{100} / 100 / 1.00 & 86 / 100 / 0.93 & 88 / 100 / 0.95 & 2 / 10 / 0.24 \\
\midrule
\toprule
 Model & \multicolumn{5}{C}{\textbf{BEATs-finetuned}}
\\
\midrule
 \algoname{AND} (Module A: ICL) & \multicolumn{5}{C}{52 / 56 / 0.68} \\
 \midrule
 \algoname{AND} (Module A: TAB) & 50 / 80 / 0.64 & 66 / 94 / 0.77

 & 52 / 80 / 0.67 & 68 / 94 / 0.77 & \textbf{74} / 92 / 0.82  \\
  \midrule
\algoname{AND} (Module A: DB) & \textbf{76} / 100 / 0.83 & 74 / 96 / 0.82 & 56 / 90 / 0.71 & 62 / 94 / 0.75 & 2 / 10 / 0.23 \\
\midrule
\toprule
 Model & \multicolumn{5}{C}{\textbf{BEATs-frozen}}
\\
\midrule
\algoname{AND} (Module A: ICL) & \multicolumn{5}{C}{16 / 16 / 0.37 } \\
 \midrule
 \algoname{AND} (Module A: TAB) & 14 / 34 / 0.35 & 38 / 76 / 0.56 & 18 / 48 / 0.42 & \textbf{46} / 72 / 0.60 & 42 / 80 / 0.60  \\
  \midrule
\algoname{AND} (Module A: DB) & \textbf{58} / 82 / 0.69 & 46 / 82 / 0.62 & 36 / 72 / 0.56 & 34 / 68 / 0.53 & 2 / 10 / 0.23 \\
\midrule

\end{tabular}
\end{sc}
\end{small}
\end{center}
\vskip -0.1in
\end{table}

%% file: supp/F_audio_machine_unlearning_example.tex
\section{Audio Machine Unlearning Example}
\label{sup sec: audio_machine_unlearning_example}
\begin{figure*}[h]
    \centering
    \includegraphics[width=\textwidth]{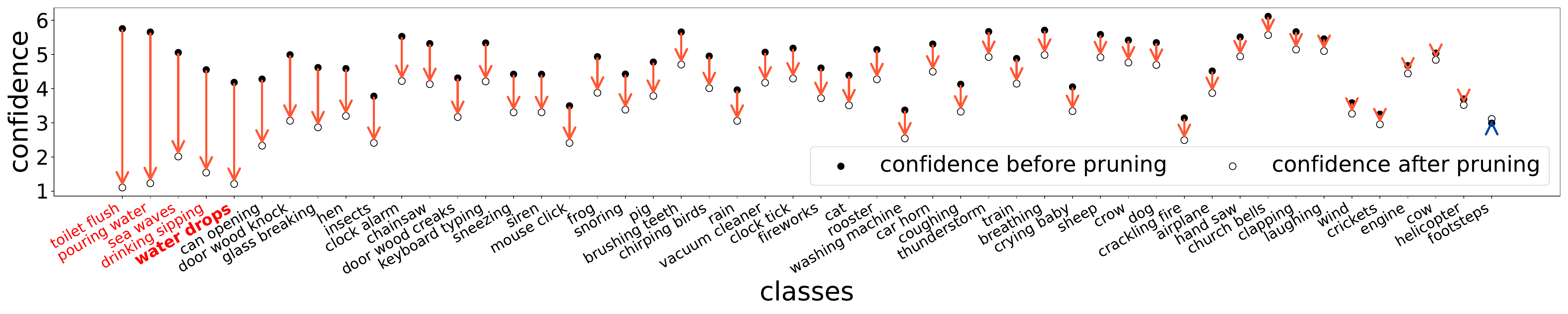}
      \caption{Change of model confidence when neurons associated with \textbf{``water drops"} are ablated. Confidence in recognizing water-related audio (with class names labeled in red) decreases, while other sounds are not significantly affected.}
      \label{fig:confidence-drop-figure}
\end{figure*}

In~\cref{subsec: Audio Machine Unlearning}, we discuss the potential use case for audio machine unlearning by \algoname{AND}. We provide an example  in~\cref{fig:confidence-drop-figure}. When we use OCP (module C) to prune out neurons in BEATs-finetuned associated with ``water drops", the classification abilities of BEATs-finetuned  on water-related concepts, such as ``toilet flush" and ``pouring water", are heavily affected, with a smaller influence on other unrelated concepts. 

%% file: supp/G_neuron_interpretability.tex
\section{Neuron Interpretability}
\label{sup sec: neuron interpretability}
\subsection{Algorithm}
\label{sup subsec: interpretability algorithm}

\begin{algorithm}
\caption{ \texttt{GET-UNINTERPRETABLE-NEURONS}}
    \begin{algorithmic}[1]
    \State \textbf{Input}: Probing dataset $D_p$, target network $F(\cdot)$, threshold $\tau$
    \State Initialize sentence pool $S$, interpreatable neuron pool $N_{i}$, and uninterpreatable neuron pool $N_{u}$ to be empty
    \For {description $d_i \in D_p$}
        \For {sentence $s_j \in d_i$} 
            \State Add $s_j$ to $S$
        \EndFor 
    \EndFor
    \State Train a K-means model $K(\cdot)$ with $S$
    \For{neuron $f(\cdot) \in F(\cdot)$}

        \State Get $S_{h}={d_1, \dots, d_k}$ of $f(\cdot)$ 
        \For {$d_{i} \in S_{h}$}
            \State Initialize an empty multiset $C_i$  
            \For {$s_j \in d_i$}
                \State Add $K(s_j)$ to $C_i$
            \EndFor
        \EndFor         
    \If {$|\bigcap^k_{i=1} C_i| \geq \tau$}
        \State Add $f(\cdot)$ into $N_i$
    \Else
        \State Add $f(\cdot)$ into $N_u$
    \EndIf
    \EndFor
    \State \Return $N_i$, $N_u$
    \end{algorithmic}
\end{algorithm}

\newpage
\subsection{Clusters in ESC50}
\label{sup subsec: clustering}
After applying K-means clustering, audio captions $D_d$ generated by SALMONN~\cite{tang2023salmonn} are categorized into 11 distinct groups. To delve into the underlying mechanism, we leverage wordcloud package~\cite{oesper2011wordcloud} to visualize word distribution of these groups, as presented in~\cref{fig:wordcloud-esc50}. The wordclouds are generated based on the word frequency within each cluster's corpus: the more frequently a word appears, the larger its font size. The numbers following the cluster id refer to the count of audio captions / sentences in the corresponding group.

For instance, cluster 1 exhibits a strong association with repetitive and high-pitched sounds, such as \textit{clock alarm}, \textit{ringing}, and \textit{bell}. Cluster 2 encompasses various words related to traffic noise, such as \textit{engine}, \textit{motorcycle}, \textit{track}. These wordcloud illustrations vividly depict the effectiveness and rationality of the clustering process.

\noindent\begin{tabularx}{\textwidth}{@{}c|c@{}}
\toprule
   Cluster 1 (119 / 565) & Cluster 2 (144 / 722)
  \\ 
  \midrule
    \includegraphics[width=0.49\textwidth]{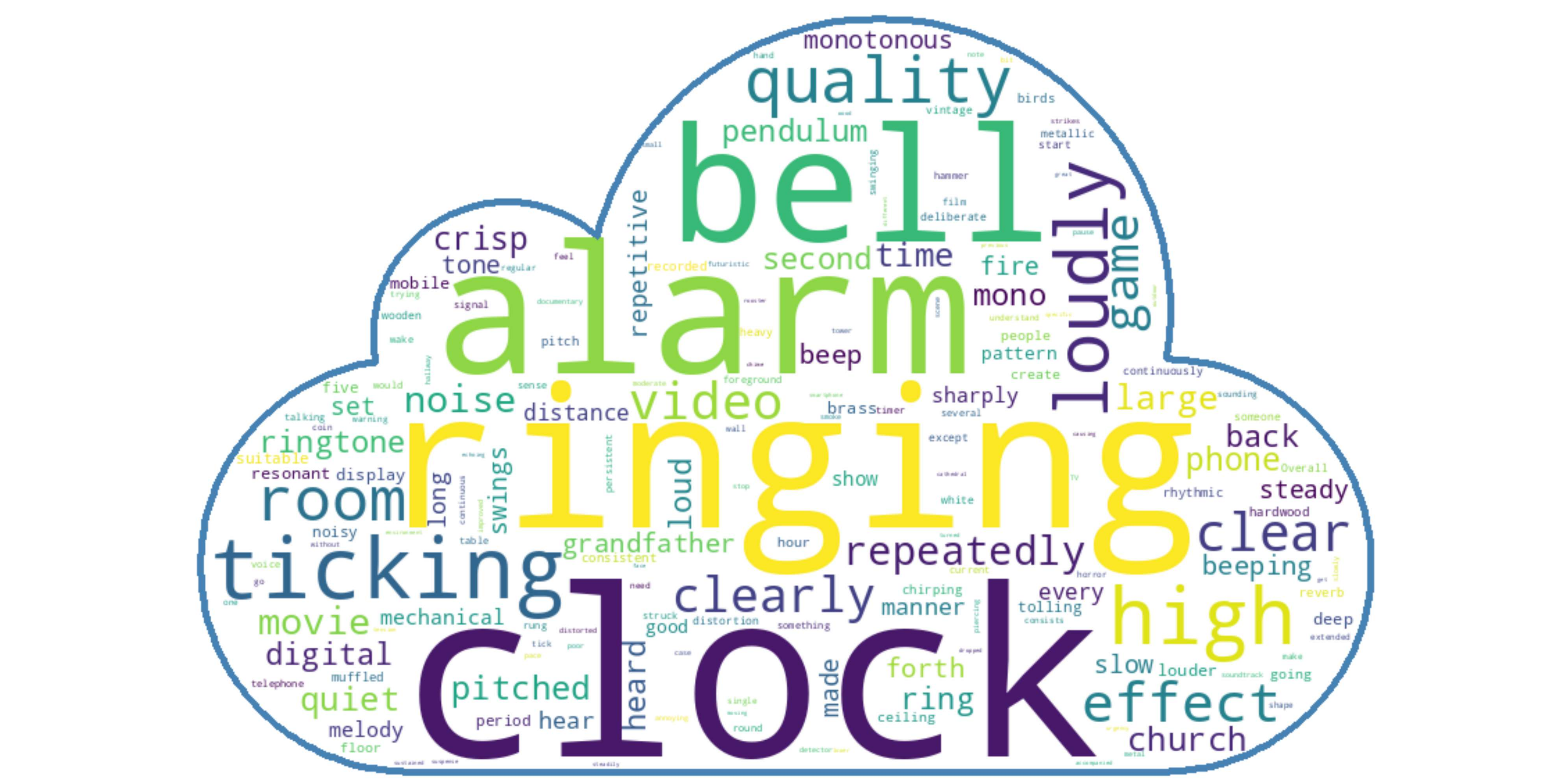} & \includegraphics[width=0.49\textwidth]{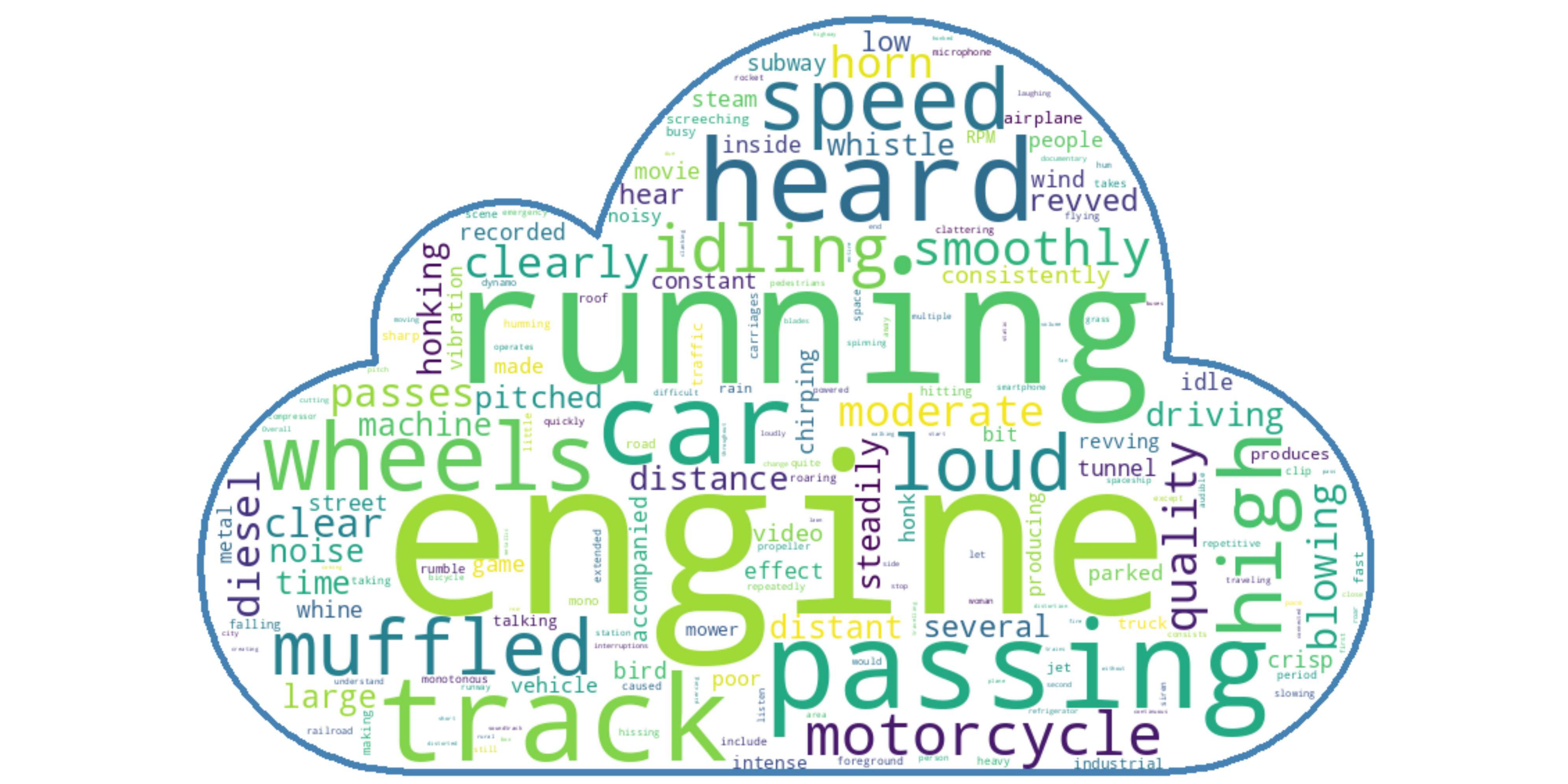 } 
  \\
  \midrule
\end{tabularx}

\noindent\begin{tabularx}{\textwidth}{@{}c|c|c@{}}  \midrule
Cluster 3 (424 / 636) & Cluster 4 (819 / 1207) & Cluster 5 (493 / 1449) 
  \\ 
  \midrule 
    \includegraphics[width=0.32\textwidth]{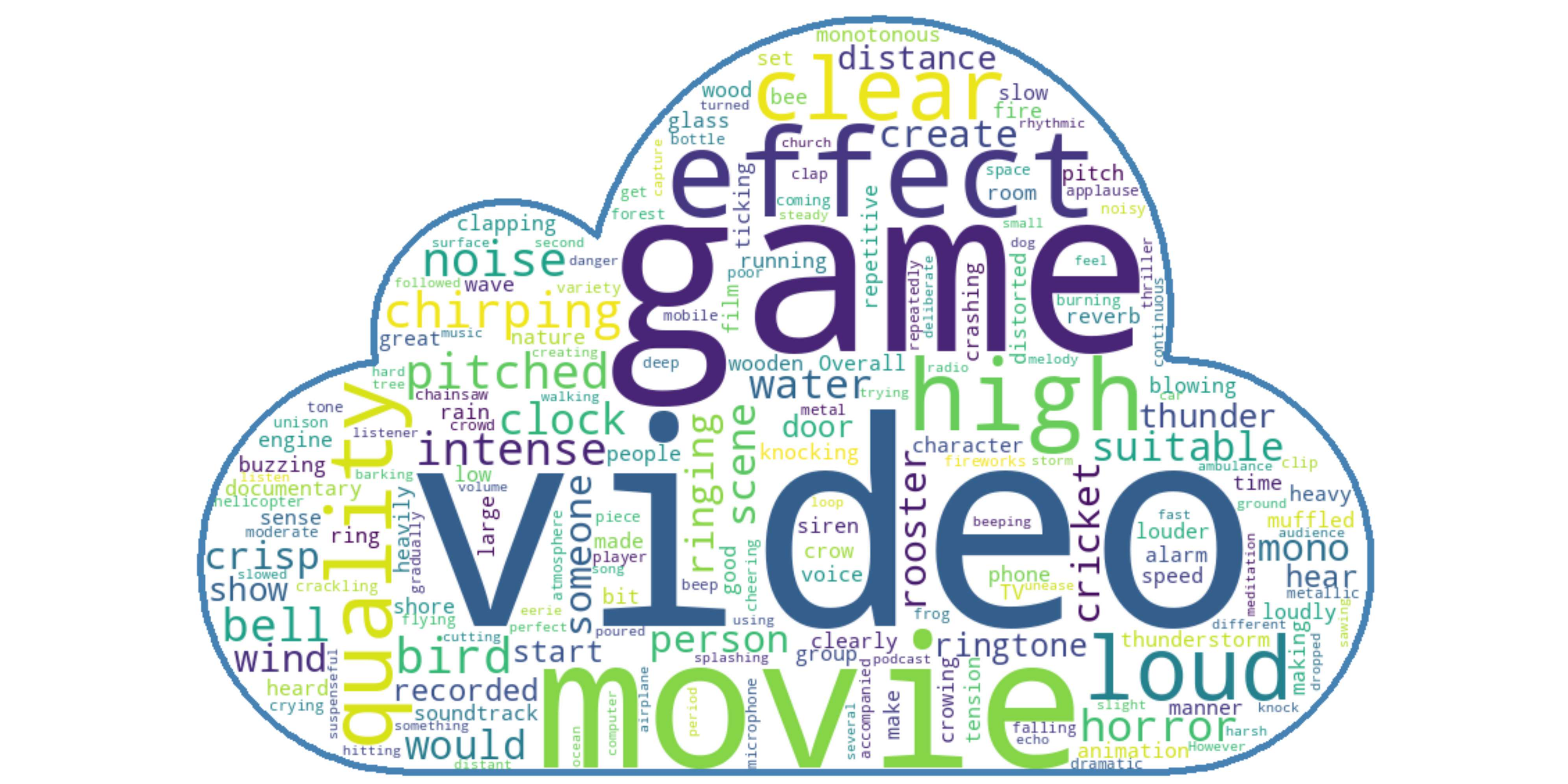} & \includegraphics[width=0.32\textwidth]{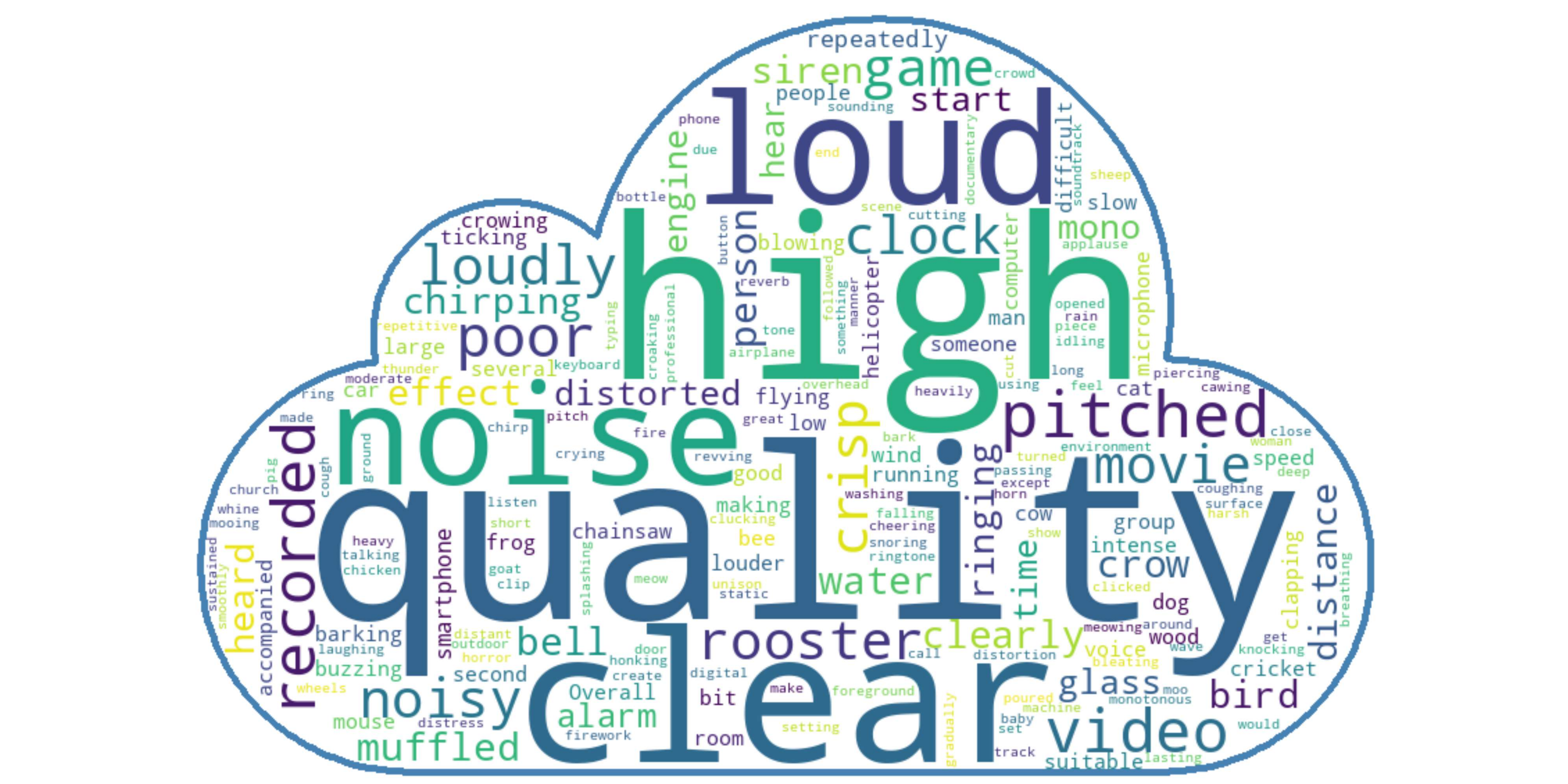 } &
    \includegraphics[width=0.32\textwidth]{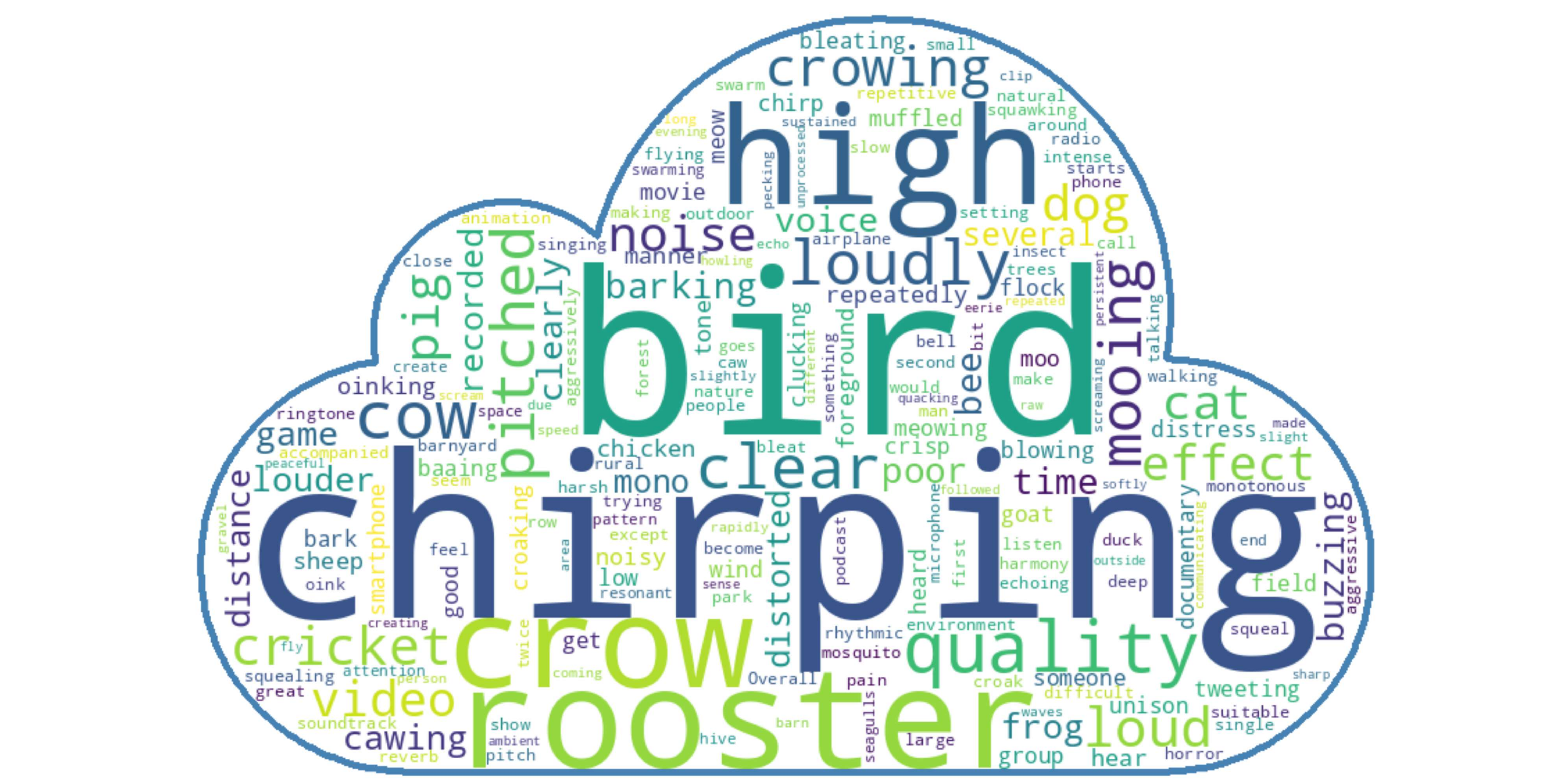 } 
  \\
  \bottomrule
\end{tabularx}

\noindent\begin{tabularx}{\textwidth}{@{}c|c|c@{}}  \midrule
Cluster 6 (532 / 651) & Cluster 7 (91 / 582)& Cluster 8 (357 / 977)
  \\ 
  \midrule 
    \includegraphics[width=0.32\textwidth]{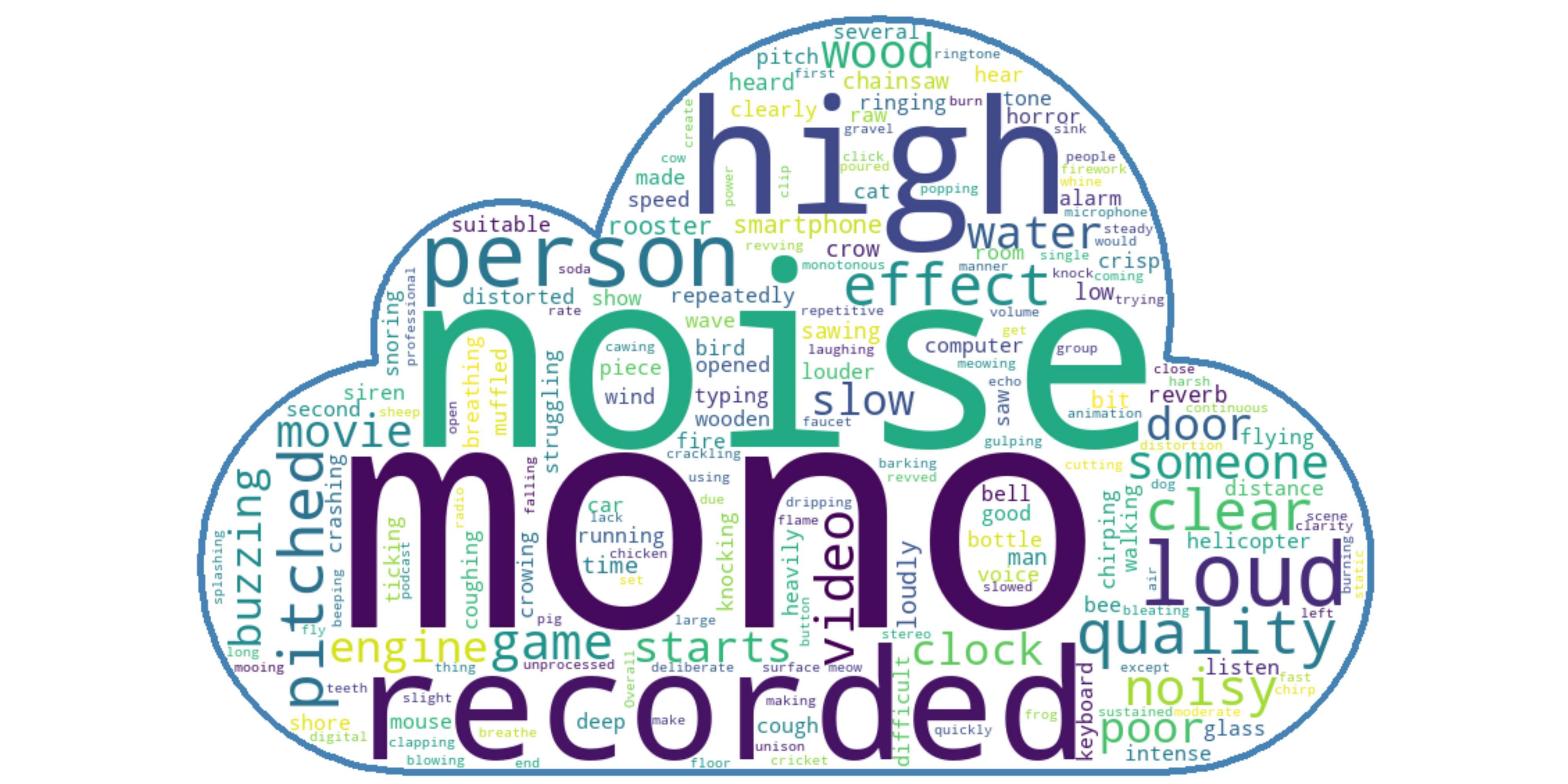} & \includegraphics[width=0.32\textwidth]{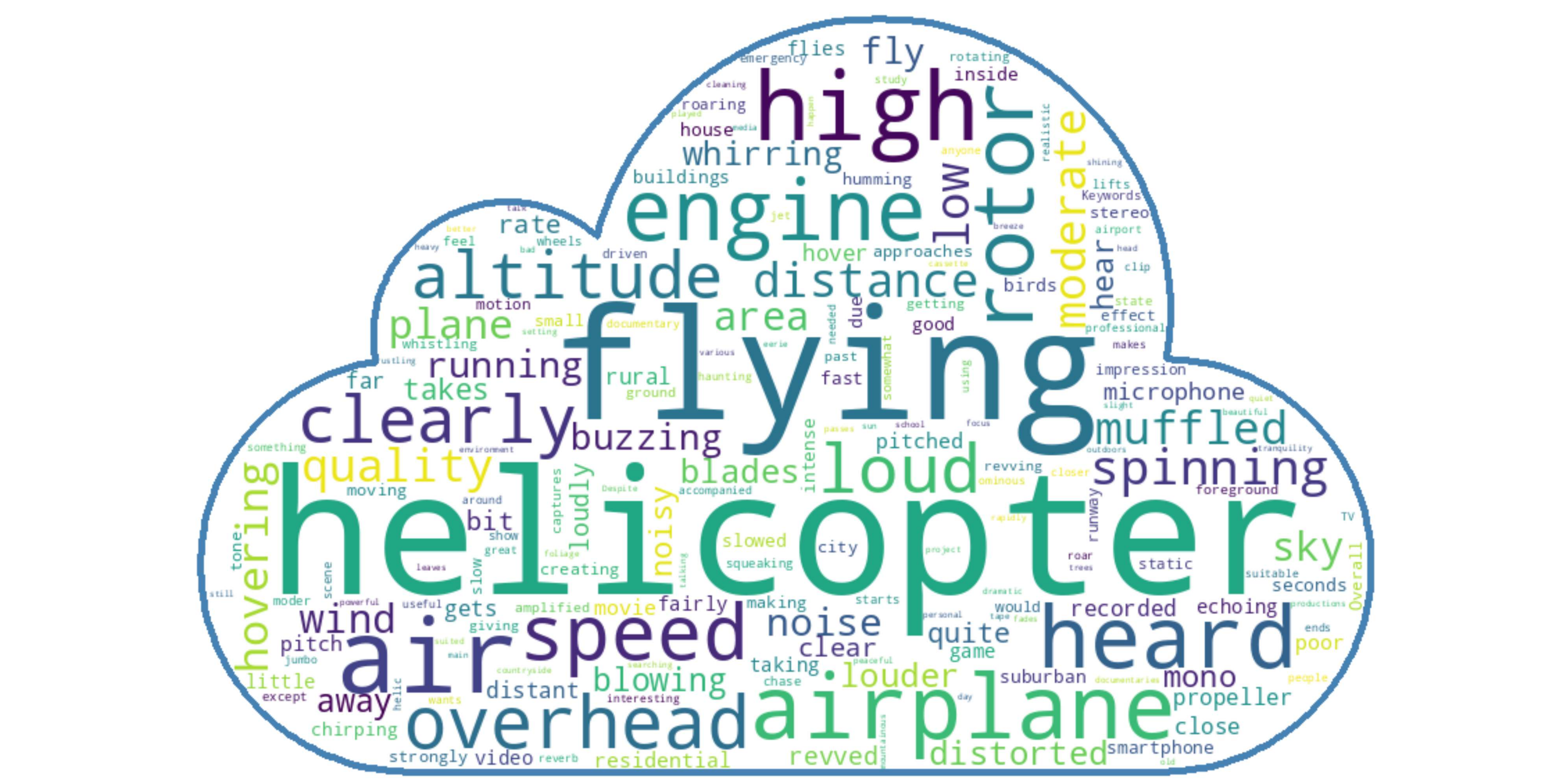} &
    \includegraphics[width=0.32\textwidth]{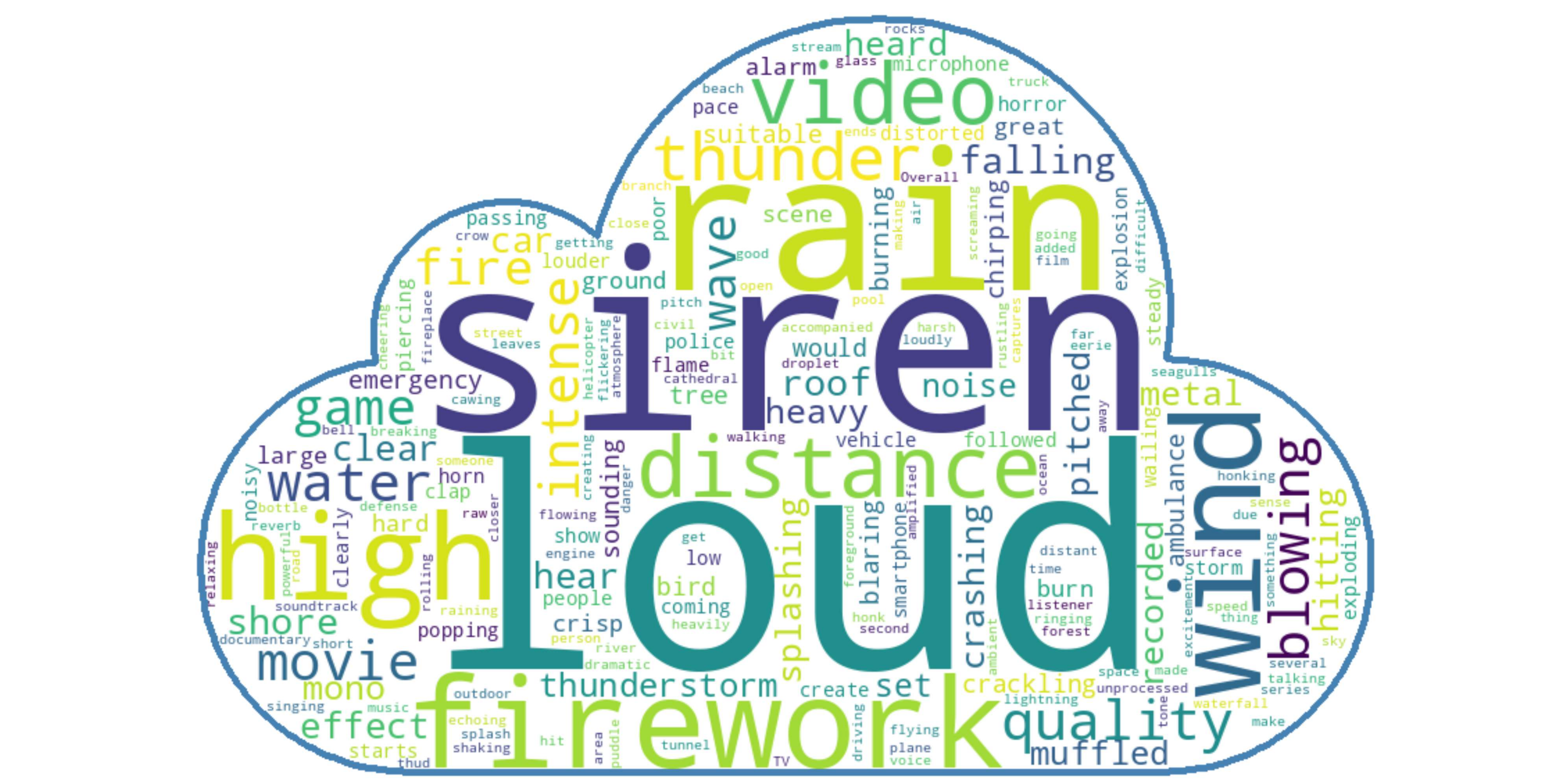} 
  \\
  \bottomrule
\end{tabularx}

\noindent
\begin{table}[ht]
\vspace{-2ex}
\begin{tabularx}{\textwidth}{@{}c|c|c@{}}  \midrule
 Cluster 9 (891 / 2084) & Cluster 10 (43 / 235) & Cluster 11 (281 / 747)
  \\ 
  \midrule 
    \includegraphics[width=0.32\textwidth]{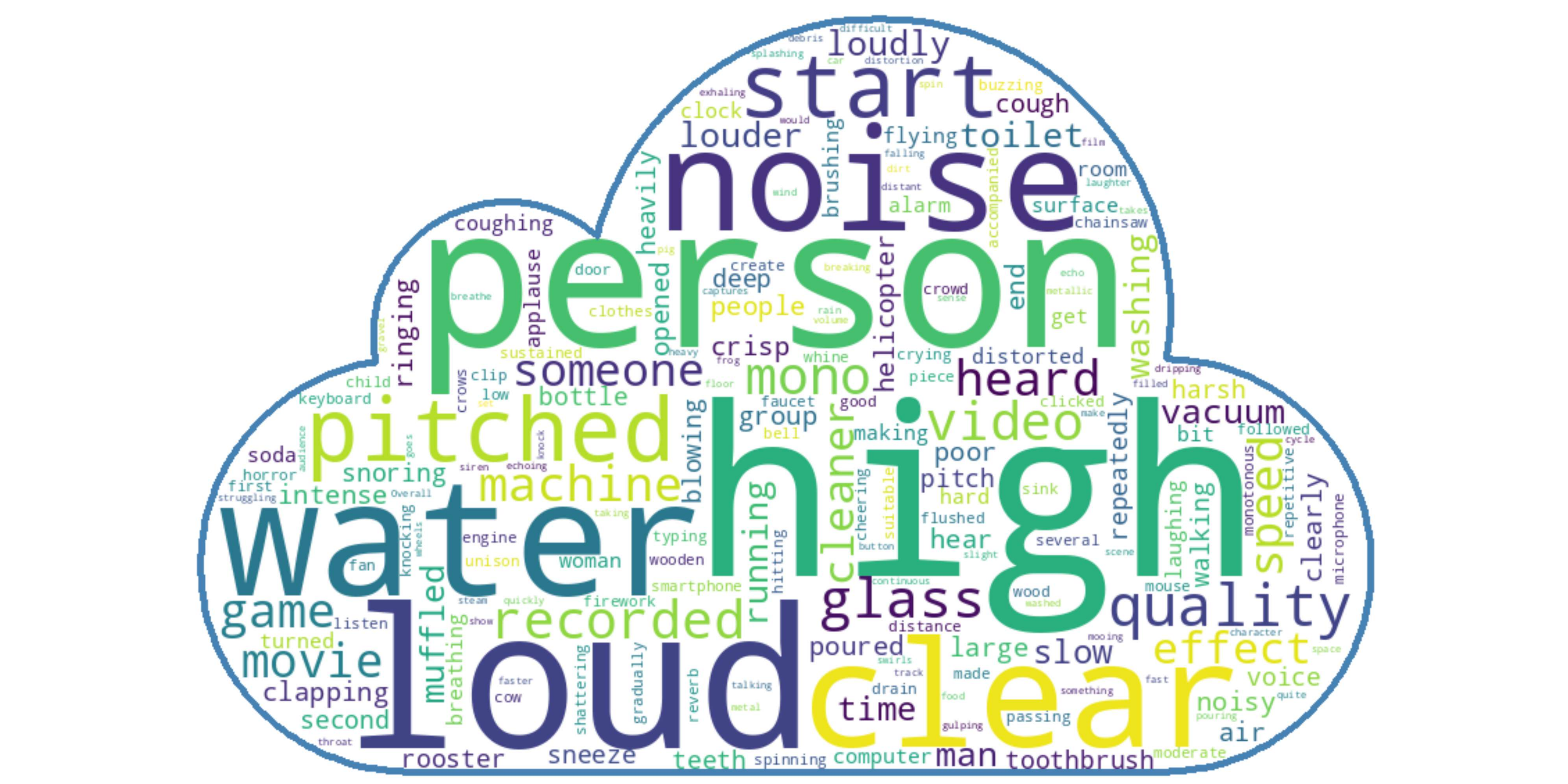} & \includegraphics[width=0.32\textwidth]{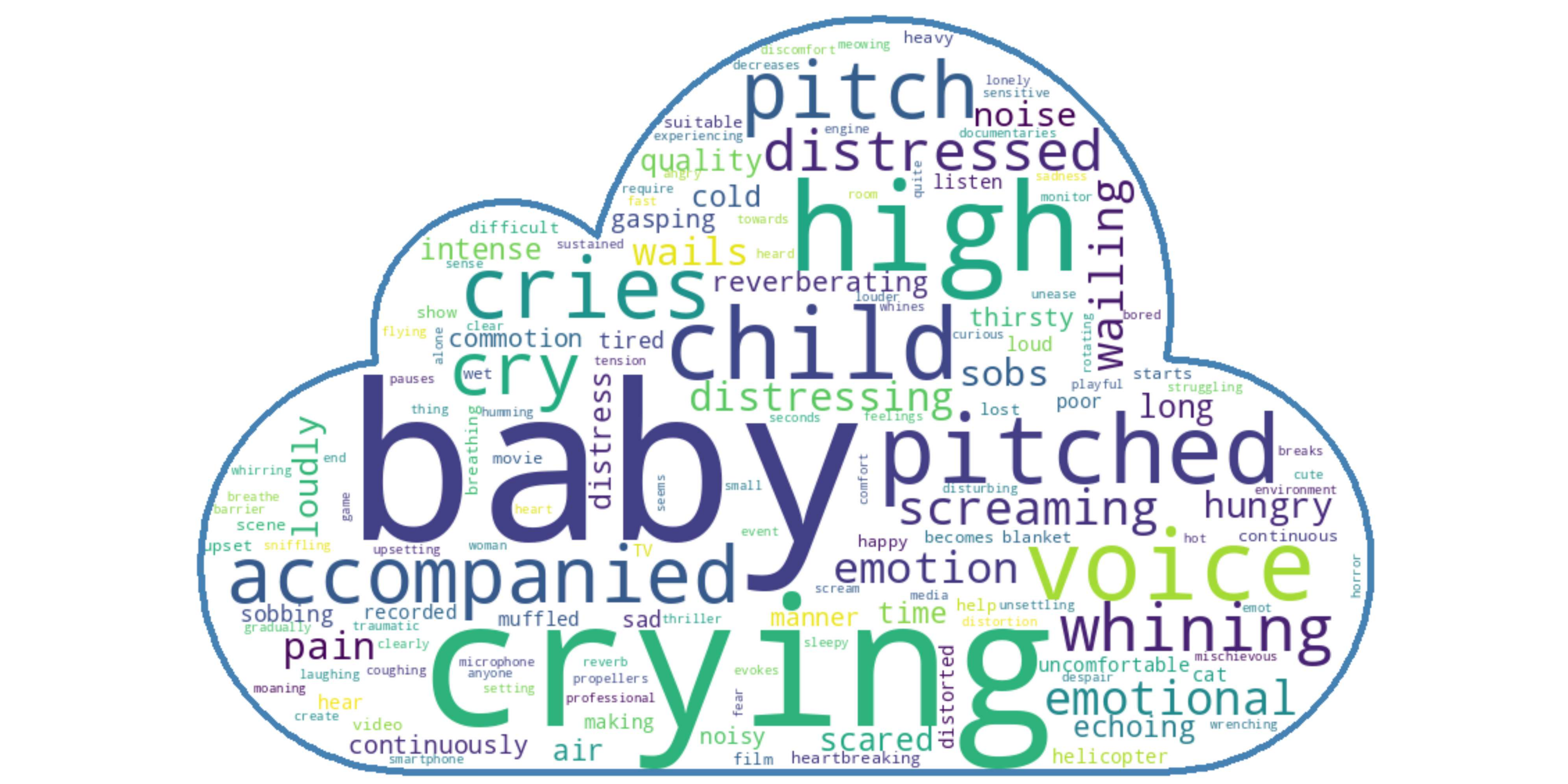} &
    \includegraphics[width=0.32\textwidth]{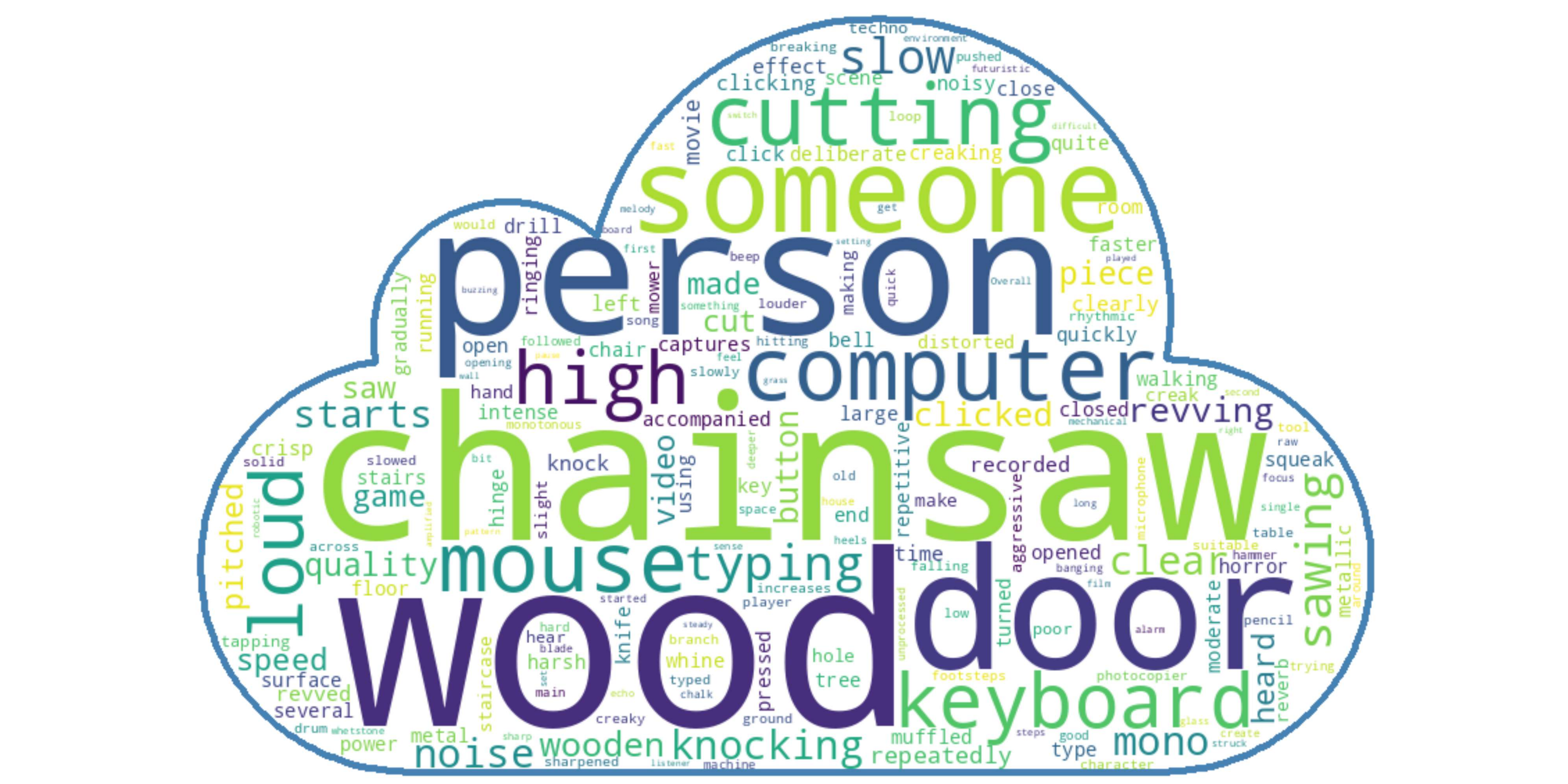} 
  \\
  \bottomrule
\end{tabularx}
  \caption{Wordclouds of different clusters. The numbers in parentheses refers to the count of audio descriptions / sentences in this cluster.}
\label{fig:wordcloud-esc50}
\end{table}

\newpage
\subsection{Measuring Neuron Interpretability under different $\tau$ and top-$K$}
\label{sup subsec: interpretability threshold}
\begin{figure}[ht]
\begin{subfigure}{.3\textwidth}
  \centering
  \includegraphics[width=\linewidth]{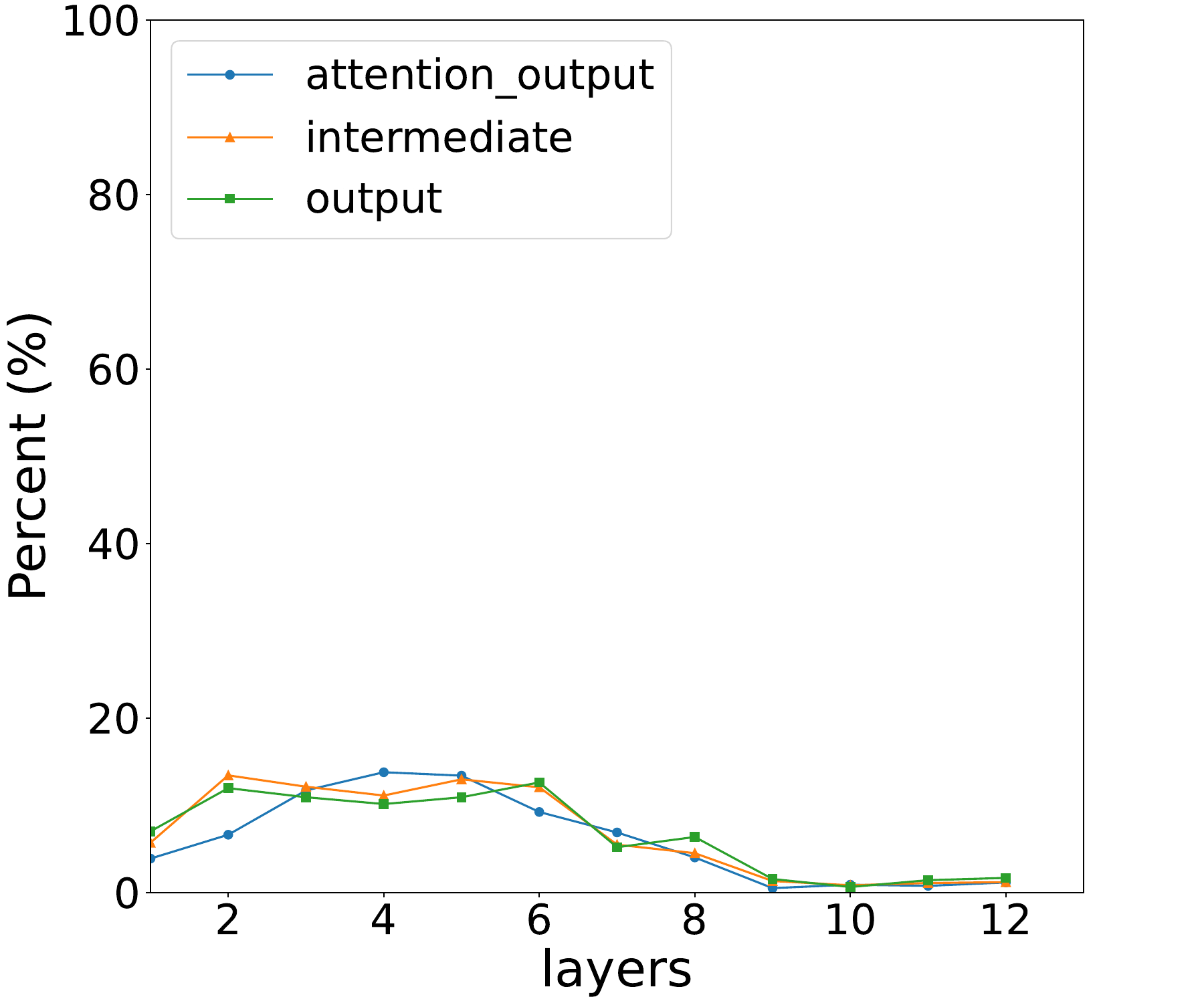}
  \caption{AST, $\tau = 3$}
\end{subfigure}%
\hfill
\begin{subfigure}{.3\textwidth}
  \centering
    \includegraphics[width=\linewidth]{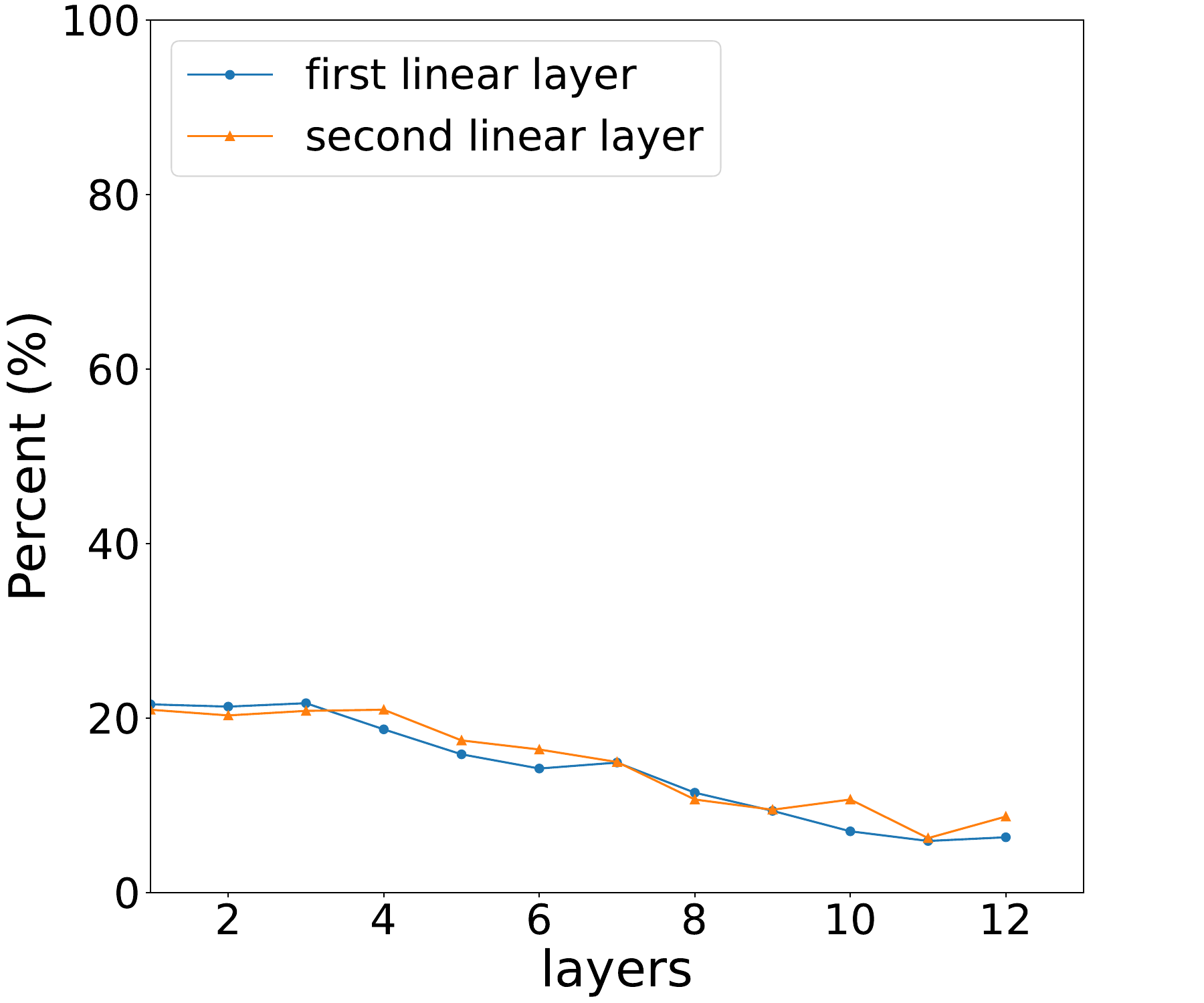}
  \caption{BEATs-finetuned, $\tau = 3$}
\end{subfigure}%
\hfill
\begin{subfigure}{.3\textwidth}
  \centering
  \includegraphics[width=\linewidth]{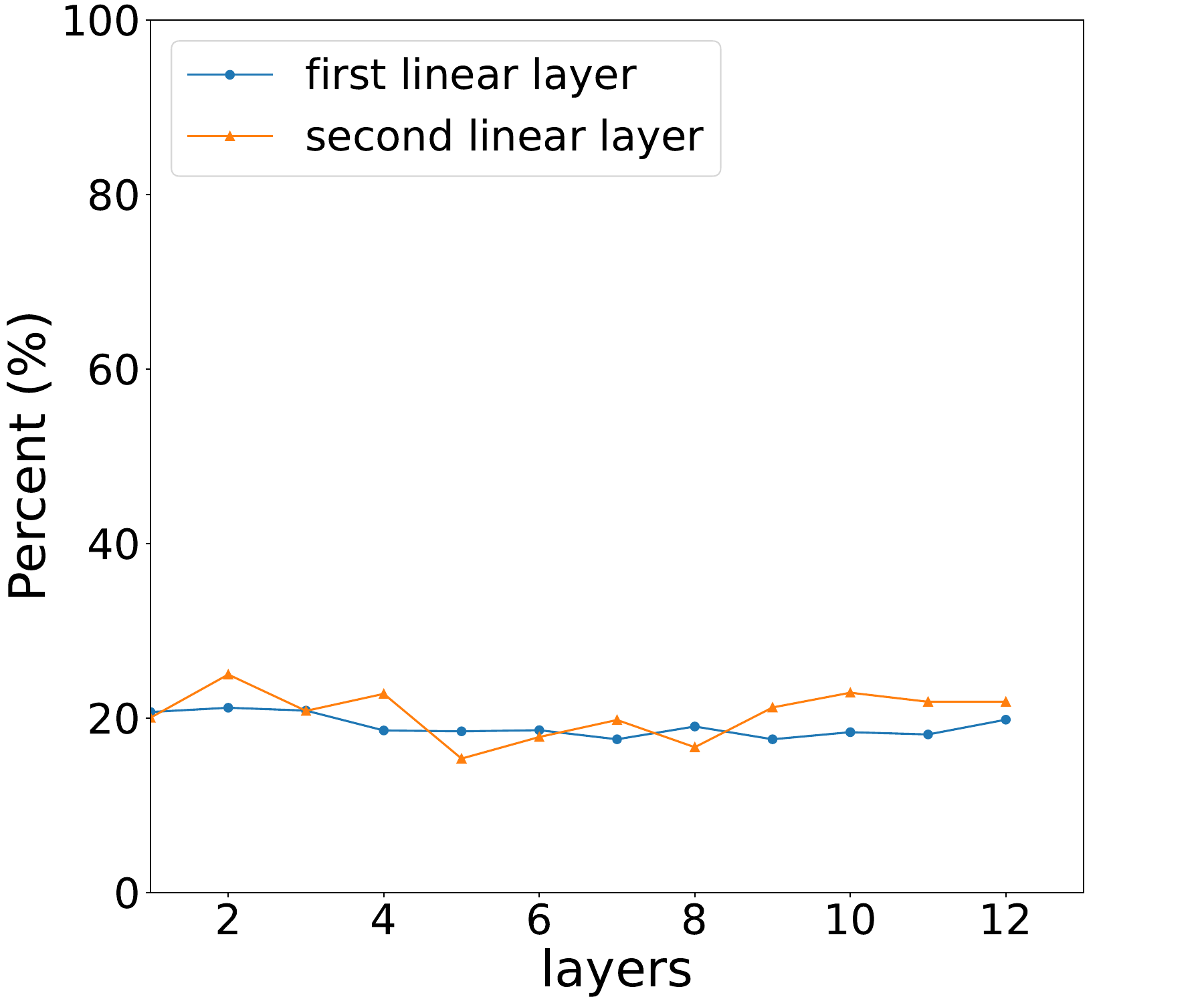}
  \caption{BEATs-frozen, $\tau = 3$}
\end{subfigure}

\vspace{30px}

\begin{subfigure}{.3\textwidth}
  \centering
  \includegraphics[width=\linewidth]{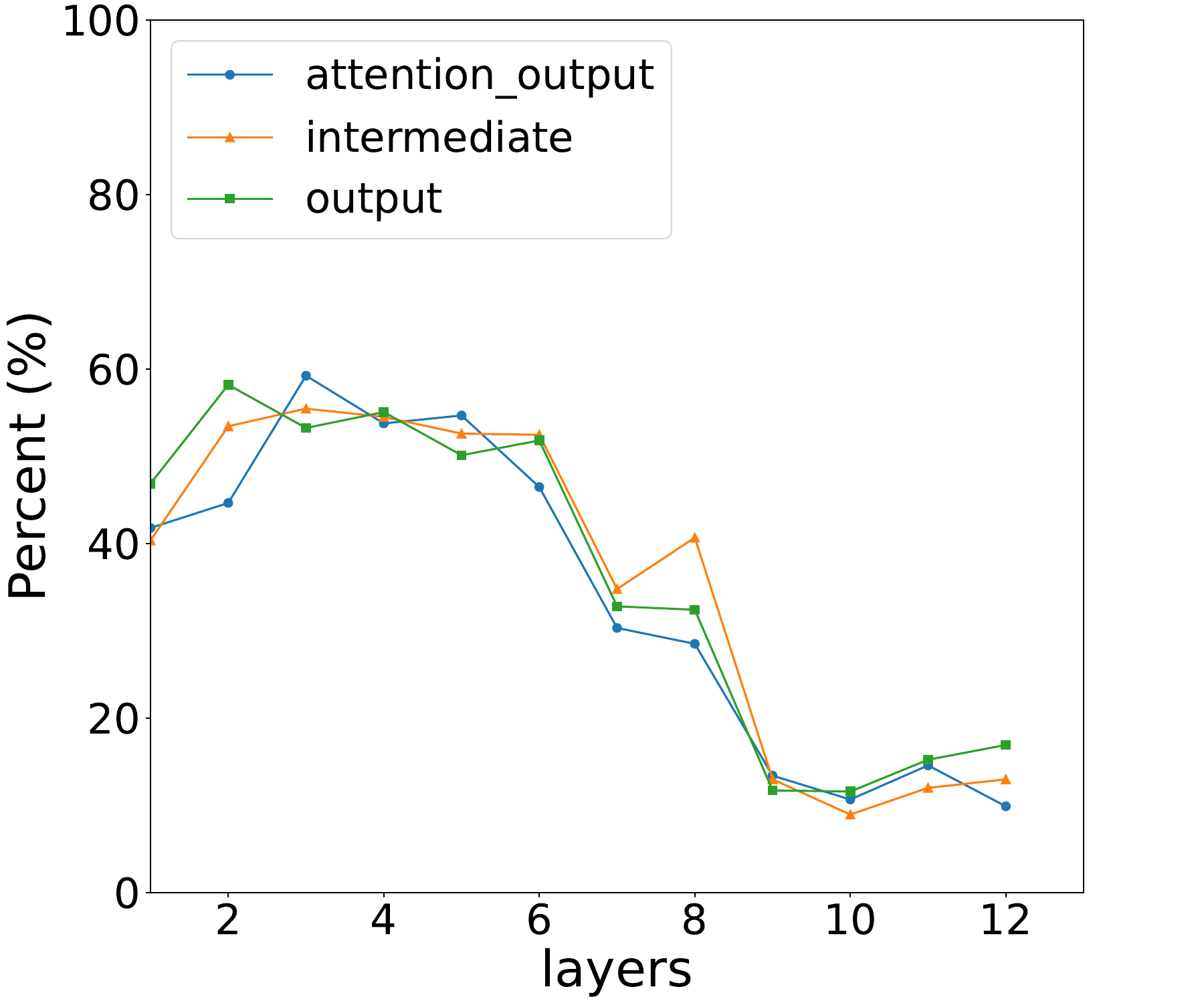}
  \caption{AST, $\tau = 4$ }
\end{subfigure}%
\hfill
\begin{subfigure}{.3\textwidth}
  \centering
  \includegraphics[width=\linewidth]{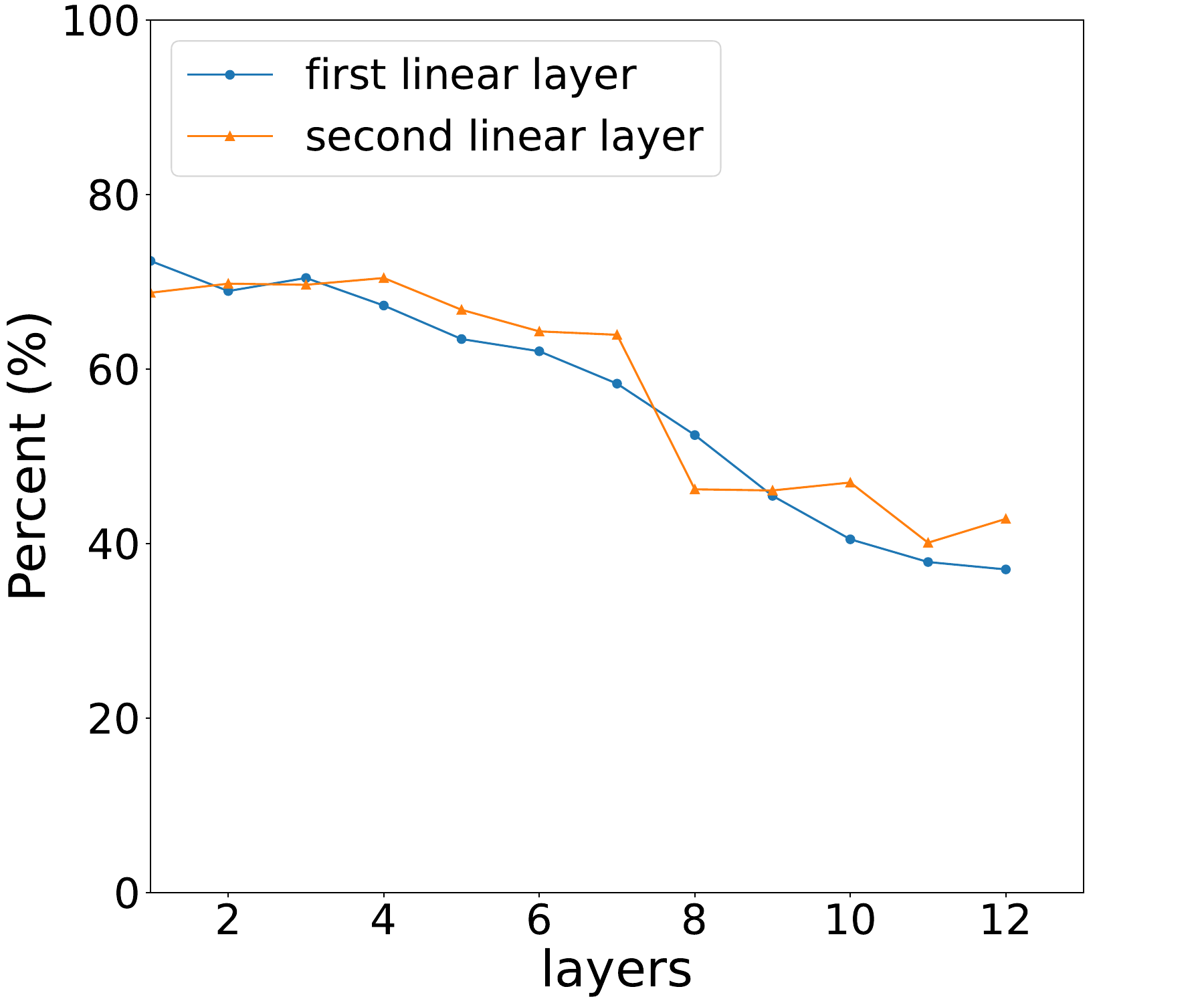}
  \caption{BEATs-finetuned, $\tau = 4$}
\end{subfigure}%
\hfill
\begin{subfigure}{.3\textwidth}
  \centering
  \includegraphics[width=\linewidth]{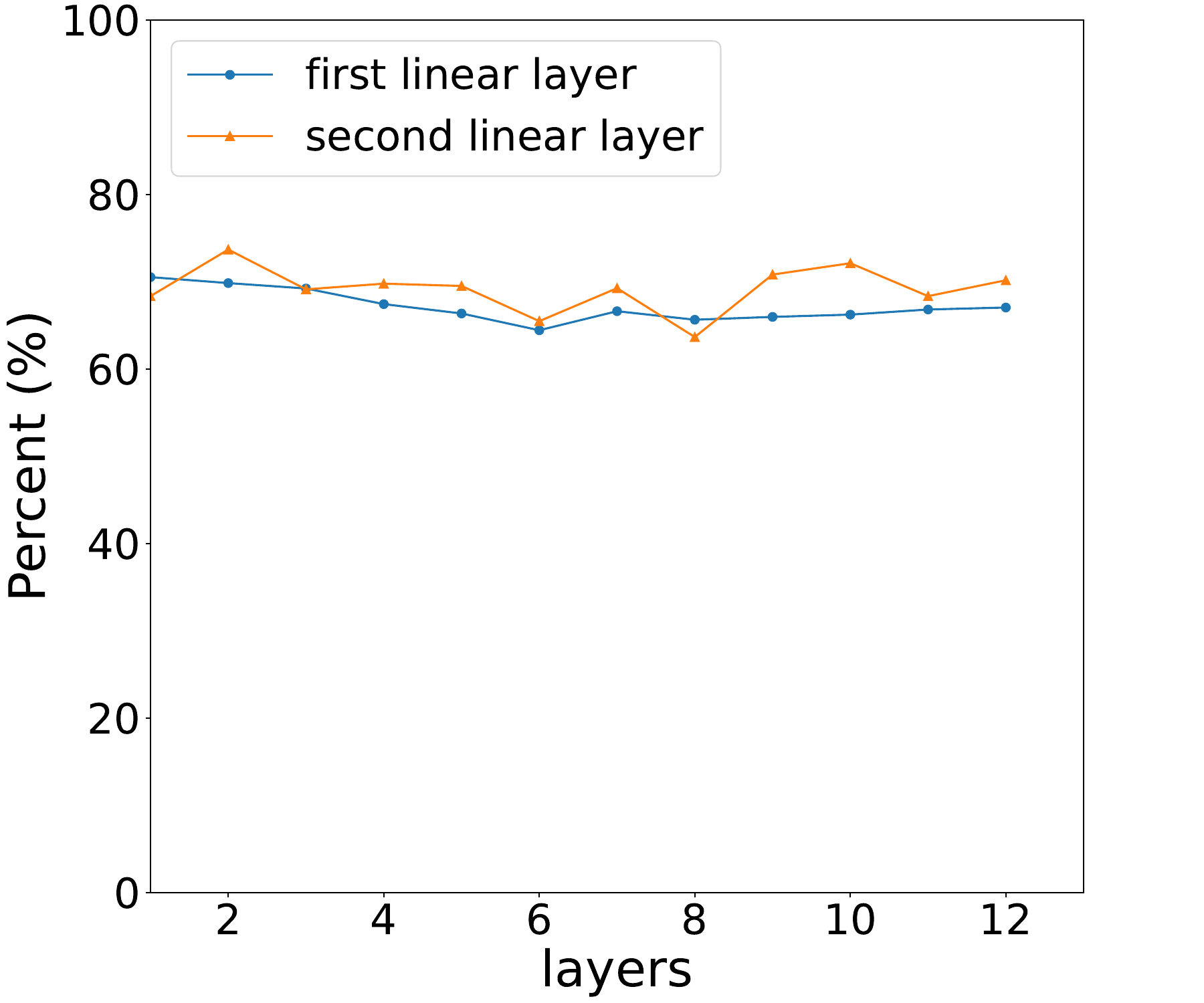}
  \caption{BEATs-frozen, $\tau = 4$}
\end{subfigure}

\vspace{30px}
\begin{subfigure}{.3\textwidth}
  \centering
  \includegraphics[width=\linewidth]{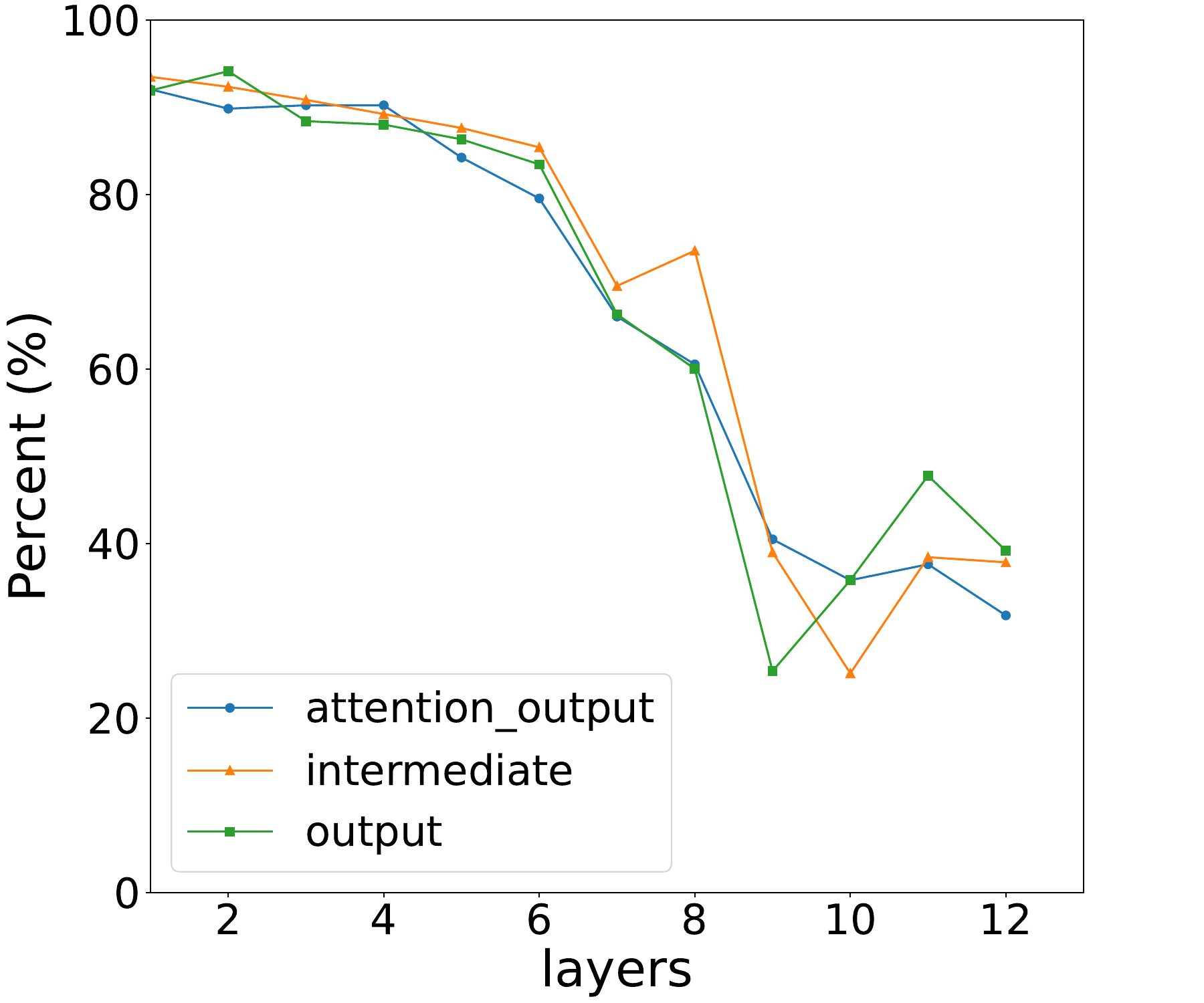}
  \caption{AST, $\tau = 5$ }
\end{subfigure}%
\hfill
\begin{subfigure}{.3\textwidth}
  \centering
  \includegraphics[width=\linewidth]{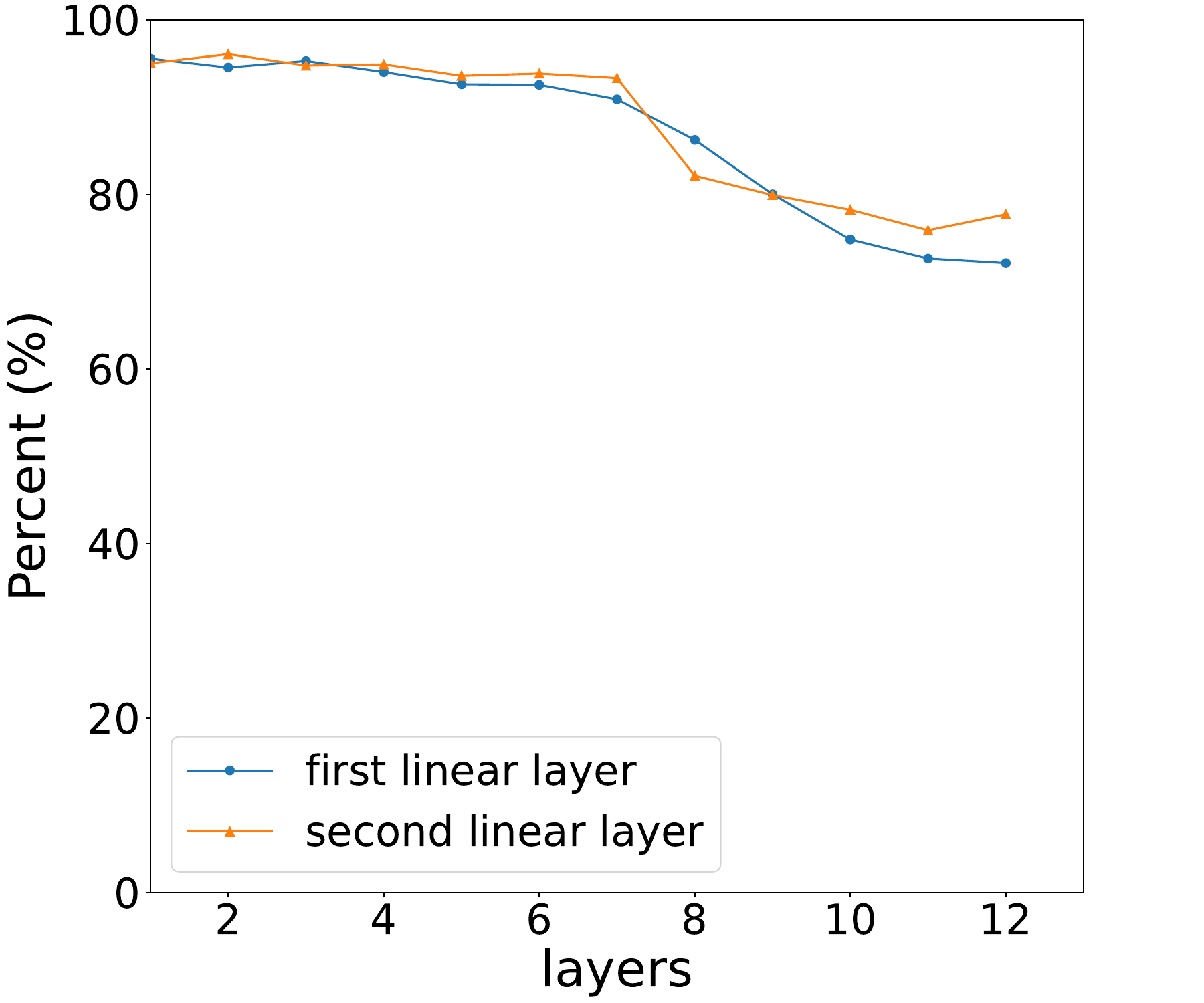}
  \caption{BEATs-finetuned, $\tau = 5$}
\end{subfigure}%
\hfill
\begin{subfigure}{.3\textwidth}
  \centering
  \includegraphics[width=\linewidth]{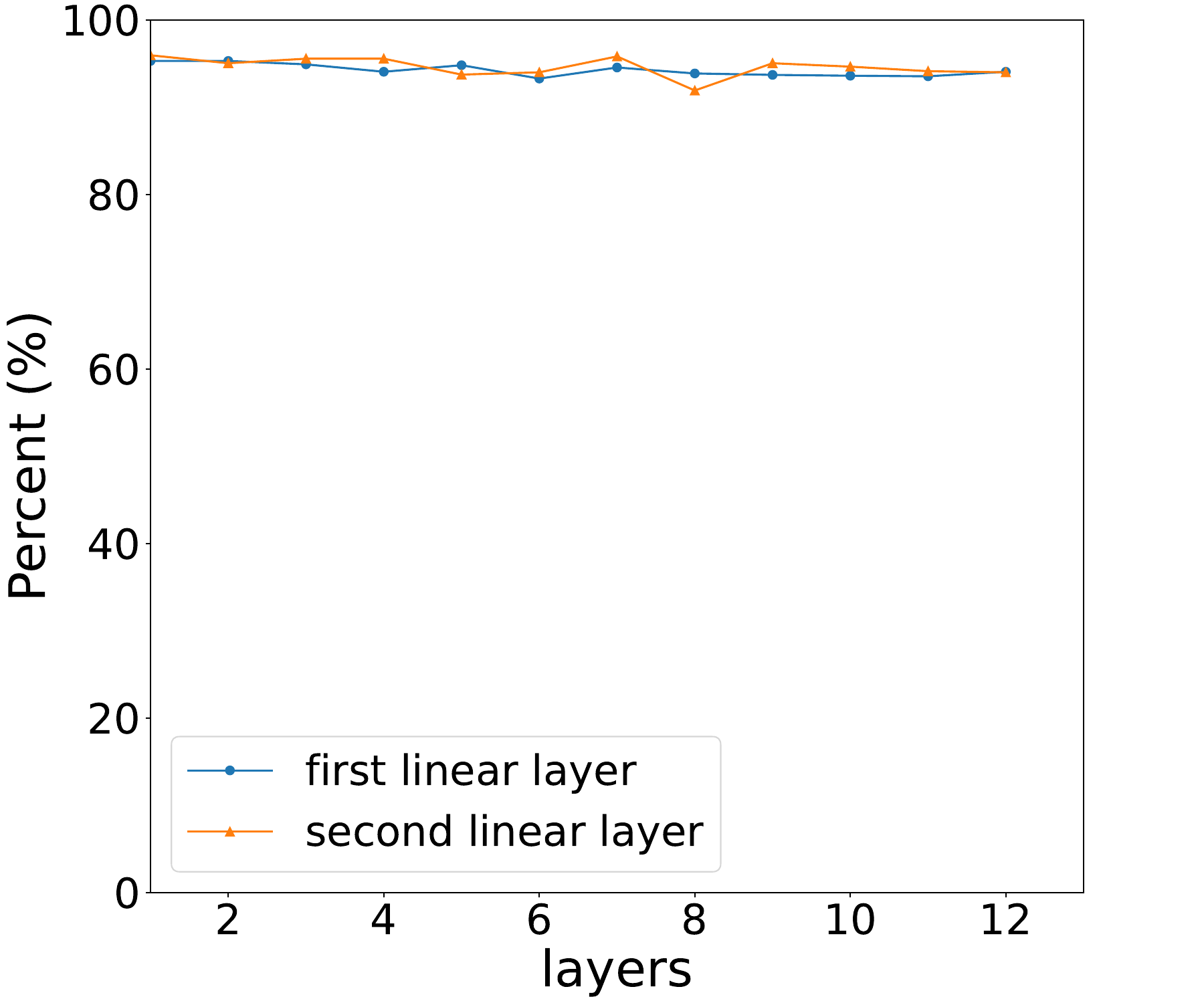}
  \caption{BEATs-frozen, $\tau = 5$}
\end{subfigure}
\caption{Percentage of polysemantic  neurons when adopting $\tau = 3, 4, 5$ and top-$K = 5$}
\label{fig:hyperparameters_top5}
\end{figure}

\begin{figure}[ht]
\begin{subfigure}{.3\textwidth}
  \centering
  \includegraphics[width=\linewidth]{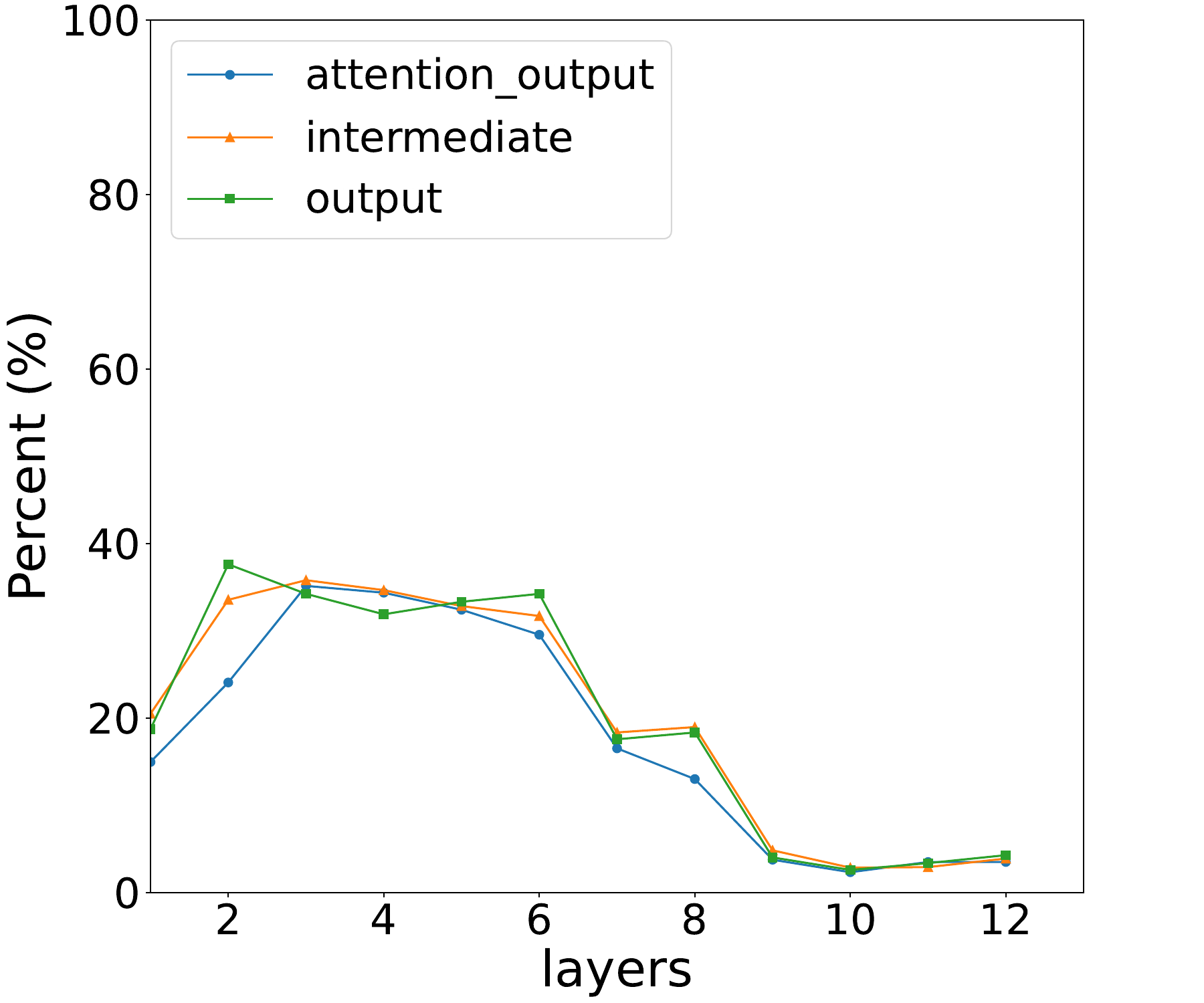}
  \caption{AST, $\tau = 6$}
\end{subfigure}%
\hfill
\begin{subfigure}{.3\textwidth}
  \centering
    \includegraphics[width=\linewidth]{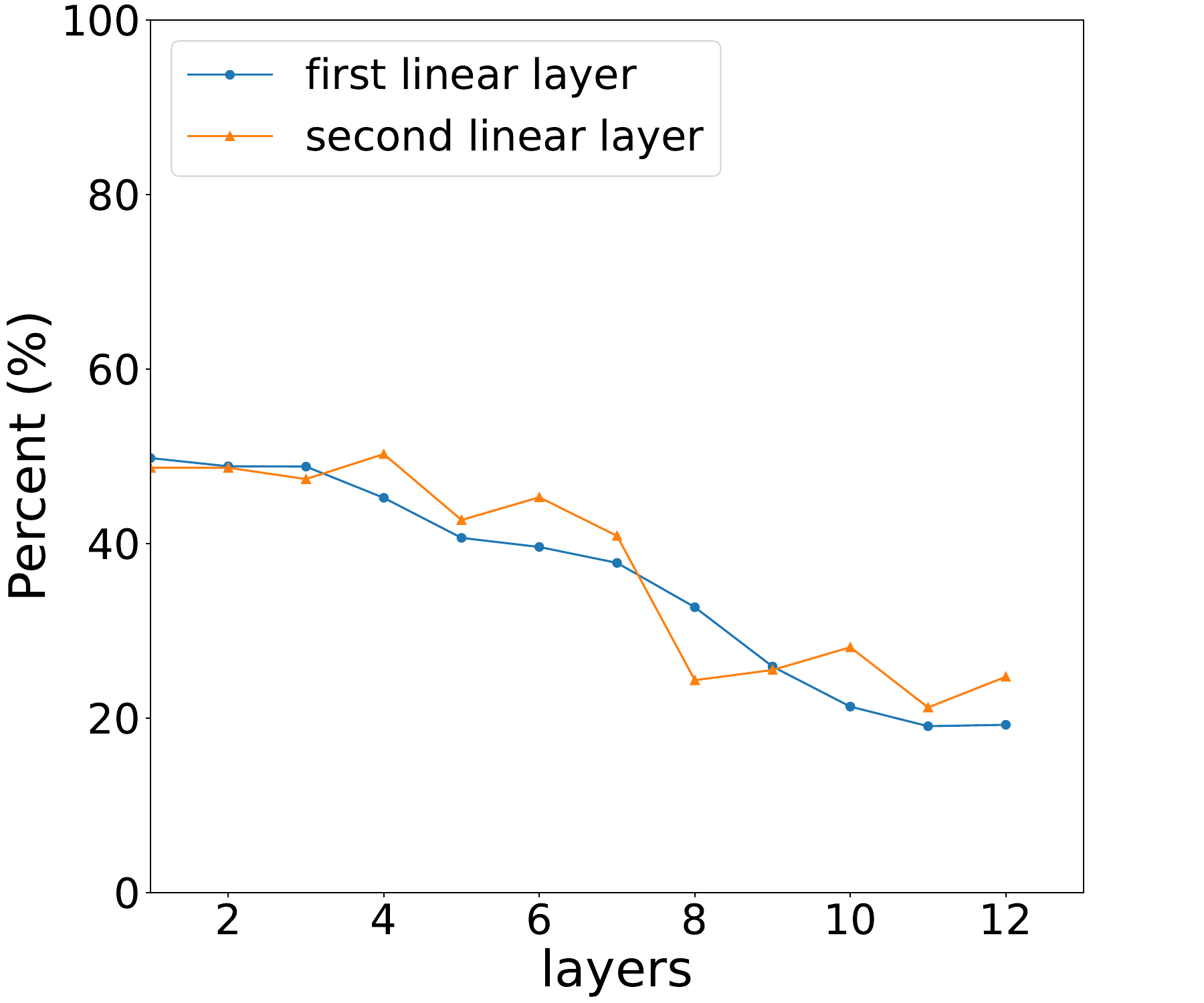}
  \caption{BEATs-finetuned, $\tau = 6$}
\end{subfigure}%
\hfill
\begin{subfigure}{.3\textwidth}
  \centering
  \includegraphics[width=\linewidth]{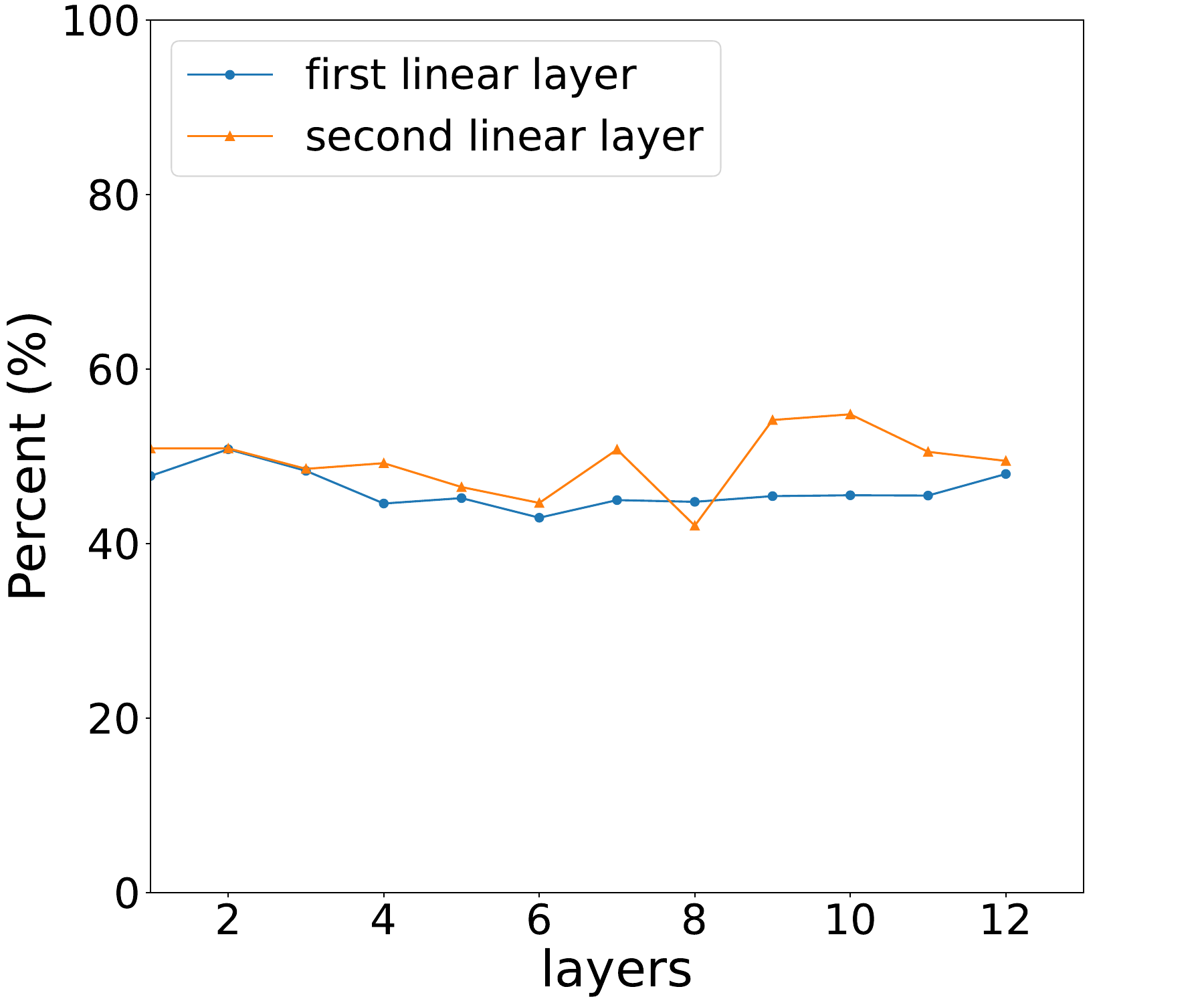}
  \caption{BEATs-frozen, $\tau = 6$}
\end{subfigure}

\vspace{30px}

\begin{subfigure}{.3\textwidth}
  \centering
  \includegraphics[width=\linewidth]{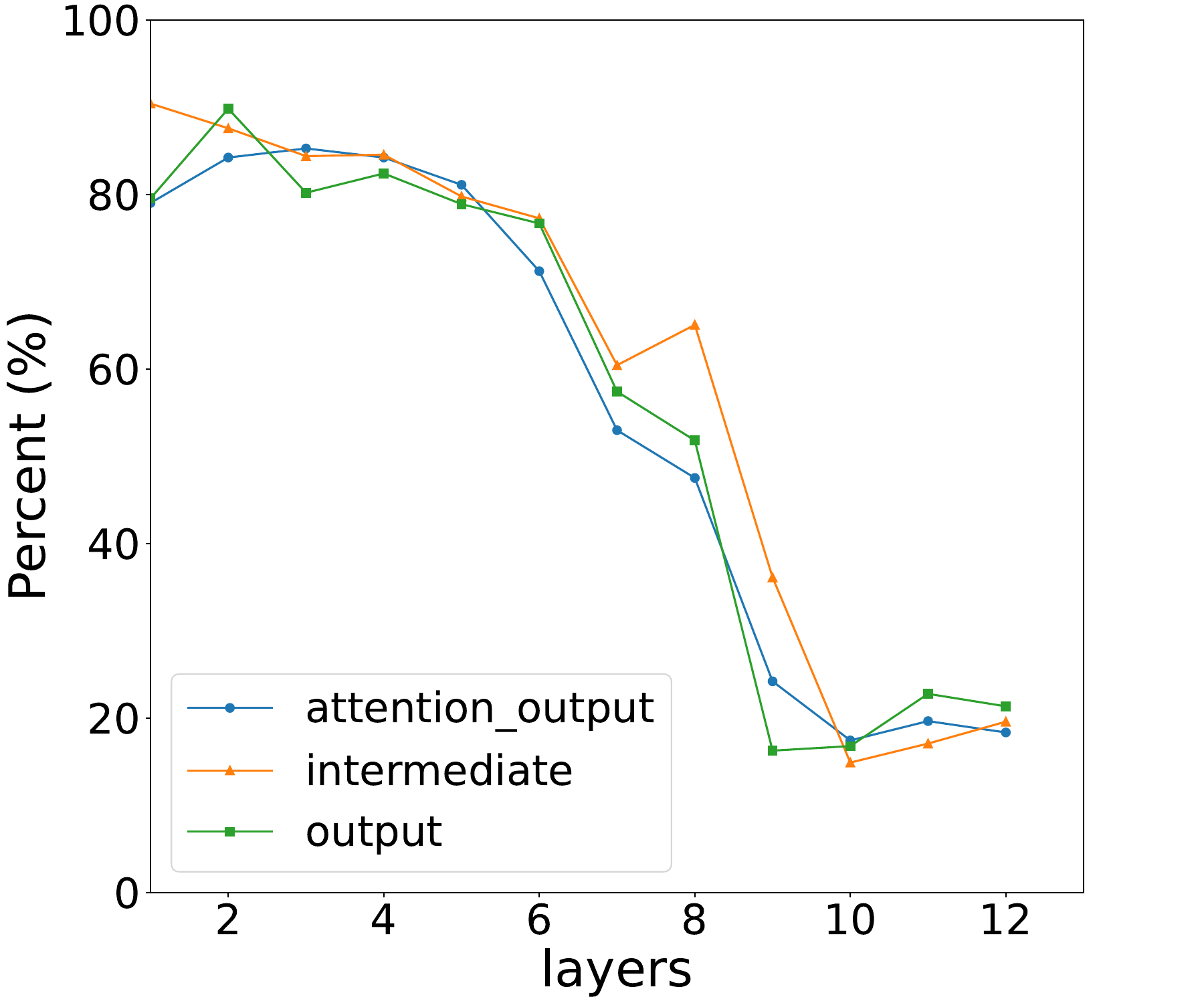}
  \caption{AST, $\tau = 8$ }
\end{subfigure}%
\hfill
\begin{subfigure}{.3\textwidth}
  \centering
  \includegraphics[width=\linewidth]{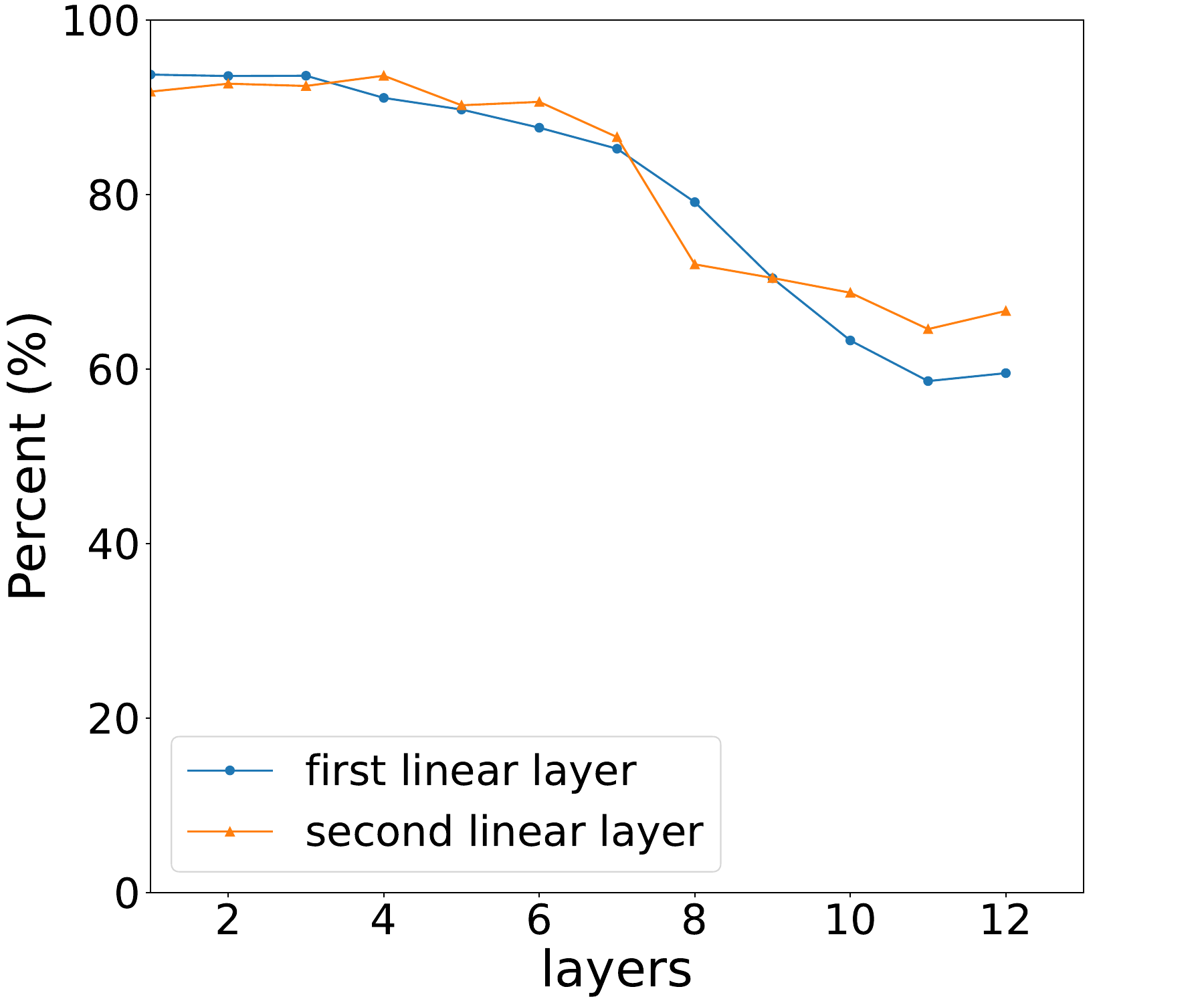}
  \caption{BEATs-finetuned, $\tau = 8$}
\end{subfigure}%
\hfill
\begin{subfigure}{.3\textwidth}
  \centering
  \includegraphics[width=\linewidth]{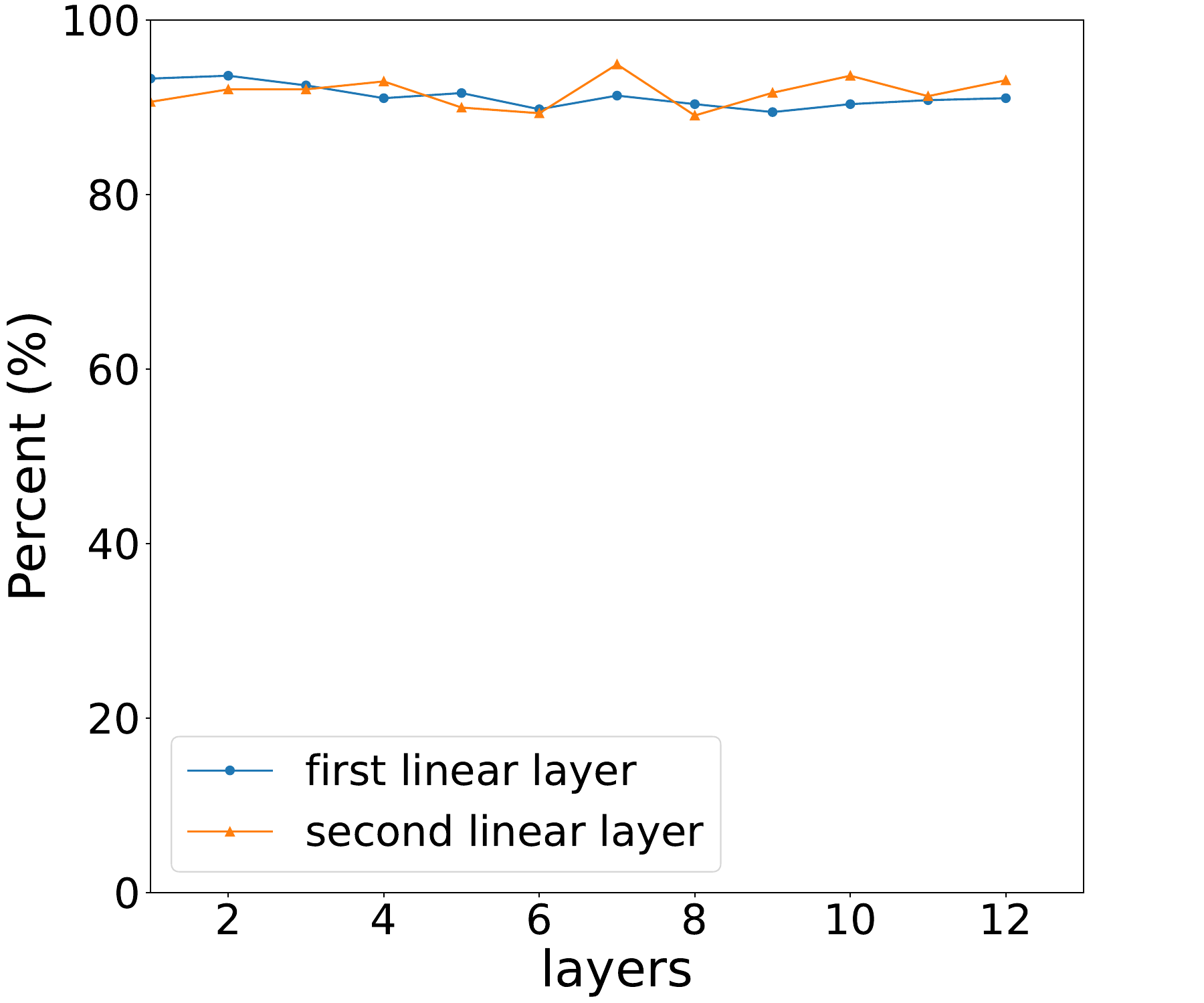}
  \caption{BEATs-frozen, $\tau = 8$}
\end{subfigure}

\vspace{30px}
\begin{subfigure}{.3\textwidth}
  \centering
  \includegraphics[width=\linewidth]{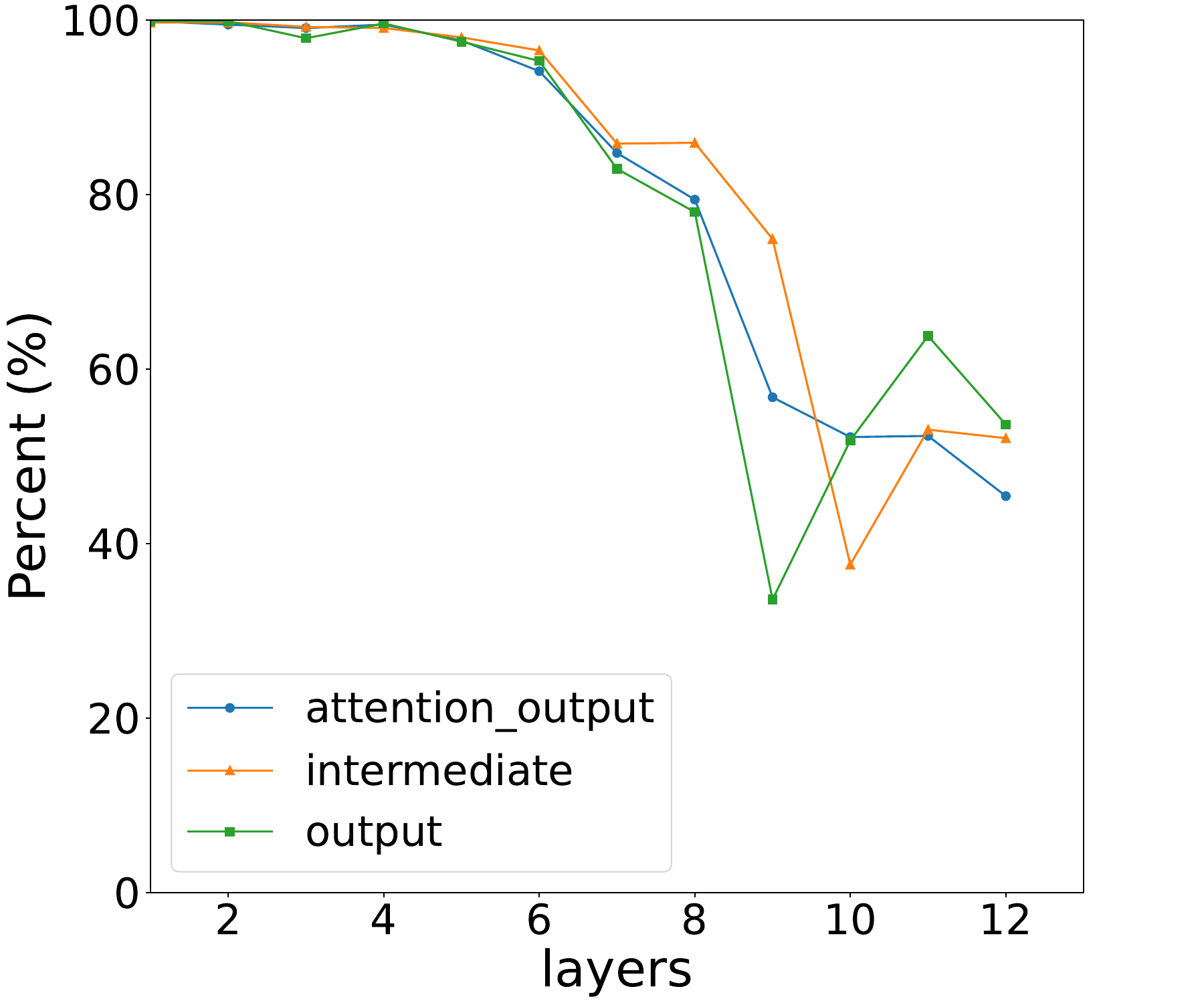}
  \caption{AST, $\tau = 10$ }
\end{subfigure}%
\hfill
\begin{subfigure}{.3\textwidth}
  \centering
  \includegraphics[width=\linewidth]{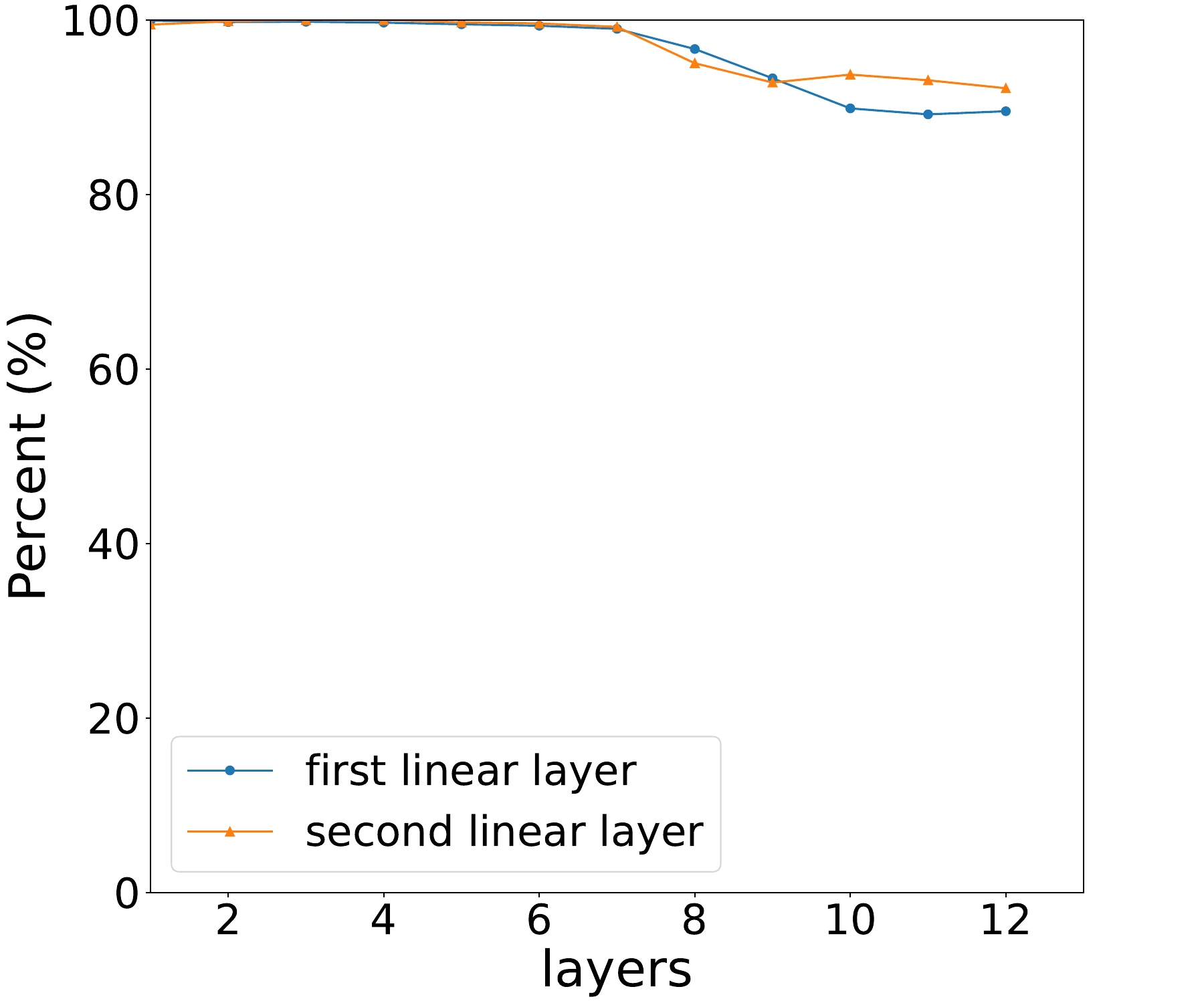}
  \caption{BEATs-finetuned, $\tau = 10$}
\end{subfigure}%
\hfill
\begin{subfigure}{.3\textwidth}
  \centering
  \includegraphics[width=\linewidth]{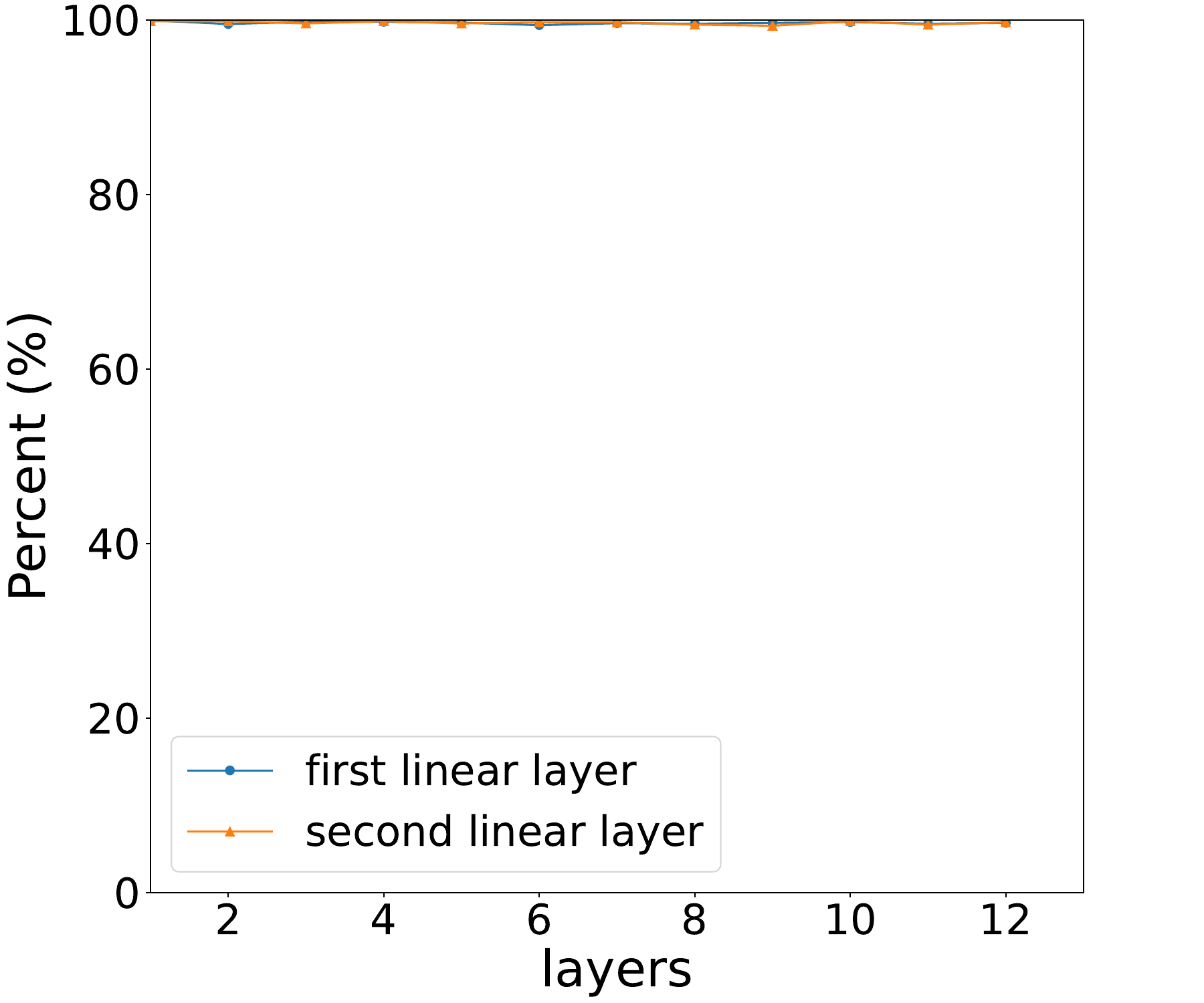}
  \caption{BEATs-frozen, $\tau = 10$}
\end{subfigure}
\caption{Percentage of polysemantic neurons when adopting $\tau = 6, 8, 10$ and top-$K = 10$.}
\label{fig:hyperparameters_top10}
\end{figure}

\begin{figure}[ht]
\begin{subfigure}{.3\textwidth}
  \centering
  \includegraphics[width=\linewidth]{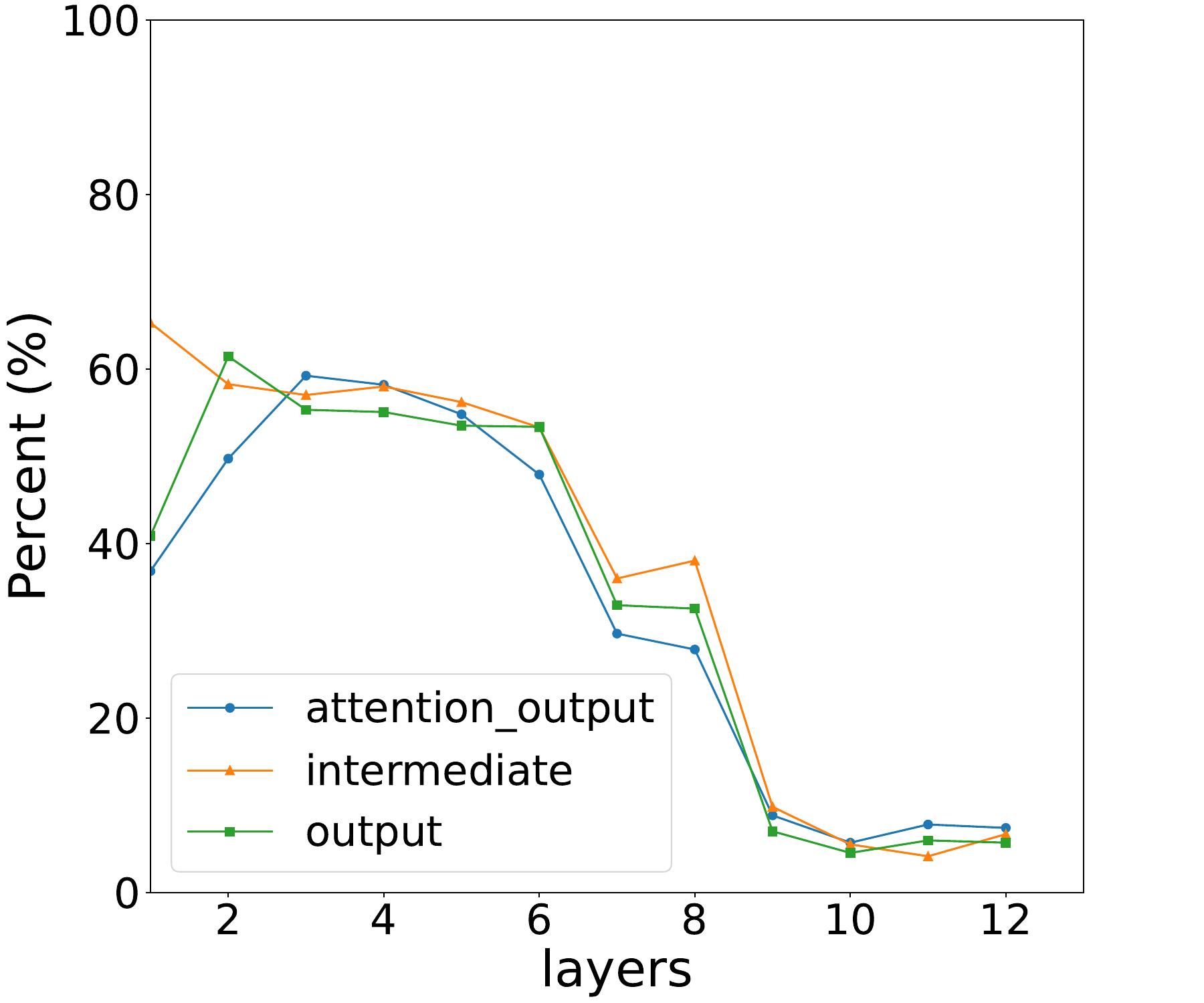}
  \caption{AST, $\tau = 12$}
\end{subfigure}%
\hfill
\begin{subfigure}{.3\textwidth}
  \centering
    \includegraphics[width=\linewidth]{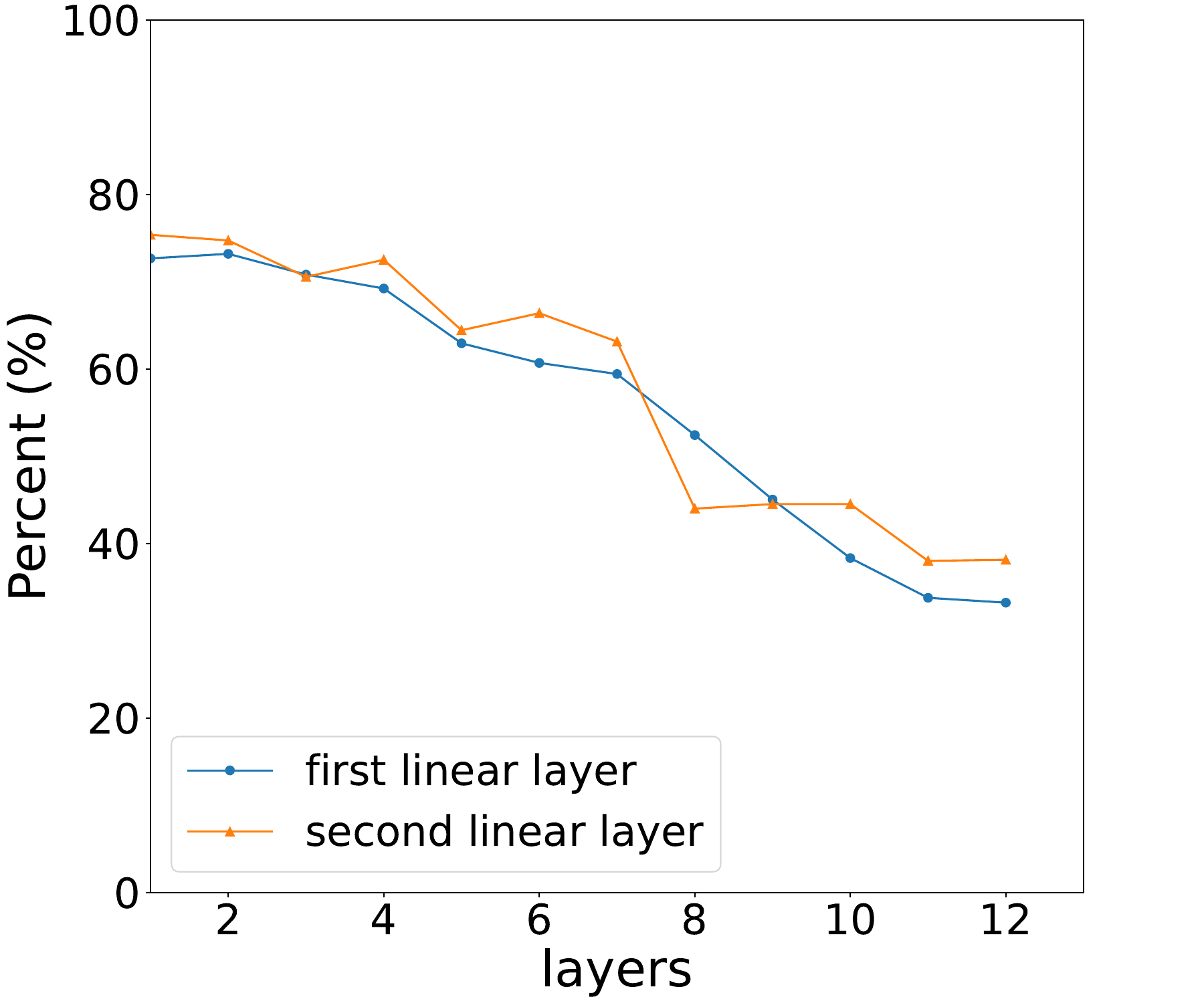}
  \caption{BEATs-finetuned, $\tau = 12$}
\end{subfigure}%
\hfill
\begin{subfigure}{.3\textwidth}
  \centering
  \includegraphics[width=\linewidth]{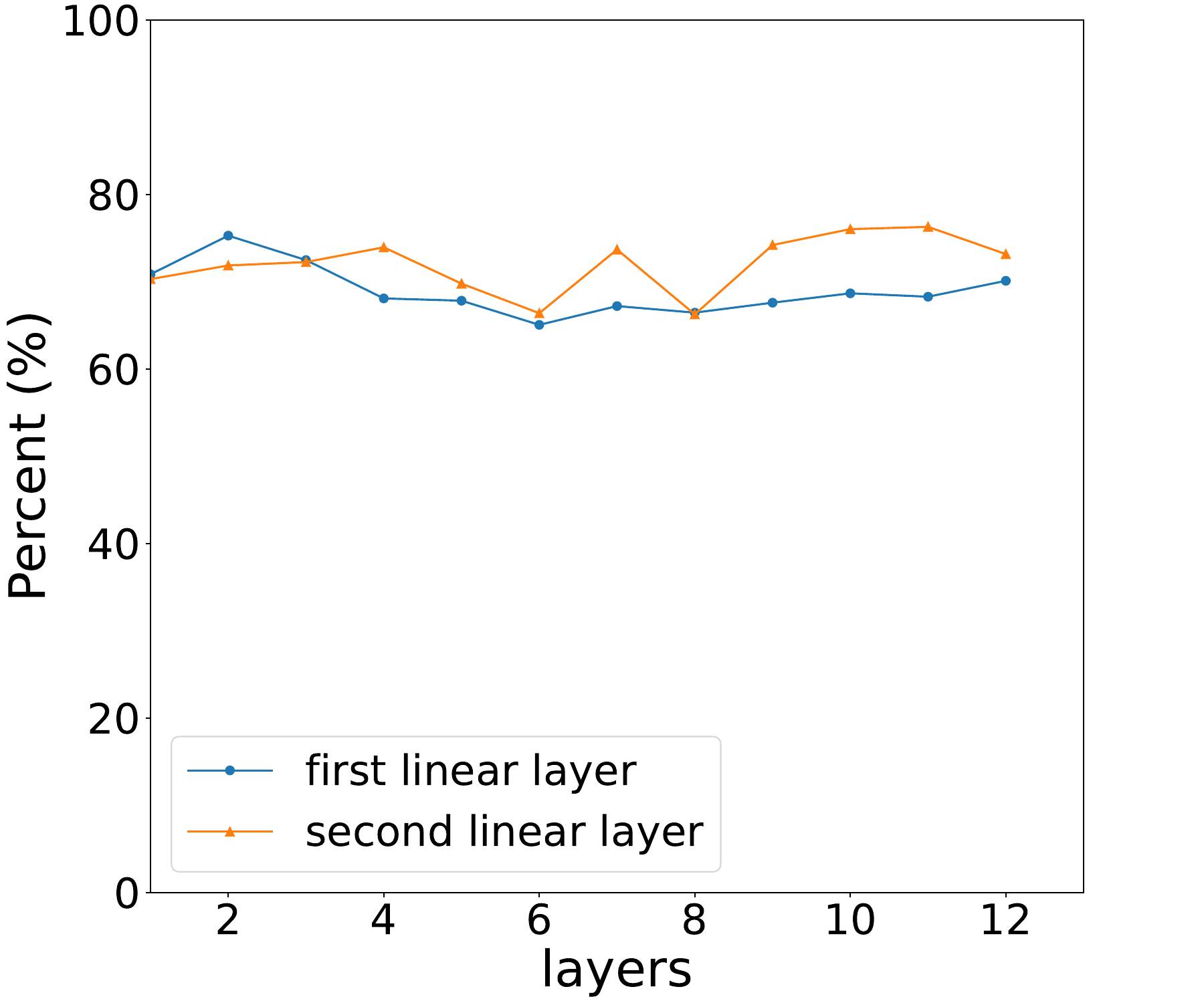}
  \caption{BEATs-frozen, $\tau = 12$}
\end{subfigure}

\vspace{30px}

\begin{subfigure}{.3\textwidth}
  \centering
  \includegraphics[width=\linewidth]{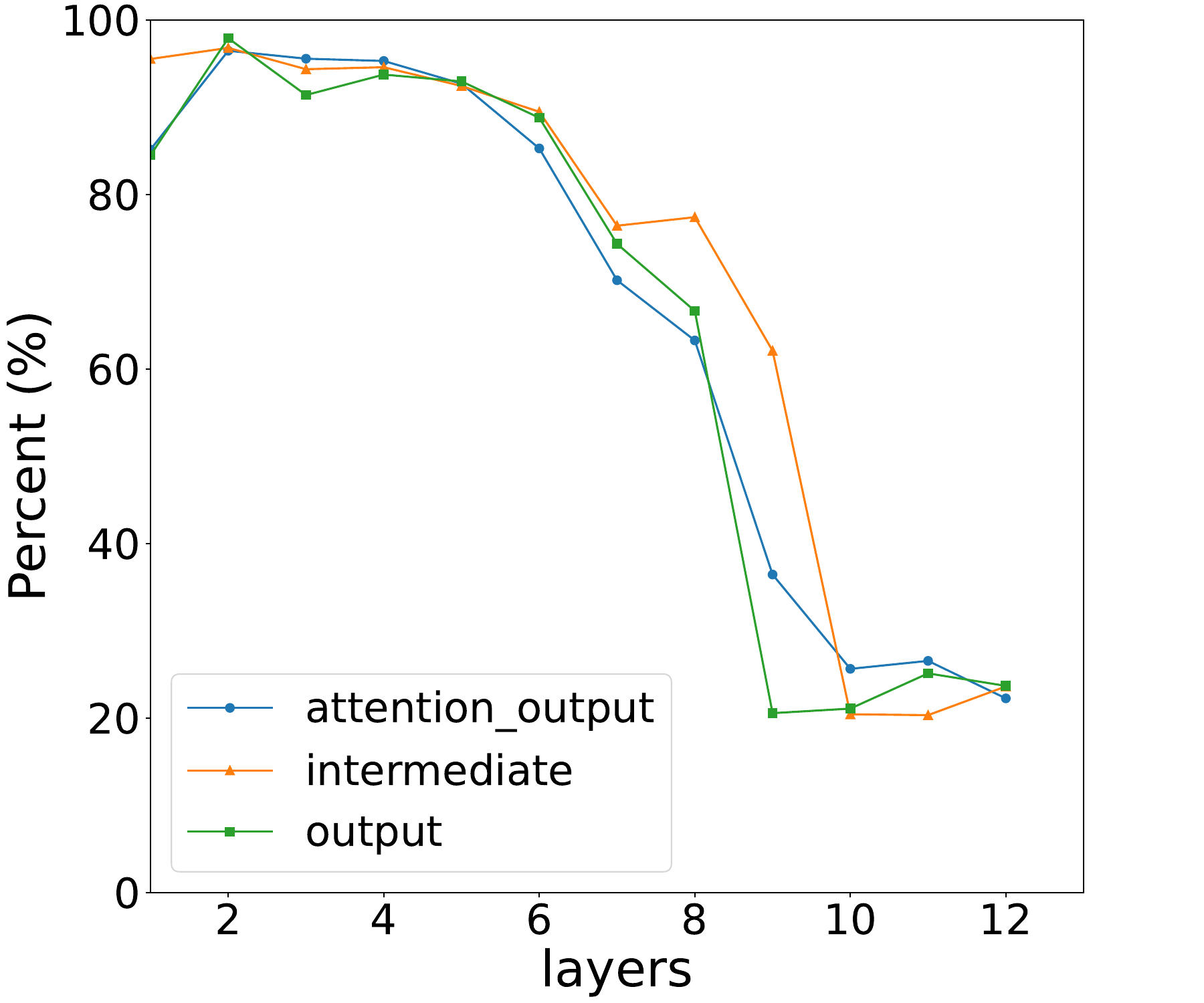}
  \caption{AST, $\tau = 16$ }
\end{subfigure}%
\hfill
\begin{subfigure}{.3\textwidth}
  \centering
  \includegraphics[width=\linewidth]{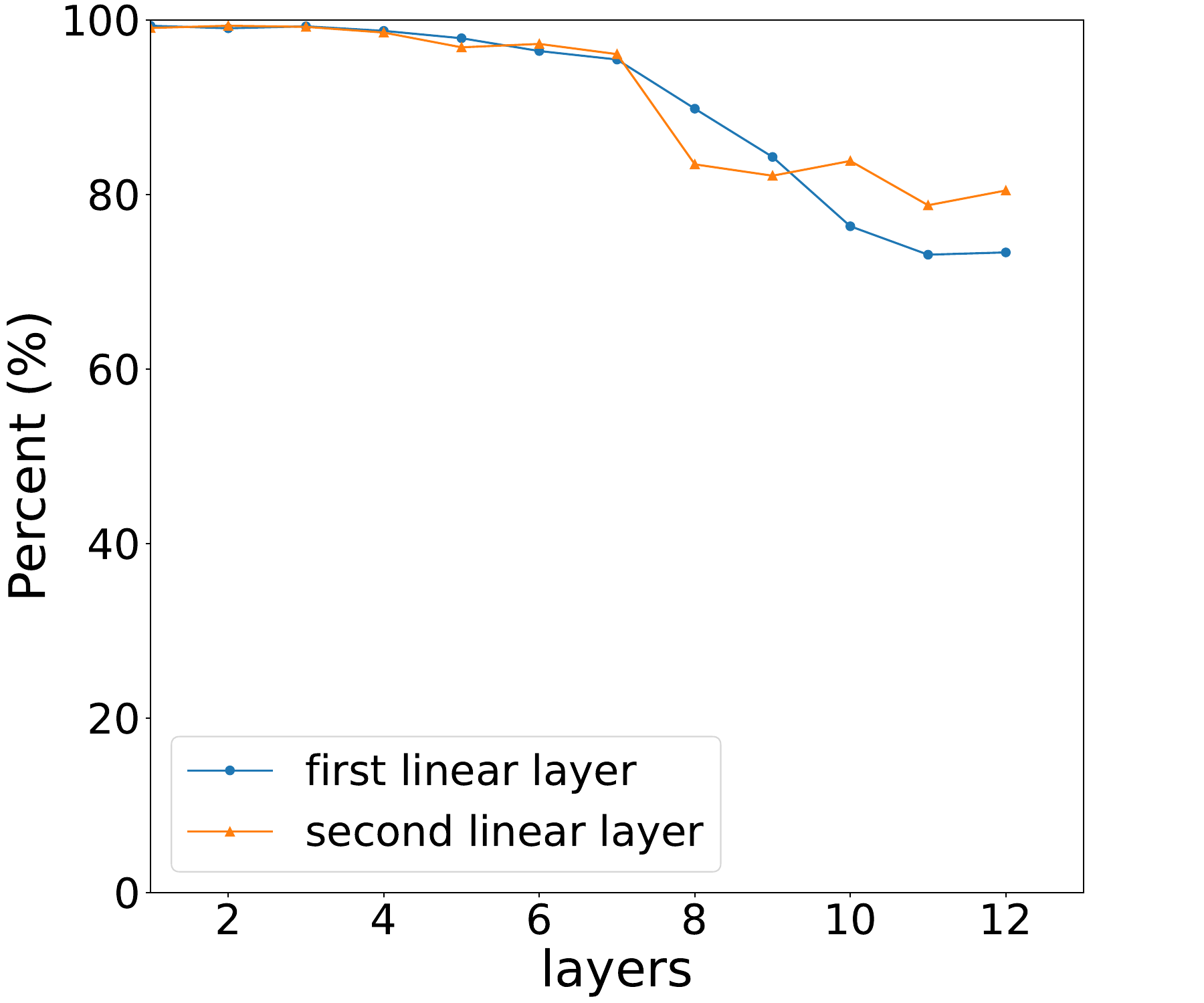}
  \caption{BEATs-finetuned, $\tau = 16$}
\end{subfigure}%
\hfill
\begin{subfigure}{.3\textwidth}
  \centering
  \includegraphics[width=\linewidth]{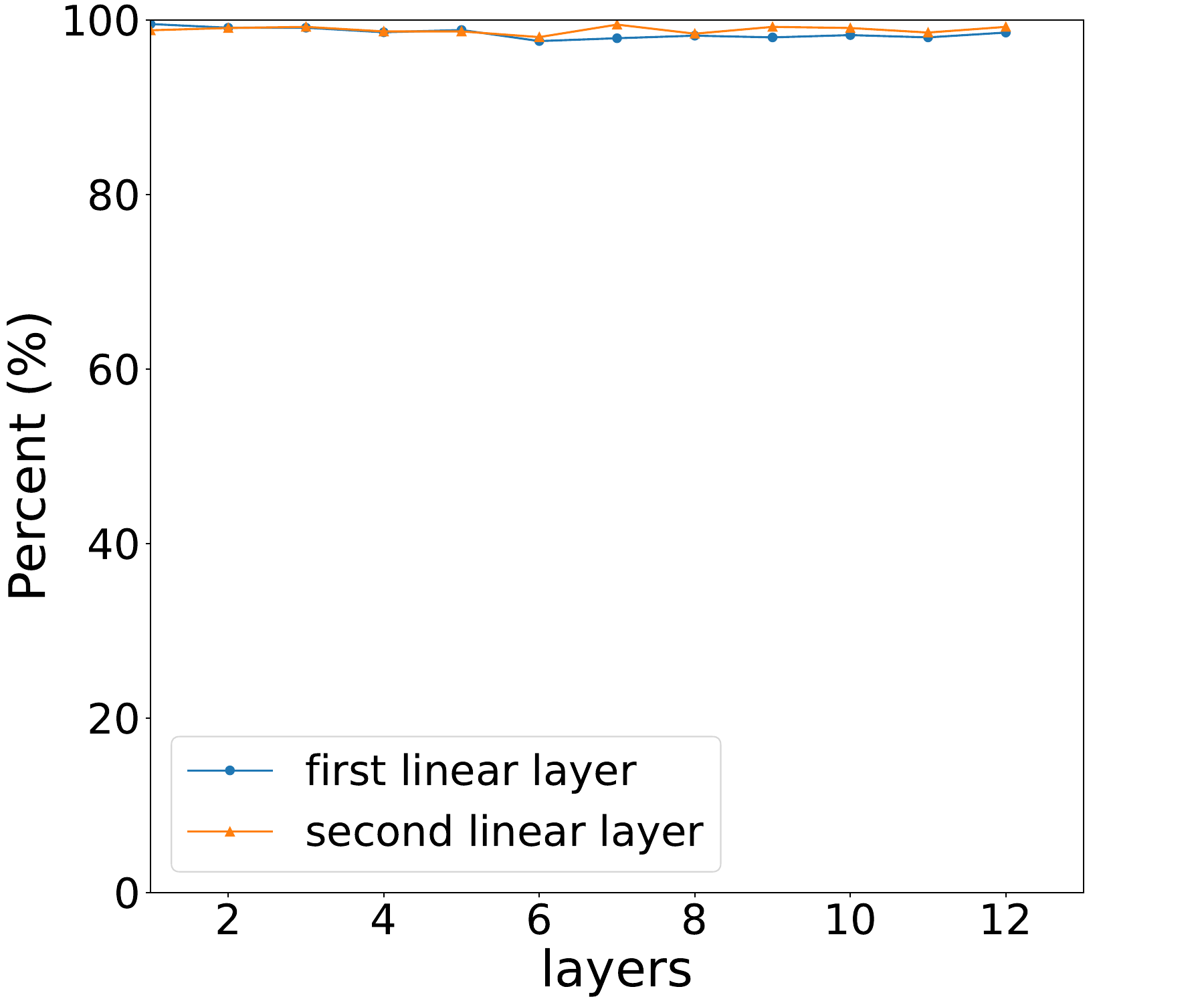}
  \caption{BEATs-frozen, $\tau = 16$}
\end{subfigure}

\vspace{30px}
\begin{subfigure}{.3\textwidth}
  \centering
  \includegraphics[width=\linewidth]{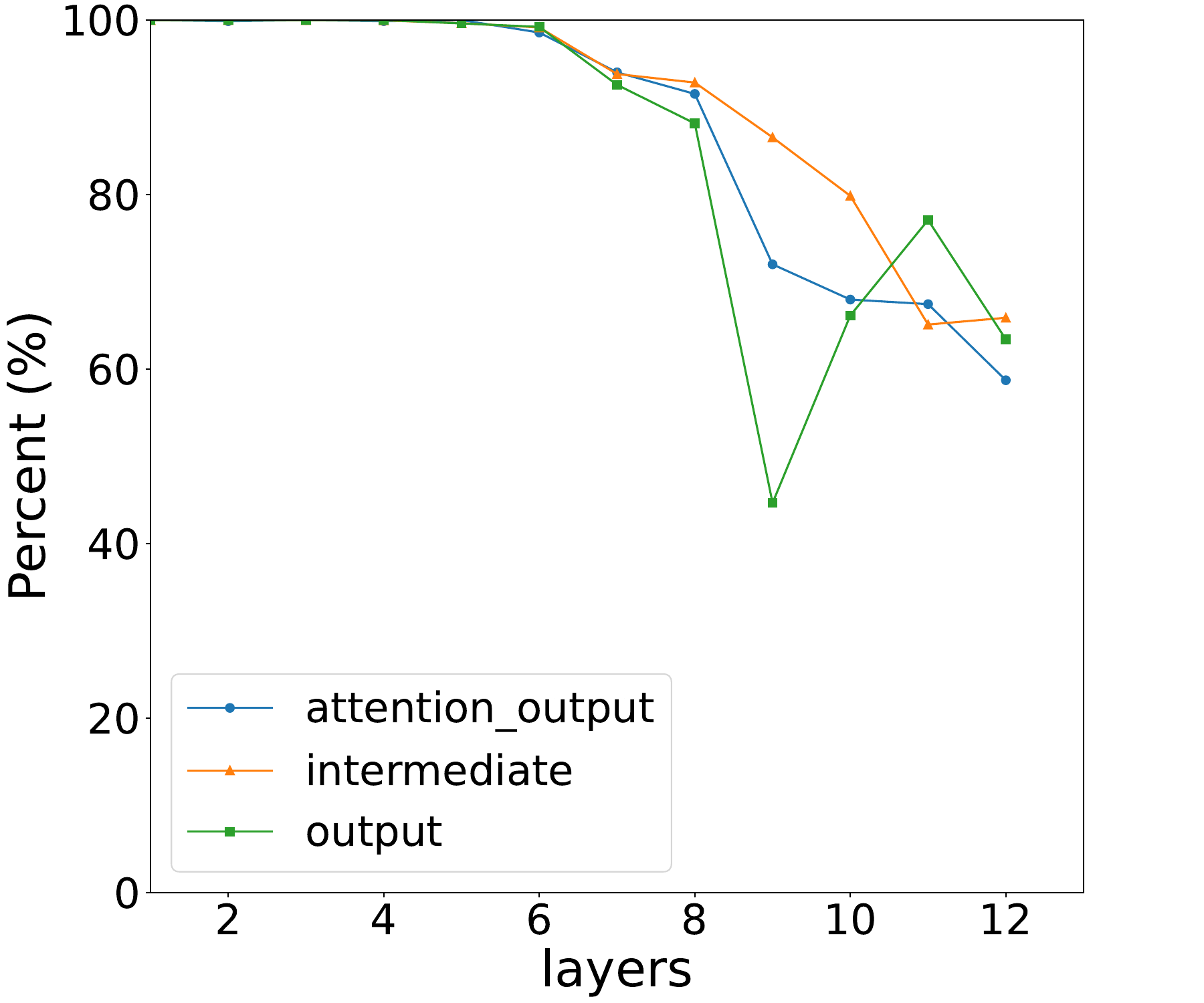}
  \caption{AST, $\tau = 20$ }
\end{subfigure}%
\hfill
\begin{subfigure}{.3\textwidth}
  \centering
  \includegraphics[width=\linewidth]{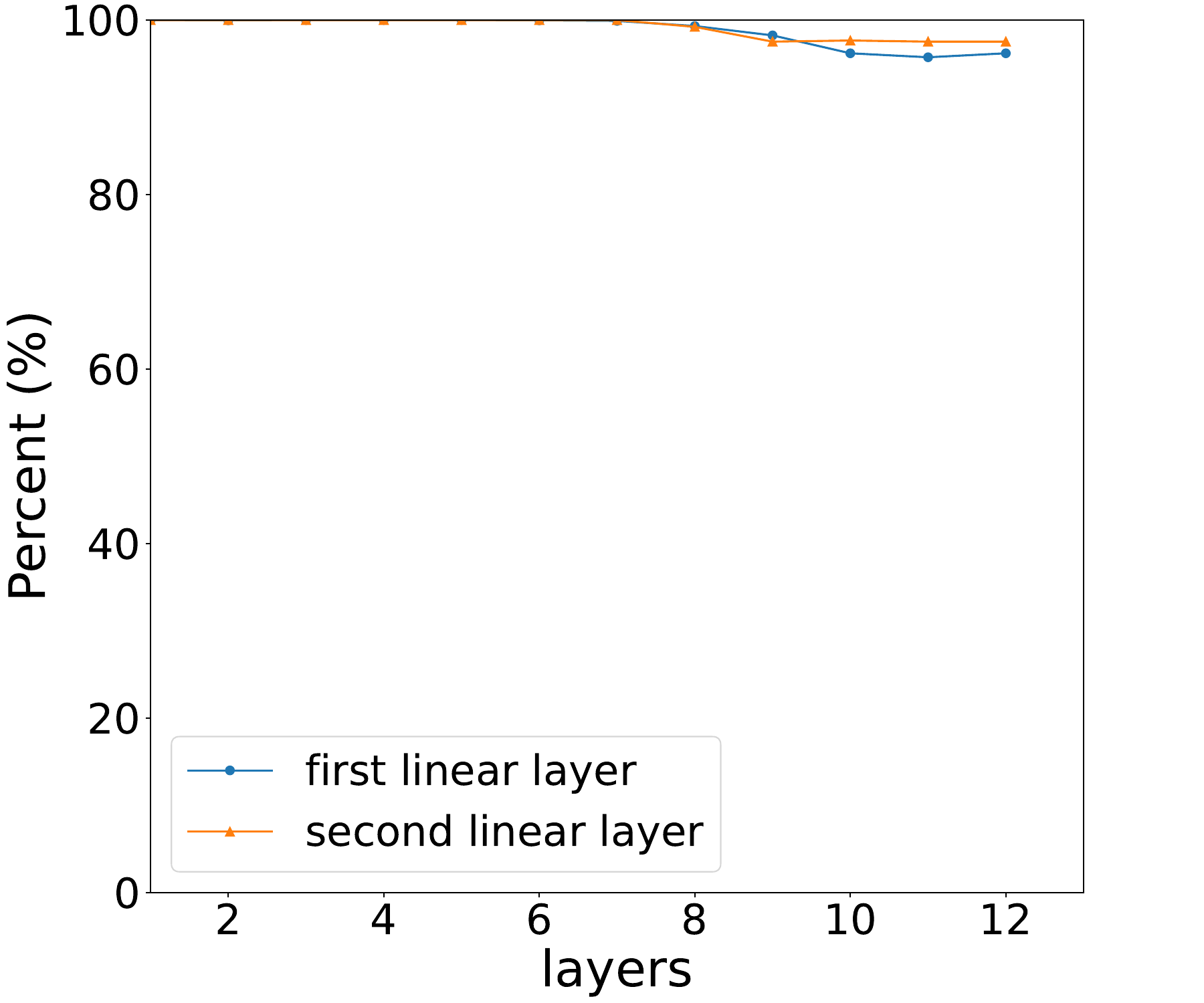}
  \caption{BEATs-finetuned, $\tau = 20$}
\end{subfigure}%
\hfill
\begin{subfigure}{.3\textwidth}
  \centering
  \includegraphics[width=\linewidth]{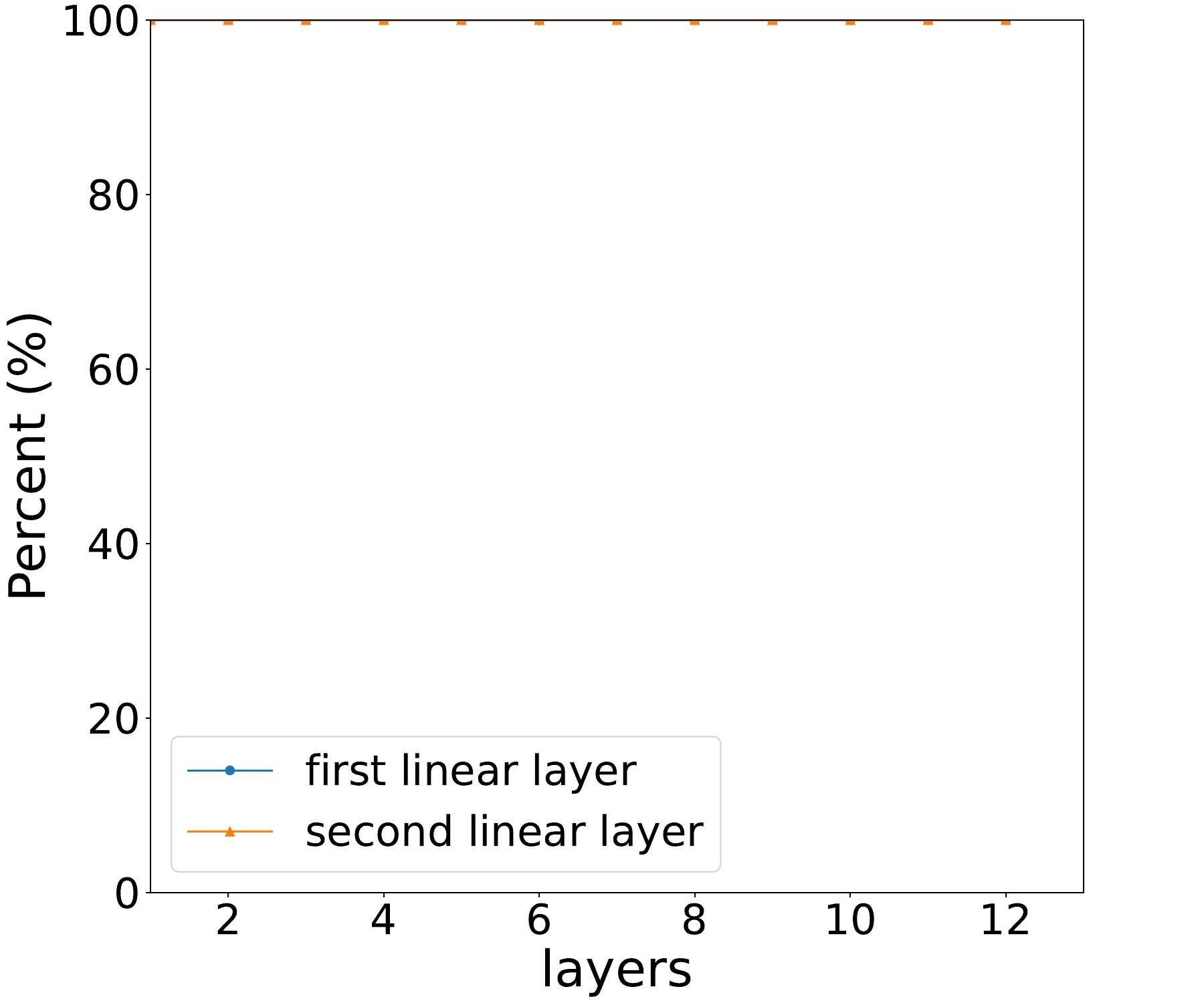}
  \caption{BEATs-frozen, $\tau = 20$}
\end{subfigure}
\caption{Percentage of polysemantic neurons when adopting $\tau = 12, 16, 20$ and top-$K = 20$.}
\label{fig:hyperparameters_top20}
\end{figure}

In \algoname{AND}, a threshold $\tau$ introduced in~\cref{subsec: Training Strategy Affects Neuron Interpretability} is used to help determine the polysemanticity of neurons. Also, there are some uncertainties about whether selecting top-$K$ highly activated samples causes different trends. Hence, this section conducts further experiments to evaluate the impact of varying $\tau$ and top-$K$. The results are shown in~\cref{fig:hyperparameters_top5},~\cref{fig:hyperparameters_top10}, and~\cref{fig:hyperparameters_top20}, with top-5, top-10, and top-20 samples selected, respectively. Although the percentage of polysemantic neurons varies with different $\tau$ and top-$K$ values, consistent trends as in~\cref{subsec: Training Strategy Affects Neuron Interpretability} are observed.

\vspace{10px}

\vspace{10px}

\vspace{10px}

\clearpage

\subsection{Neuron Interpretability under Different Training Strategies - GTZAN Music Genre}
\label{sup subsec: gtzan neuron Interpretability}
\begin{figure}[ht]
\vspace{10px}
\begin{subfigure}{.3\textwidth}
  \centering
  \includegraphics[width=\linewidth]{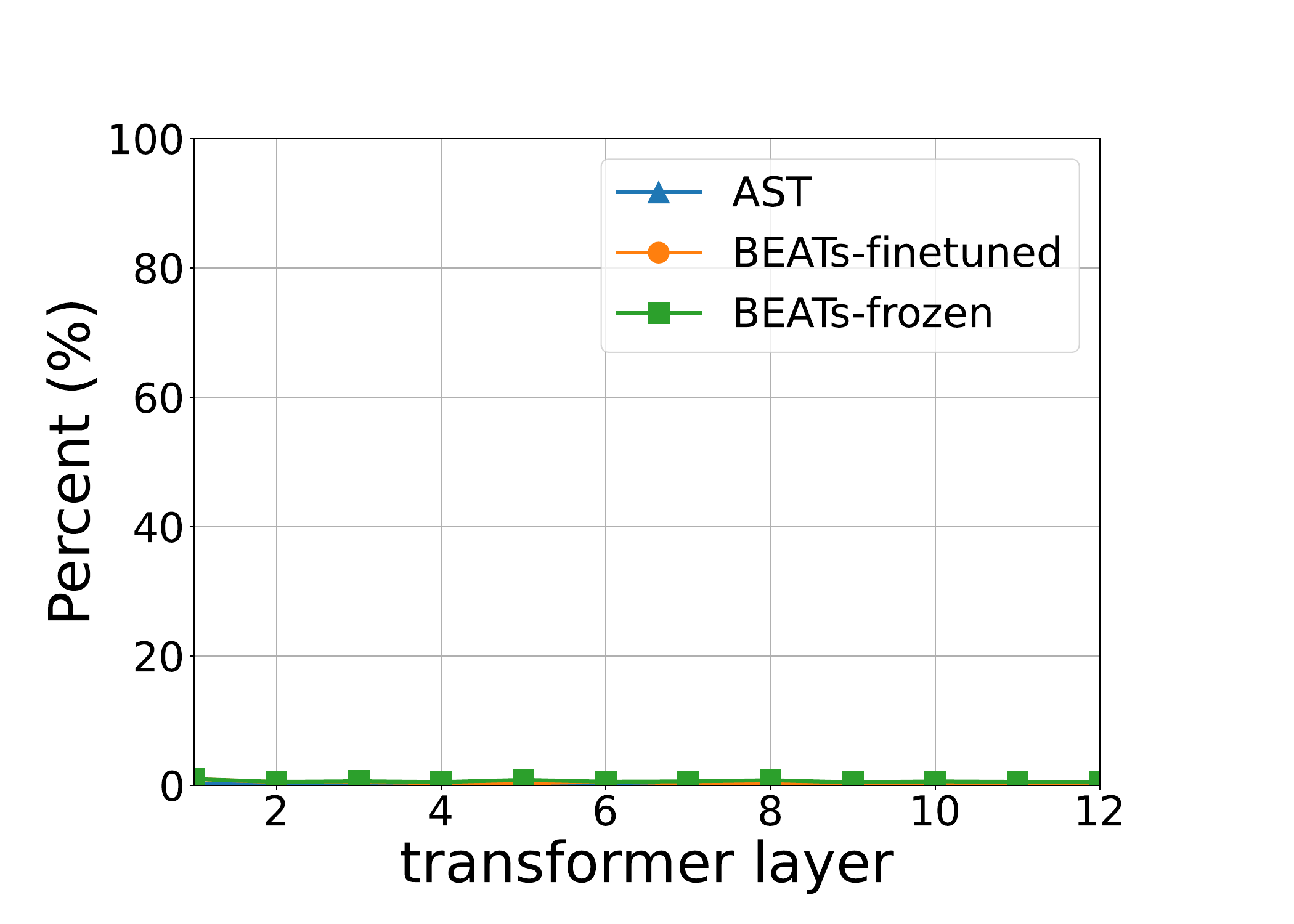}
  \caption{$\tau = 3$ }
\end{subfigure}%
\hfill
\begin{subfigure}{.3\textwidth}
  \centering
  \includegraphics[width=\linewidth]{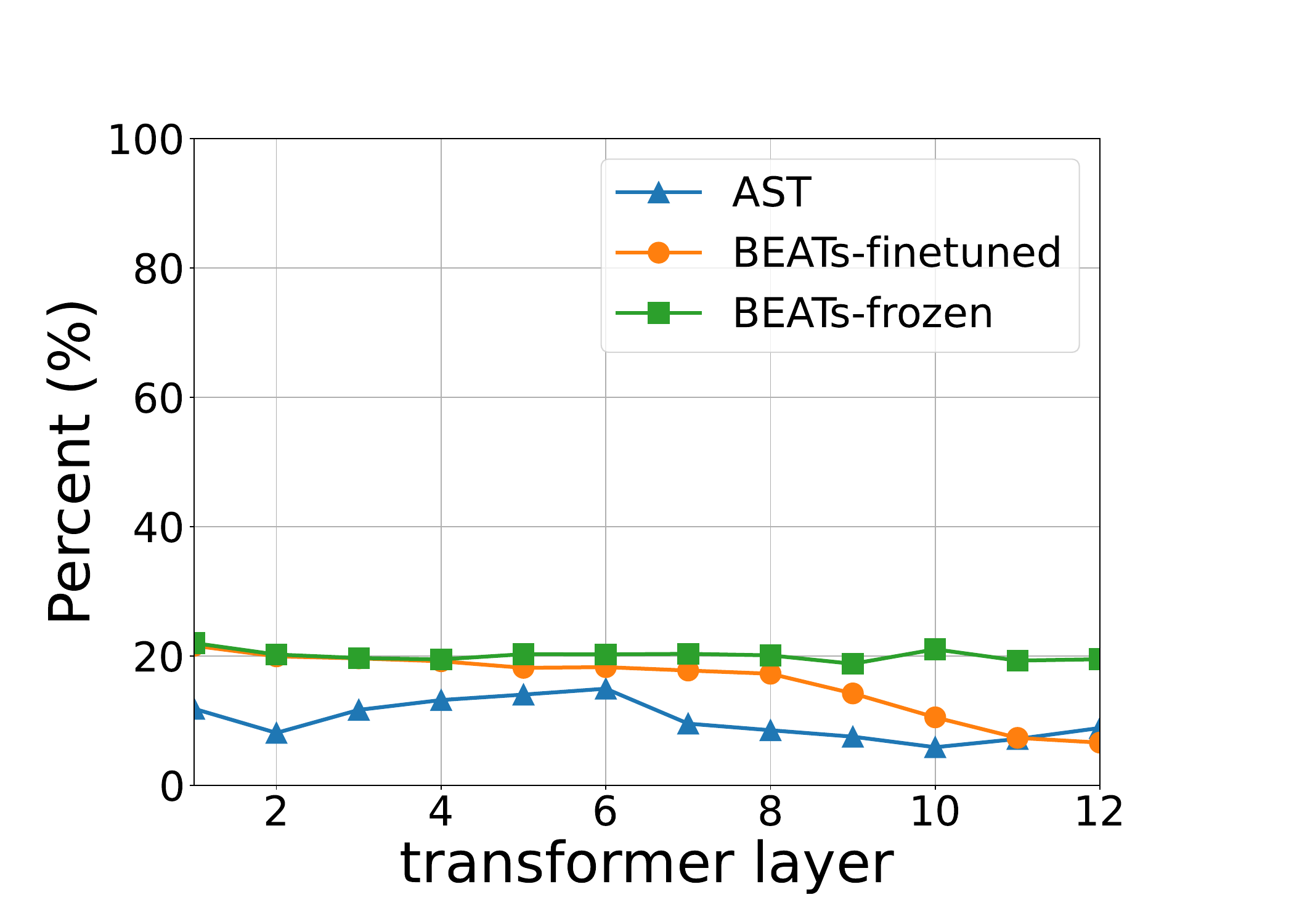}
  \caption{$\tau = 4$}
\end{subfigure}%
\hfill
\begin{subfigure}{.3\textwidth}
  \centering
  \includegraphics[width=\linewidth]{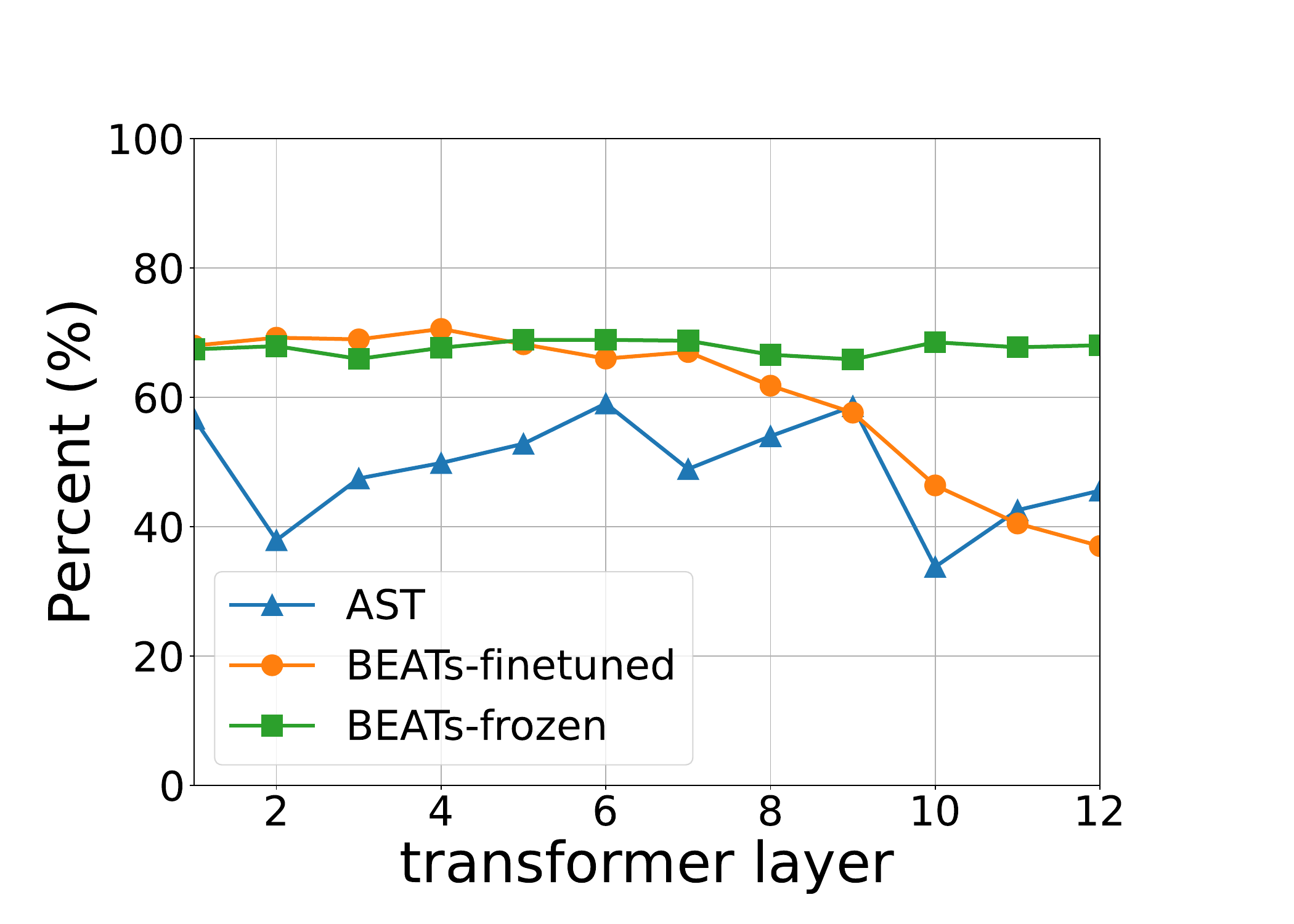}
  \caption{$\tau = 5$}
\end{subfigure}
\caption{Percentage of polysemantic neurons when adopting $\tau = 3, 4, 5$ and top-$K = 5$, with GTZAN Music Genre being the training dataset and probing dataset of AST, BEATs-finetuned, and BEATs-frozen.}
\label{fig: gtzan_results}
\end{figure}

This section conducts the neuron interpretability experiments as in~\cref{subsec: Training Strategy Affects Neuron Interpretability} but adopts GTZAN Music Genre~\cite{1021072} dataset as the probing dataset and training dataset of AST, BEATs-finetuned, and BEATs-frozen. The results are shown in~\cref{fig: gtzan_results}. Since the GTZAN Music Genre is a simpler dataset with only 10 classes, neuron behaviors are more explainable when responding to samples in this dataset, with no more than 20\% neurons classified as ``uninterpretable'' for $\tau = 4$. Strengthening the criterion to $\tau = 5$ produces a clearer trend, with BERTs-frozen being always diverse, and AST being gradually concentrating, coinciding with the findings on the ESC50 dataset.

